\documentclass[twocolumn,aps,prd,preprintnumbers,showpacs,superscriptaddress,nofootinbib,amsmath,amssymb,floats,floatfix,showkeys,notitlepage]{revtex4}

\usepackage{graphicx}
\usepackage{dcolumn}
\usepackage{bm}
\usepackage{color}
\usepackage[normalem]{ulem} 
\usepackage[dvipsnames]{xcolor} 
\usepackage{palatino}
\usepackage{changes}
\usepackage{hyperref}
\hypersetup{colorlinks=true,linkcolor=blue,urlcolor=blue,citecolor=blue}
\usepackage{mathrsfs}
\usepackage[utf8]{inputenc}
\usepackage{mathtools}
\usepackage{doi}
\usepackage{amsmath}
\usepackage{amssymb}

\begin{document}

\title{Quasiperiodic oscillations, weak field lensing and shadow cast around\\ black holes in symmergent gravity}

\author{Javlon Rayimbaev}
\email{javlon@astrin.uz}
\affiliation{Ulugh Beg Astronomical Institute, Astronomy St. 33, Tashkent 100052, Uzbekistan}
\affiliation{Akfa University, Milliy Bog Street 264, Tashkent 111221, Uzbekistan}
\affiliation{Tashkent Institute of Irrigation and Agricultural Mechanization Engineers, Kori Niyoziy 39, Tashkent 100000, Uzbekistan}
\affiliation{Tashkent State Technical University, Tashkent 100095, Uzbekistan}
\affiliation{National University of Uzbekistan, Tashkent 100174, Uzbekistan}
\author{Reggie C. Pantig}
\email{reggie.pantig@dlsu.edu.ph}
\affiliation{Physics Department, De La Salle University, 2401 Taft Avenue, Manila, 1004 Philippines}

\author{Ali \"Ovg\"un}
\email{ali.ovgun@emu.edu.tr}
\affiliation{Physics Department, Eastern Mediterranean University, Famagusta, 99628 North Cyprus via Mersin 10, Turkey}

\author{Ahmadjon~Abdujabbarov}
\email{ahmadjon@astrin.uz}
\affiliation{Shanghai Astronomical Observatory, 80 Nandan Road, Shanghai 200030, China}
\affiliation{Ulugh Beg Astronomical Institute, Astronomy St.  33, Tashkent 100052, Uzbekistan}
\affiliation{Tashkent Institute of Irrigation and Agricultural Mechanization Engineers, Kori Niyoziy 39, Tashkent 100000, Uzbekistan}
\affiliation{National University of Uzbekistan, Tashkent 100174, Uzbekistan}
\affiliation{Institute of Nuclear Physics, Ulugbek 1, Tashkent 100214, Uzbekistan}

\author{Durmu\c{s}~Demir}
\email{durmus.demir@sabanciuniv.edu }
\affiliation{Faculty of Engineering and Natural Sciences, Sabanc{\i} University, 34956 Tuzla, \.{I}stanbul, Turkey}

\begin{abstract}
In this work, we perform a systematic study of the symmergent gravity in a black hole environment. The symmergent gravity, an emergent gravity model in which gravity emerges in a way restoring the gauge symmetries and stabilizing the Higgs boson mass, possesses a quadratic-curvature term with a loop-induced coupling proportional to the boson-fermion number difference. We investigated boson-fermion number difference and black hole parameters in symmergent gravity by utilizing the values of various observables. In this regard, we investigated particle dynamics and obtained Keplerian frequencies describing the harmonic oscillations. We determined quasiperiodic oscillations about such orbits by utilizing relativistic precession, warped disk, and epicyclic resonant models. We studied weak deflection angle and planetary perihelion shift and determined bounds on symmergent gravity parameters. We computed shadow radius for both static and co-moving observers near and far from the symmergent black hole and revealed the sensitivity of these observers to the model parameters. We conclude the work by giving an overall discussion of the bounds, and giving future prospects concerning other possible analyses of the symmergent gravity. 

\end{abstract}

\date{\today}

\maketitle


\section{Introduction}
Black holes, whose horizons are defined as the shell of the points of no return, are one of the most fundamental predictions of general relativity (GR). The first direct detection of gravitational waves in 2015 by the LIGO-Virgo collaboration \cite{LIGOScientific:2016aoc} and the first-ever observation of an image of the black hole in 2019 by the Event Horizon Telescope (EHT) \cite{Event2021a,Event2021b} signaled the start of a new era in gravitational physics. These experiments open up a new methodology to test theories of gravity in the strong field (high curvature) domains.

In the present work, we will test a recent emergent gravity model in appropriate black hole spacetimes. In general, the reconciliation of quantum matter and gravitation is a thorny fundamental problem, and one plausible approach is provided by emergent gravity models \cite{sakharov,visser,verlinde}. Indeed, quantum field theories (QFT) by formulation are intrinsic to flat spacetime \cite{incompatible,wald} because they rest on a unique Poincar\'e-invariant vacuum state. Flat spacetime means the sheer absence of gravity. The incorporation of gravity requires QFT to be taken to curved spacetime. This, however, is simply impossible because Poincar\'e symmetry gets broken in curved geometry and special states like vacuum get banned by general covariance \cite{dyson,wald}. In view of this impasse and view also of the absence of a quantum theory of gravity \cite{thooft}, one is left with essentially emergent gravity \cite{sakharov,visser} as the most plausible option. In recent years, Sakharov's setup has been extended by the inclusion of gauge sector  \cite{demir1}, and it has been shown \cite{demir1}, that gravity could emerge in a way erasing loop-induced gauge boson masses and stabilizing scalar masses, which leads to the naturalization of the effective field theories regarding their sensitivities to high scales \cite{weinberg,eff-action2}. This mechanism leads to a new framework \cite{demir1,demir2} in which one obtains the usual Einstein gravity (Einstein-Hilbert term) plus quadratic curvature terms plus a ${\overline{MS}}$-renormalized QFT. It possesses sui generis features that can be searched at the collider, astrophysical and cosmological experiments and observations. This emergent gravity, termed as ``symmergent gravity" for its gauge symmetry-restoring nature \cite{demir2,demir3}, has all its couplings induced by the flat spacetime quantum loops. Symmergent gravity is not an effective field theory constructed in curved spacetime \cite{visser,birrel}. It is, in fact, flat spacetime effective field theory taken to curved spacetime in a way restoring gauge symmetries \cite{demir1,demir2}. Its various couplings are related to the flat spacetime QFT so that its curvature sector enables one to extract information about the particle spectrum, mass spectrum, and other details of the QFT.  Since the new particles do not have to couple to the already known particles, one ends up with a wide spectrum of masses and couplings \cite{demir1,demir4}. As a matter of fact, quadratic curvature terms in symmergent gravity, different from those in the other approaches \cite{visser,birrel}, have been shown to lead to successful Starobinsky inflation \cite{irfan}. In addition, the same quadratic curvature terms have recently been shown to give distinctive signatures on black hole shadow \cite{symmergent-bh}. In the present work, we shall probe the symmergent gravity model in the quasi-periodic oscillations, weak field lensing, and shadow cast around the black holes. Our work can be viewed are ``doing particle physics via black hole properties" as it will reveal salient features of the QFT sector through various black hole features. 

Since the verification of the GR as the relativistic theory of gravity by Eddington in 1919,  gravitational lensing has remained one of the main tools for observing and understanding various phenomena in astronomy and cosmology (like galaxies, dark matter, dark energy and the evolution of the Universe \cite{Bozza:2002af}). Over the decades, various methods have been devised to calculate the deflection angle by black holes \cite{Virbhadra:1999nm,Virbhadra:2002ju,Bozza:2001xd,Bozza:2002zj,Hasse:2001by,Perlick:2003vg,He:2020eah,Bozza2008}. In 2008, Gibbons and Werner devised a different method to determine the deflection angle in weak fields using the Gauss-Bonnet theorem (GBT) on the optical geometries in asymptotically flat spacetimes \cite{Gibbons:2008rj}. To achieve this, the GBT needs to be integrated into an infinite domain bounded by the light ray. This method has been applied to various related phenomena  \cite{Ovgun:2018fnk,Ovgun:2019wej,Ovgun:2018oxk,Javed:2019kon,Javed:2019rrg,Javed:2019ynm,Javed:2020lsg,Javed:2019qyg,Ovgun:2018fte,Javed:2019jag}. In  Werner's work \cite{Werner_2012}, for instance, the GBT method for deflection angle was extended to stationary spacetimes in the Finsler-Randers type optical geometry on Nazim's osculating Riemannian manifolds. Following this, Ishihara and others have found a way to apply the GBT method to finite-distances differently than the use of asymptotic receiver and source \cite{Ishihara:2016vdc,Ishihara:2016sfv}. Next, Ono and others have extended Ishihara method to axisymmetric spacetimes \cite{Ono:2017pie}, with further applications to   various non-asymptotic spacetimes, such as those with dark matter contributions in their metric function \cite{Pantig:2020odu,Pantig2022,Pantig2022a,Pantig2022b}. Finally, the finite distance method has been studied by Li and others by using massive particles and Jacobi-Maupertuis Randers-Finsler metric within the GBT \cite{Li:2020dln,Li:2020wvn}.

Recently, an eloquent step has been taken to probe black hole models in modified gravity theories with the EHT, which has showcased the first image (a 2D dark zone) of the supermassive black hole M87* \cite{Event2021a,Event2021b}.  Black holes have a dark region as seen from the observer in the celestial sky, which is known as the BH shadow. It can reveal a fine structure near the horizon of the black hole, and can therefore be used to test gravity theories in the strong gravity regime. Black hole shadow was first studied by Synge in 1966 for the specific case of the Schwarzschild black hole \cite{Synge1966}. After Synge, in 1979, Luminet gave the formula for the angular radius of the shadow \cite{Luminet1979}. It is well known that the shadow of a non-rotating black hole is just a standard circle. The shadow of a rotating black hole, on the other hand,  is an elongated shape in the direction of rotation. To this end,  theoretically-predicted shadow sizes of various black hole solutions have been contrasted with the 2017 measurements of M87* and various constraints have been put using the data. The shadows of modified gravity black holes are smaller and distorted compared to the shadow of the Kerr black hole. Black hole properties have been analysed by various recent works which focus on the black hole shadow cast   \cite{Tsupko_2020,Hioki2009,Dymnikova2019,Wei2019,Xu2018a,Hou:2018avu,Bambi2019,Konoplya2019,Pantig2021,Okyay:2021nnh,Uniyal:2022vdu,Kuang:2022xjp,Kumar:2020hgm,Belhaj:2020rdb,Li2020,Ovgun:2020gjz,Ovgun:2019jdo,Ovgun:2018tua,Ovgun:2020gjz,Ling:2021vgk,Belhaj:2020okh,Abdikamalov:2019ztb,Abdujabbarov:2016efm,Atamurotov:2015nra,Papnoi:2014aaa,Abdujabbarov:2012bn,Atamurotov:2013sca,Cunha:2018acu,Gralla:2019xty,Perlick:2015vta,Nedkova:2013msa,Li:2013jra,Cunha:2016wzk,Johannsen:2015hib,Johannsen:2015mdd,Shaikh:2019fpu,Allahyari:2019jqz,Yumoto:2012kz,Cunha:2016bpi,Moffat:2015kva,Cunha:2016bjh,Zakharov:2014lqa,Tsukamoto:2017fxq,Hennigar:2018hza,Chakhchi:2022fls,pantig2022testing,Pantig2020b}.

There appears a series of peaks in the radio to X-ray band of the electromagnetic spectrum from astrophysical objects. These peaks are signatures of the  quasi-periodic oscillations (QPOs). One observes QPOs with more than two peaks in low-mass X-ray binaries (LMXBs) such that at least one compact object in the binary is a neutron star \cite{Ingram2010MNRAS,Schaab1999MNRAS,Torok2005AA}. In particular, most of the microquasars reveal twin-peak QPOs in their electromagnetic spectrum. The nature of these peaks is related to the process of matter accretion into the compact objects \cite{Ingram2016MNRAS,Stuchlik2013AA,Stella1998ApJL,Rezzolla_qpo_03a}. It turns out that the twin-peak QPOs and QPOs with more than two peaks are different in nature. In fact, various proposed models describe them as two types of QPOs \cite{Rezzolla_qpo_03b,Germana2017PhRvD}. The observation and analysis of the twin-peak QPOs in microquasars could be a useful tool to test different properties of neutron stars and black holes such as their magnetic fields, accretion disk properties, and the underlying gravity models \cite{Torok2019MNRAS,Zdunik2000AA,Klis2000ARAA}.  On the other side, depending on the frequencies, one may distinguish two types of QPOs: {\it (i)} high-frequency (HF) QPOs corresponding to the frequency range from 0.1 to 1 {\rm kHz},  and {\it (ii)} low-frequency (LF) QPOs corresponding to the frequency less than 0.1 {\rm kHz}. 

It is worth noting that there have been numerous proposals for describing the QPOs. Despite this, those proposals are not able to take into account all the astrophysical variables in the description of QPOs. Numerous observations and their detailed analyses indicate that QPOs are related to the dynamics of the test particles around the black holes, in particular, to their harmonic oscillations in the radial, vertical and azimuthal directions. Analyzing the dynamics of these test particles, especially their orbital motions and oscillations, one may test the gravity theories as well as spacetime properties around compact objects through observational data on QPOs~ \cite{Bambi17e,Stuchlik2015MNRAS, Silbergleit2001ApJ,Wagoner2001ApJL,Rayimbaev2022PDU,Vrba2021Univ,Vrba2021EPJP,Vrba2021JCAP,Rayimbaev2021EPJCQPO,Rayimbaev2021GalaxQPO,Rayimbaev2021PhRvDQPO}. Studies of models of QPOs in view of the orbital motions of the test particles enable the study of the innermost stable circular orbit (ISCO) of the particles~\cite{Stuchlik2011AA,Torok2011AA,Rayimbaev2021Galax}. 

In this paper, as mentioned before, we will perform a detailed test study of the symmergent gravity in the strong field (high-curvature) domains using the EHT observations \cite{Event2021a,Event2021b}. We will determine, in particular, the implications of the symmergent gravity setup for the quasi-periodic oscillations, weak field lensing, and shadow cast around the black holes. The paper is organized as follows: In Sec. \ref{sec2}, a summary is given of the symmergent gravity model with emphasis on its gravity (curvature) sector and associated model parameters. Sec. \ref{sec3} is devoted to study of the dynamics of test particles. The fundamental frequencies due to Keplerian orbits and harmonic oscillations are discussed in Sec. \ref{sec4}. In Sec. \ref{sec5}, orbits of quasiperiodic oscillations are explored, and the results of this exploration are used in Sec. \ref{sec6} to determine the constraints on the parameters of the symmergent gravity. Further examination of the effects of these parameters is performed through the study of the weak deflection angle in Sec. \ref{sec7}, and the shadow cast in Sec. \ref{sec8}. We close the paper with a Conclusion section in which we give a summary and discuss future prospects regarding the topics covered.   

En passant, we note that in this paper we use geometrized (natural) units $G=1$, $c=1$ and  ($-,+,+,+$)  metric signature.

\section{Black holes in symmergent gravity} \label{sec2}
Symmergent gravity is a quadratic-curvature gravity. It is a subclass of $f(R)$ gravity theories. This special emergent gravity theory has been discussed in detail in \cite{demir1} (see also \cite{demir2,demir3} for earlier analyses). The latest detailed discussion of the symmergent gravity was given in \cite{symmergent-bh} while analyzing its implications for black hole shadow. This introductory black hole article gives all the basic aspects of symmergent gravity. Below, we summarize symmergent gravity in view of the required parameters and relevant features of the model (and kindly refer the reader to \cite{demir1,symmergent-bh} for other details). 

The symmergent gravity is governed by the action 
\begin{eqnarray}
\label{curvature-sector}
\int d^4x \sqrt{-g}\left\{
\frac{R(g)}{16\pi G} +\frac{c_S}{4} S^2 R(g) - \frac{c_\text{O}}{16} R(g)^2\right\}
\end{eqnarray}
in which $R(g)$ is the scalar curvature and $c_i$'s, and $G_N$ are loop induced quantities, that is, quantities produced by flat spacetime matter loops. Here $S$ is a generic scalar field, including the SM Higgs field. Two relevant parameters are given by, 
\begin{eqnarray}
\label{params}
\frac{1}{G} = \frac{{\rm str}\left[{\mathcal{M}}^2\right]}{8 \pi}\ ,\;\; c_\text{O} = \frac{n_\text{B} - n_\text{F}}{128 \pi^2}\ ,
\end{eqnarray}
where $n_\text{B}$ ($n_\text{F}$) stands for the total number of bosons (fermions) in Nature. The $n_\text{B}$ bosons plus $n_\text{F}$ fermions involve the known particles (quarks, leptons, photon, gluon and weak bosons) as well as completely new particles (massive as well as massless) that do not have to couple to the known particles non-gravitationally. The graded trace  ${\rm str}[{\mathcal{M}}^2] = \sum_s (-1)^s\, {\rm tr}[{\mathcal{M}}^2]_s$, $s$ being the particle spin, is a sum over the mass-suqared matrix ${\mathcal{M}}^2$ of fields. The constant $c_S$ is around 0.019 when $S$ is the usual Higgs field \cite{demir1,demir2}, and can take on different values for different scalars beyond the SM. 

\begin{figure}[ht!]
   \centering
\includegraphics[width=\linewidth]{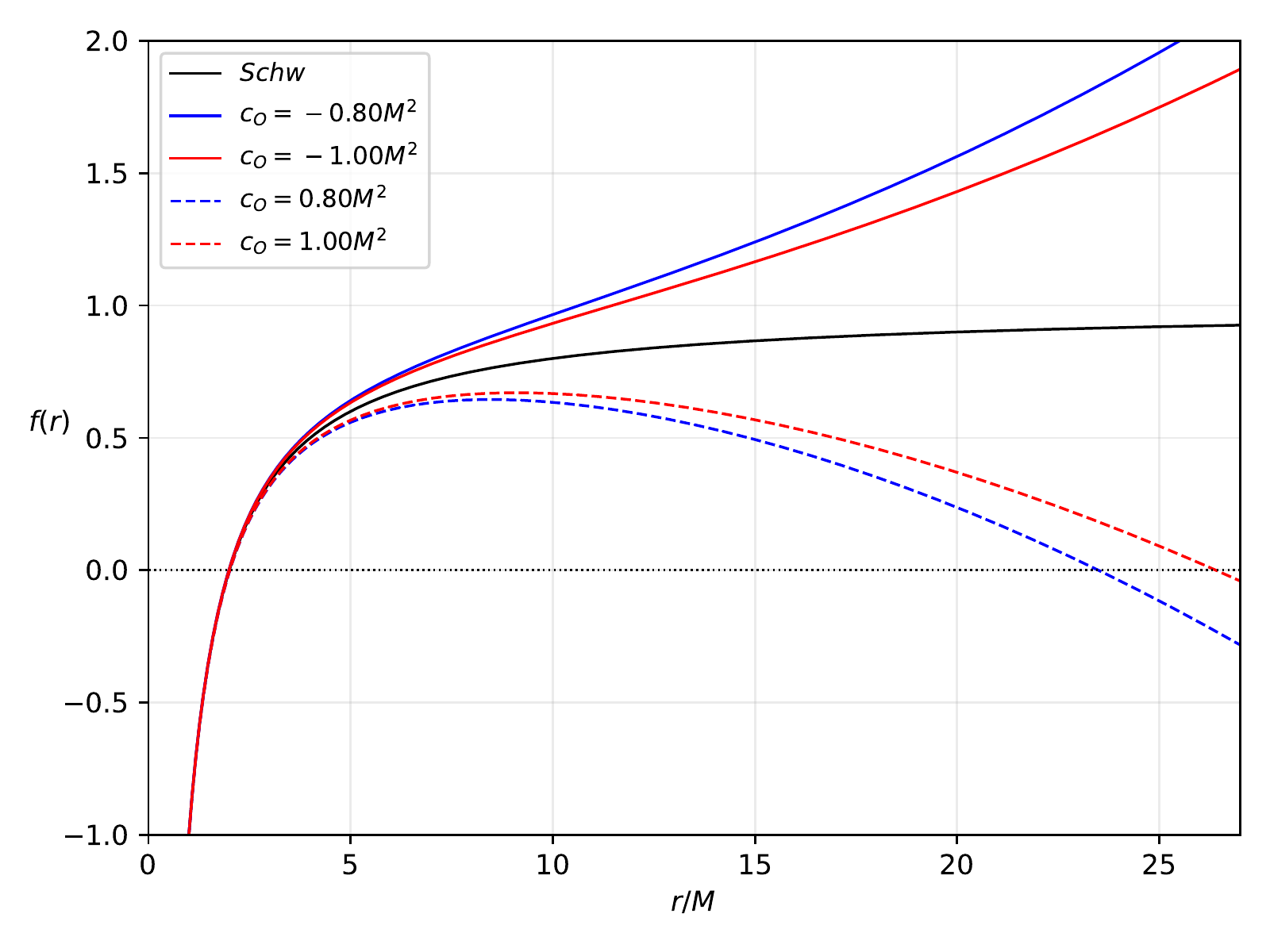}
\caption{The lapse function $f(r)$ for $a=0.90$, and $c_{\rm O} = \pm 0.8$ and $\pm 1.0$. The full curves 
correspond to $c_{\rm O} <0$, and mimic, an AdS ($c_{\rm O} <0 \Rightarrow n_\text{B}<n_\text{F}$) structure. The dashed curves, on the other hand,  correspond to $c_{\rm O} >0$, mimic a dS ($c_{\rm O} >0 \Rightarrow n_\text{B}>n_\text{F}$) structure, and develop a cosmological-like horizon at large $r$. \label{lapse}}
\end{figure}

The curvature sector of the symmergent gravity, as follows from the action  (\ref{curvature-sector}), contains linear and quadratic curvature terms. Their coefficients are loop-induced quantities defined in (\ref{params}). The theory is a special case of generic $f(R)$ gravity theories, where instead of the standard Einstein-Hilbert action with a linear curvature, $R(g)$ the theory contains a covariant function of the curvature $f(R)$ as
\begin{eqnarray}
S=\frac{1}{16 \pi G} \int d^4 x \sqrt{-g}  (R - \pi G c_\text{O} R^2)\ ,
\label{fr-action}
\end{eqnarray}
after leaving aside the irrelevant scalar field sector. This action leads to the equations of motion 
\begin{eqnarray} 
\label{f1}
\mathit{R}_{\mu \nu} \mathit{F(R)}-\frac{1}{2}g_{\mu \nu}\mathit{f(R)}+[g_{\mu \nu}\Box -\nabla_\mu \nabla_\nu]\mathit{F(R)}= 0\ ,
\end{eqnarray}
where $\Box$ is the d'Alembertian operator, and
\begin{eqnarray}
\label{FR}
F(R) = \frac{d f(R)}{d R} = 1 - 2 \pi G c_\text{O} R
\end{eqnarray}
is the derivative of $f(R)$ with respect to its argument. Making use of its trace, equation (\ref{f1}) leads to a traceless motion equation
\begin{eqnarray} 
\label{f3ss}
\mathit{R}_{\mu \nu} \mathit{F(R)}-\frac{g_{\mu \nu}}{4}\mathit{ R}\mathit{F(R)}+\frac{g_{\mu \nu}}{4}\Box\mathit{F(R)} -\nabla_\mu \nabla_\nu\mathit{F(R)}=0
\end{eqnarray}
whose solution is directly tied to the loop factors $G$ and $c_O$, and thus, varies with the particle spectrum of the underlying QFT. Now, using the ansatz,  
\begin{eqnarray}
\label{FRsoln0}
F[R(r)] = a + b r 
\end{eqnarray}
the geometry of the black hole spacetime  in spherical coordinates $x^{\alpha}=\{t,r,\theta,\phi\}$ is found to have the form
\begin{eqnarray}\label{metric}
ds^2=-f(r)dt^2 + \frac{1}{f(r)}dr^2+r^2 \left(d\theta^2+\sin^2\theta d\phi^2 \right),
\end{eqnarray}
with the lapse function 
\begin{eqnarray} 
\label{smetric}
    f(r)=1-\frac{2 G M}{r}+\frac{a-1}{24 \pi G c_\text{O}} r^2
\end{eqnarray}
becoming the Schwarzschild black hole spacetime  when $a\to1$ or/and $c_\text{O}\to \infty$. From now on, we set $G=1$. We plot the lapse function in Fig. \ref{lapse} as a function of $r/M$. 
For the Schwarzschild case, $r_h = 2M$ and hence $f(r_h) = 0$. Inclusion symmergent gravity parameters makes the horizon shift slightly toward or away from the Schwarzschild case. In Figure \ref{lapse}, full curves 
correspond to $c_{\rm O} <0$, and mimic an AdS ($c_{\rm O} <0 \Rightarrow n_\text{B}<n_\text{F}$) behaviour. The dashed curves, on the other hand,  correspond to $c_{\rm O} >0$, mimic a dS ($c_{\rm O} >0 \Rightarrow n_\text{B}>n_\text{F}$) structure, and develop a second horizon at large $r$. We observe that as $r/M$ gets too large, deviation from the Schwarzschild case is pronounced.

\subsection{Horizon properties}

The radius of the event horizon is determined by solving $g^{rr}=0$ or, equivalently, $f(r)=0$. It is easy to see from Eq.~(\ref{smetric}) that these equations possess three distinct solutions:
\begin{eqnarray}
    r_1=2(A-B)\ ,\;\,  r_{2,3}=(B-A) \pm i\sqrt{3}(A+B)
    \label{horizons}
\end{eqnarray}
in which 
\begin{eqnarray}
    A=\frac{\sqrt[3]{\pi {\cal A}}}{a-1}, \qquad B=\frac{2\pi c_\text{O}}{\sqrt[3]{\pi {\cal A}}}\ ,
\end{eqnarray}
with ${\cal A}=(a-1)^{3/2} c_\text{O} \sqrt{9 (a-1) M^2+8 \pi  c_\text{O}}+3 (a-1)^2 c_\text{O} M$.

\begin{figure}[ht!]
   \centering
\includegraphics[width=0.95\linewidth]{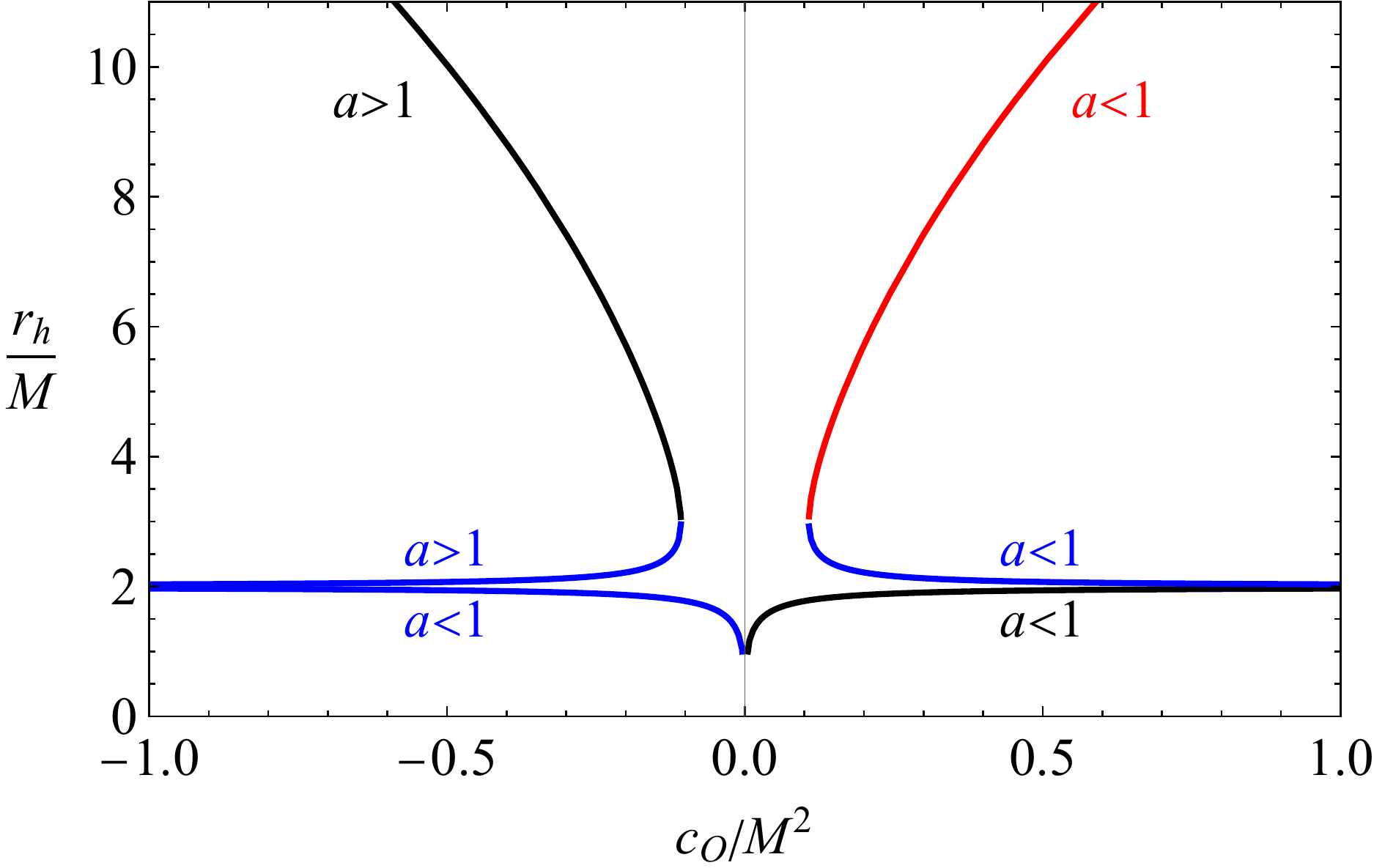}
\caption{Horizon radius $r_h$ as a function of the loop factor  $c_\text{O}$ for two representative parameter values $a=0.3$ and $a=1.7$. The black line shows $r_1$ solution, while the blue and red lines imply $r_2$ and $r_3$ ones, respectively.   \label{rvsp}}

\end{figure}
The three roots in (\ref{horizons}) are plotted in Figure~\ref{rvsp} as the Cauchy horizon, event horizon and the cosmological-like horizon. The  black, blue and red curves correspond, respectively, to the solutions $r_1$, $r_2$, $r_3$ for both $a>1$ and $a<1$ ranges. 
In fact, numerical and graphical analyses show that one can distinguish various types of horizons depending on the values of $a$ and $c_\text{O}$ (see Figure \ref{rvsp}):
\begin{itemize}
    \item The root $r_1$ comes to mean the inner (Cauchy) horizon for $c_\text{O}>0$ and cosmological-like horizon for $c_\text{O}<0$ \& $a>1$. It turns to negative and loses its physical meaning when $a<1$ \& $c_\text{O}>0$. 
    
    \item The root $r_2$ corresponds to the outer (event) horizon for $c_\text{O}<0$ \& $a>1$ or $c_\text{O}>0$ \& $a<1$. It turns to inner (Cauchy) horizon for $a<1$ \& $c_\text{O}<0$. 
    
    \item The root $r_3$ corresponds to cosmological-like horizon for $c_\text{O}>0$ \& $a<1$.
    
    \item For all the remaining cases, the roots $r_1, \ r_2, \ r_3$ turn out to be a complex number of no physical meaning.
\end{itemize}
Fig.~\ref{rvsp} is rich enough to suggest further properties:
\begin{itemize}
    \item The event (cosmological-like) horizons tend to decrease (increase) as $c_{\cal O}$ increases,
    \item The horizon move to $r_h=2 M$ (infinity) in the pure Schwarzschild limit of $c_{\cal O}\to \pm \infty$, 
    \item The Cauchy horizon increases as $c_{\cal O}$ grows, and eventually coincides with the Schwarzschild solution,
    \item The distance between the outer horizon and cosmological-like horizon decreases with increasing (decreasing) $a$ in the case of negative (positive) $c_{O}$. 
\end{itemize}

\begin{figure}[ht!]
   \centering
\includegraphics[width=0.95\linewidth]{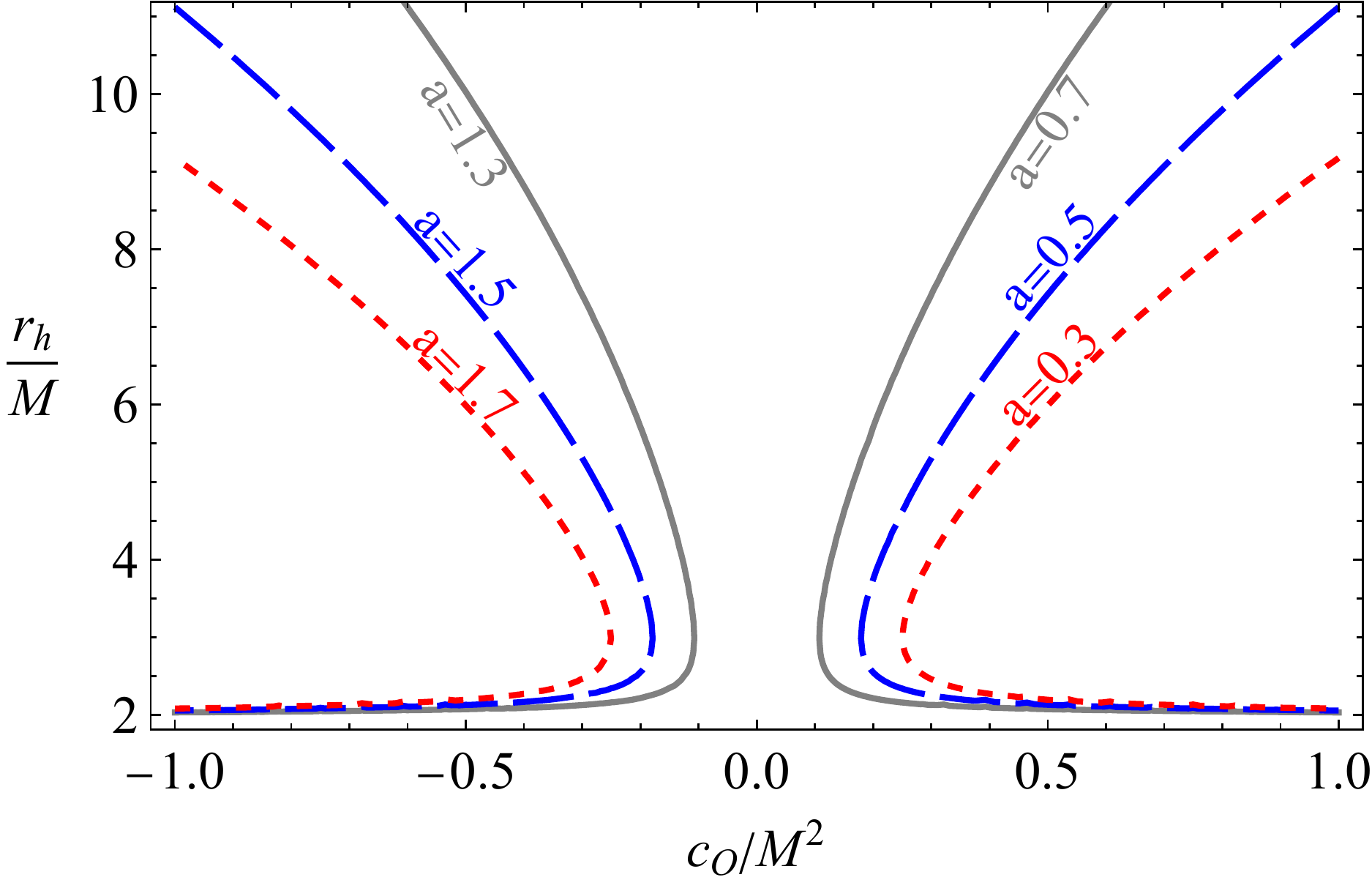}
\caption{The dependence of the event horizon radius $r_h$ on the loop factor $c_O$ for the different values of the parameter $a$ ($r_h=2$ is the Schwarzschild limit).  \label{rvsp1}}
\end{figure}

\begin{figure}[ht!]
   \centering
\includegraphics[width=0.95\linewidth]{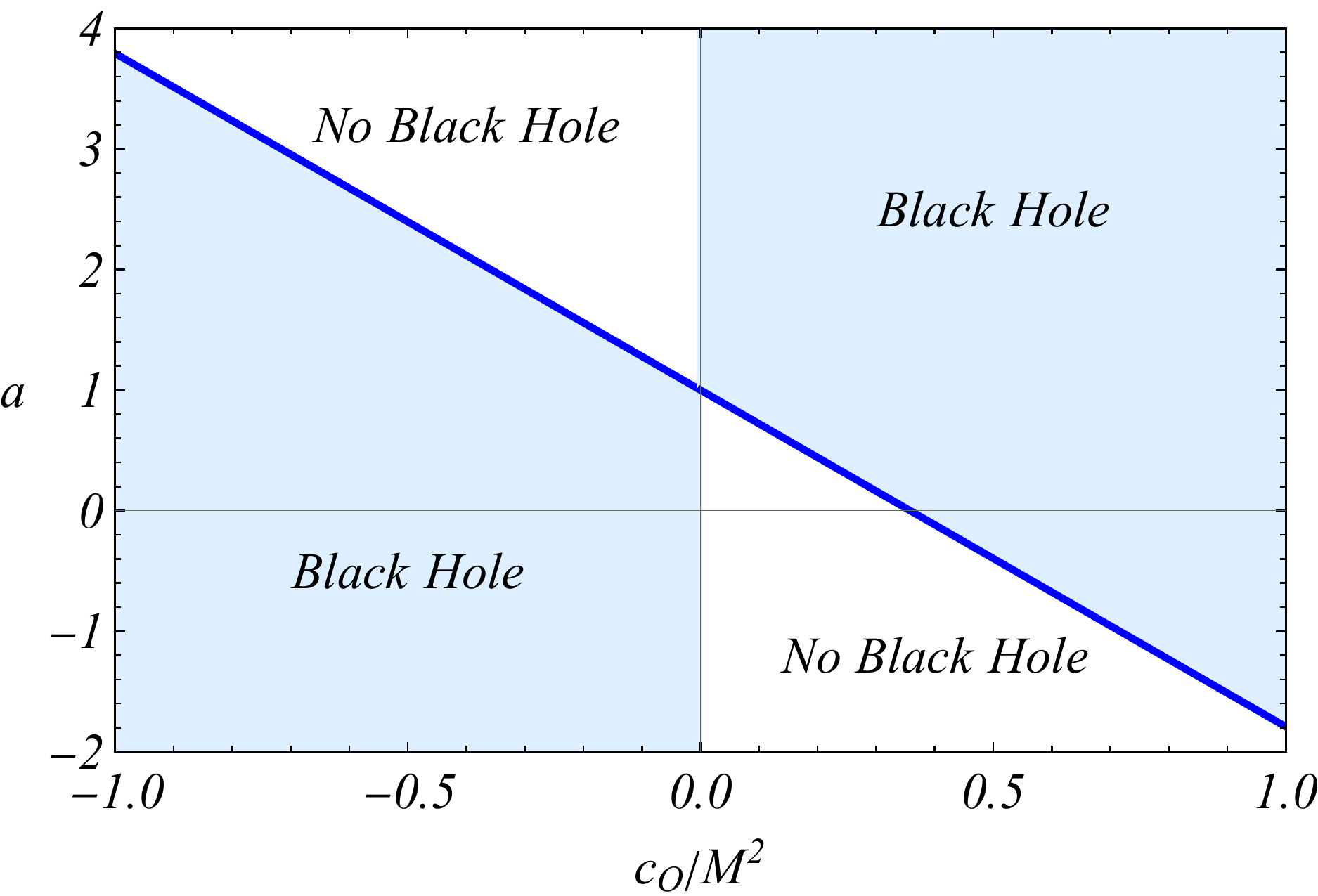}
\caption{The blue (white) regions in $c_\text{O}$--$a$ plane are the domains in which a symmergent black hole solution does (does not) exist.  The thick-blue line corresponds to the criticality relation (\ref{critical}). \label{avschor}}
\end{figure}

To extract information on the ranges of the parameters $a$ and $c_O$, it proves useful to define a critical value. To this end, $f(r)=0$ and $f'(r)=0$ prove useful as two independent conditions. These conditions ascribe critical values to the symmergent gravity parameters $a$ and $c_\text{O}$ as
\begin{equation}
    a_{cr}= 1-\frac{8 \pi (c_\text{O})_{cr}}{9M^2}
    \label{critical}
\end{equation}
for which the lapse function (\ref{smetric}) vanishes with vanishing slope. The critical values $a_{cr}$ and $(c_\text{O})_{cr}$ set the thick-blue line in Figure~\ref{avschor}. The light-blue regions make the metric (\ref{metric}) a true black hole metric. The white regions in Figure~\ref{avschor} are the ones which do not lead to black hole solutions. 

\section{Dynamics of test particles} \label{sec3}
In this section, we give a detailed analysis of the motions of test particles in the black hole spacetime (\ref{metric}) of the symmergent gravity. 

\subsection{Derivation of the equations of motion}

Here, we study the dynamics of test particles in the spacetime of static black holes in symmergent gravity. In order to derive the integrals of motion, we start with the Lagrangian of the test particles 
\begin{eqnarray}
{\mathbb{L}}=\frac{1}{2}m g_{\mu\nu} \dot{x}^{\mu} \dot{x}^{\nu} \ , 
\end{eqnarray}
in which $m$ stands for the masses of the test particles, $x_\mu(\tau)$ denotes their worldlines, and $\dot{x}^{\mu}=dx^\mu/d\tau$ gives their four-velocities.  

\begin{figure}[ht!]
   \centering
\includegraphics[width=0.948\linewidth]{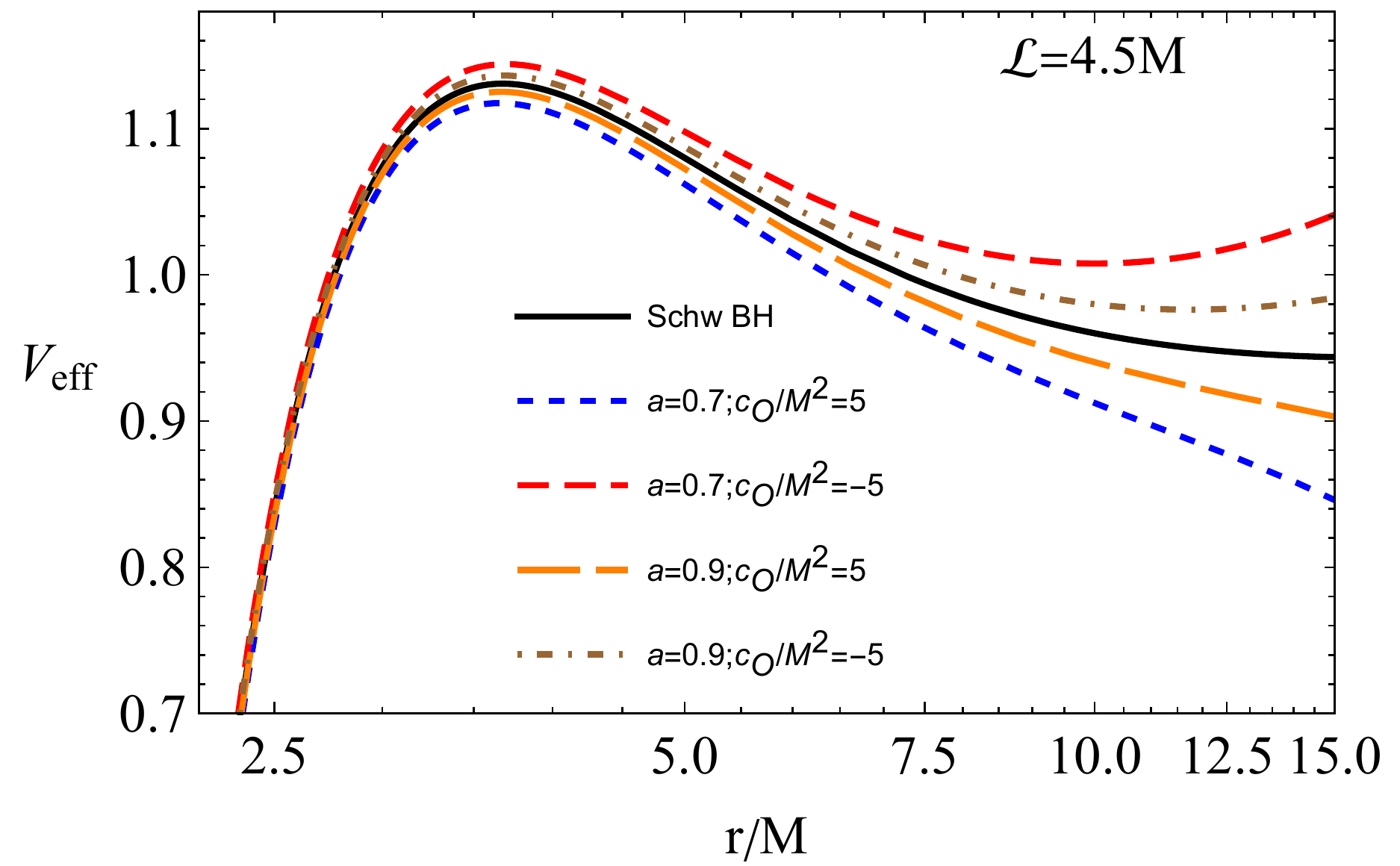}
\includegraphics[width=0.948\linewidth]{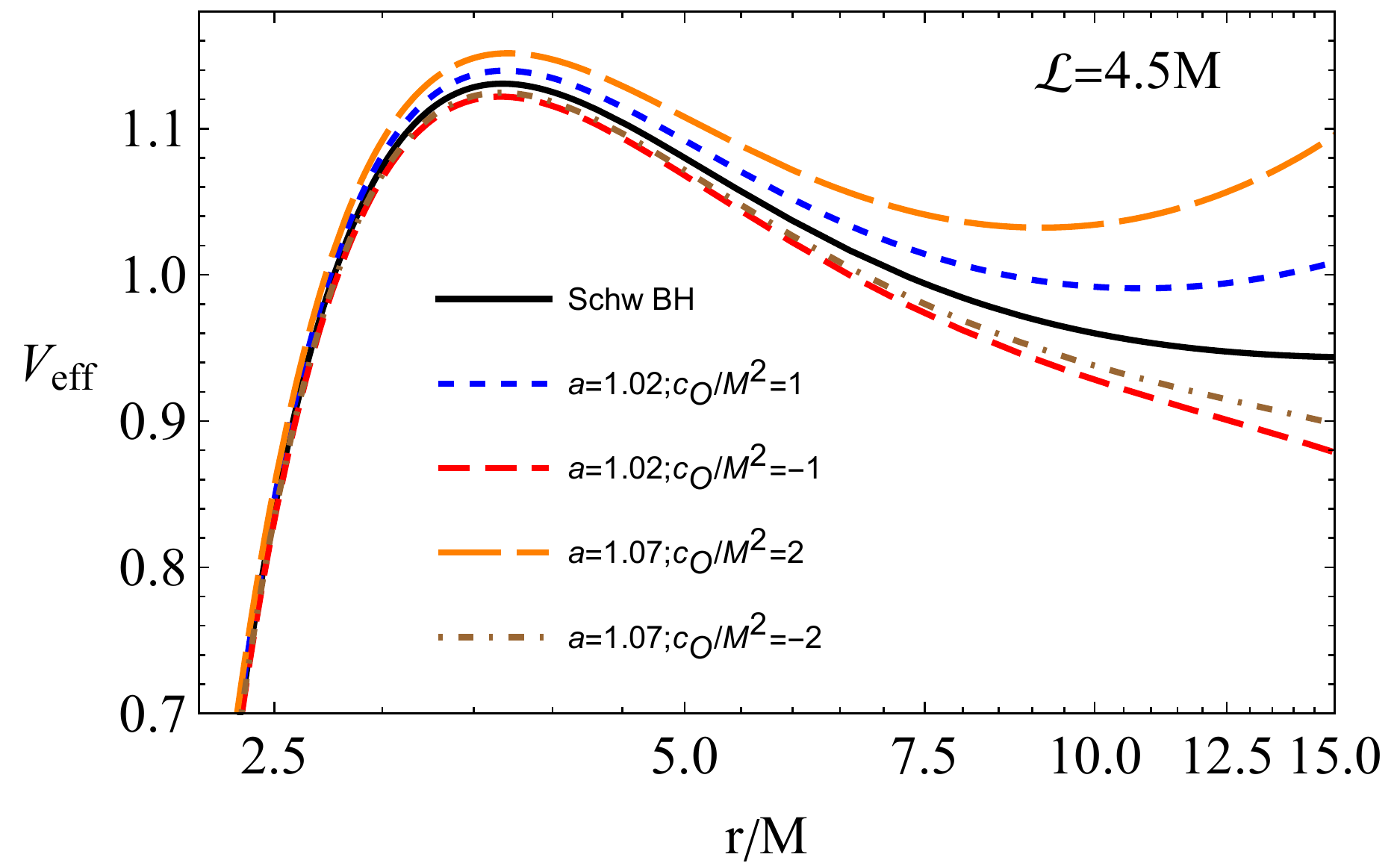}
\caption{The effective potential profile for radial motions of test particles around a static symmergent black hole for ${\cal L}/M=4.5$. The top (bottom) panel stands for $a<1$ with $c_O/M^2 =\pm 5$  ($a>1$ with $c_O/M^2 = \pm 1\  {\rm and}\ \pm 2$). The effective potential diverges significantly from the Schwarzschild solution (black curve) at large radii. 
\label{effective}}
\end{figure}

Since the metric tensor (\ref{metric}) is independent of time $t$ and azimuthal coordinate $\phi$, there arise two conserved quantities:
\begin{eqnarray}
\label{consts1}
 g_{tt}\dot{t}=-{\cal E}\ , \qquad g_{\phi \phi}\dot{\phi} = {\cal L}, 
\end{eqnarray}
where ${\cal E}=E/m$ and ${\cal L}=L/m$ are the specific energy and angular momentum of the particle of mass $m$, respectively. One can use the following normalization condition to govern the equations of motion for test  particles:
\begin{equation}\label{norm4vel}
g_{\mu \nu}\dot{x}^{\mu} \dot{x}^{\nu}=\epsilon\ 
\end{equation}
in which $\epsilon=0$ corresponds to massless particles with null geodesics,  and $\epsilon=-1$ corresponds to massive particles with time-like geodesics. The equation of motion for particles with non-zero mass can be expressed as ~\cite{Rayimbaev2020PhysRevDRGI,BokhariPhysRevD2020,JuraevaEPJC2021}
\begin{eqnarray}\label{eqmotionneutral}
\dot{t}&=&-\frac{{\cal E}}{g_{tt}}\ ,\\
\dot{r}&=&\sqrt{{\cal E}^2+g_{tt}\left(1+\frac{\cal K}{r^2}\right)}\ ,
 \\
\dot{\theta}&=&\frac{1}{g_{\theta \theta}}\sqrt{{\cal K}-\frac{{\cal L}^2}{\sin^2\theta}}\ ,
 \\
\dot{\phi}&=&\frac{{\cal L}}{g_{\phi \phi}}\ ,
\end{eqnarray}
in which ${\cal K}$ is the Carter constant, corresponding to the total angular momentum of the particle. Specializing to the  equatorial motion of the particles with  $\theta=\pi/2$ and $\dot{\theta}=0$, one is led to  ${\cal K}={\cal L}^2$. In consequence, test particles assume the following radial motion equation, 
\begin{equation}
 \dot{r}^2={\cal E}^2-V_{\rm eff}(r)\ ,
 \label{eom}
\end{equation}
where the effective potential 
\begin{equation}
\label{effpotentail}
     V_{\rm eff}(r) = f(r) \left(1 + \frac{{\cal L}^2}{r^2}\right) 
   \end{equation}
encodes effects of rotational dynamics via the motion integral ${\cal L}$. The test particles execute a radial motion around the symmergent black hole, as stated in (\ref{eom}). The radial variation of the effective potential is depicted in Fig.~\ref{effective} for various values of the parameters $a$ and $c_{\rm O}$. The top panel shows $V_{\rm eff}(r)$ for $a<1$ and $c_O/M^2 =\pm 5$. The bottom panel, on the other hand, shows $V_{\rm eff}(r)$ for $a>1$, $c_O/M^2 = \pm 1$ and $c_O/M^2 = \pm 2$. It is seen from the two panels that the effective potential decreases (increases) at large distances for $a<1$ \& $c_{\rm O}>0$ ($a<1$ \& $c_{\rm O}<0$) and $a>1$ \& $c_{\rm O}<0$ ($a>1$ \& $c_{\rm O}>0$). 

\subsection{Stable circular orbits}
To reveal the effects of the symmergent gravity parameters on the circular orbits of the test particles we go to the following extremum of the effective potential 
\begin{eqnarray} \label{conditions}
V_{\rm eff}={\cal E}, \qquad V_{\rm eff}'=0, \end{eqnarray}
in which we find 
\begin{eqnarray} \label{ang-mom}
{\cal L}^2=\frac{r^2 \left[4 \pi  M c_\text{O}-(1-a) r^3\right]}{24 \pi  c_\text{O} (r-3 M)}
\end{eqnarray}
and 
\begin{eqnarray}
\label{energy}
{\cal E}^2=\frac{\left[(a-1) r^3+24 \pi  c_\text{O} (r-2 M)\right]^2}{576 \pi ^2 c_\text{O}^2 r (r-3 M)} 
\end{eqnarray} 
as the specific angular momentum and specific energy of the particles in their circular orbits around the black hole. These circular orbits remain stable in the minima of the effective potential. 

\begin{figure}[h!]
   \centering
\includegraphics[width=0.93\linewidth]{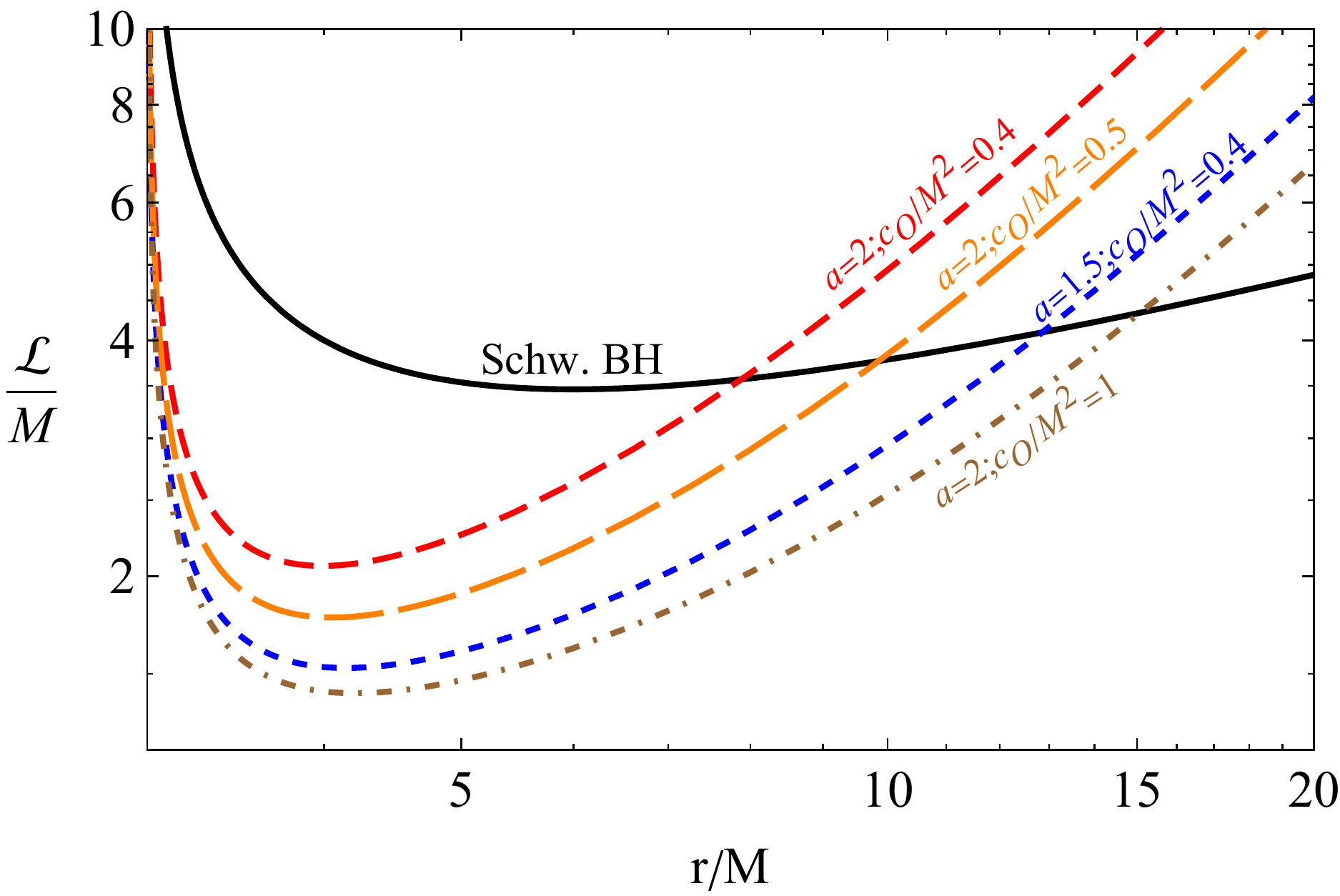}
\includegraphics[width=0.98\linewidth]{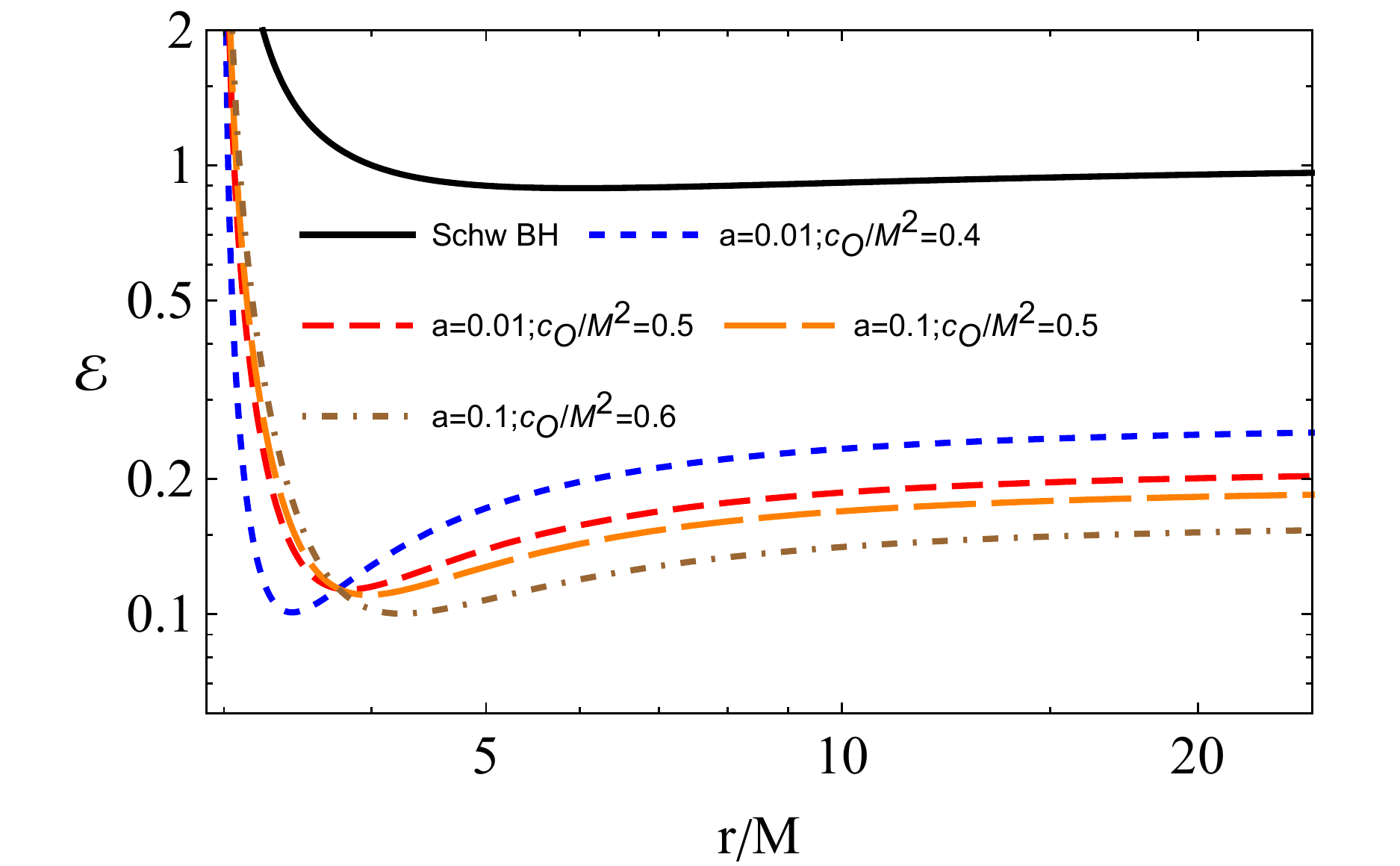}
\caption{The specific angular momentum (top panel) and specific energy (bottom panel) of test particles in their stable orbits (minimum energy) around a static black hole in symmergent gravity. The paper stable orbit varies slightly with the variations in the parameters $c_{\rm O}$ and $a$.  \label{LLL}}
\end{figure}

In Fig.~\ref{LLL} we give radial profiles of the angular momentum (top panel) and energy (bottom panel) of the test particles in their stable circular orbits around the symmergent black hole. It is clear from the figure that, away from the Schwarzschild solution, both the angular momentum and energy get minimized $r/M \simeq 3.5$. One notes that an increase of the parameter $a$ ($c_{\rm O}$) gives cause for an increase (decrease) in the energy and angular momentum. Needless to say, the stable orbit shifts slightly with the variations in  $a$ and $c_{\rm O}$. 

\subsection{Innermost stable circular orbit}

\begin{figure}[ht!]
   \centering
\includegraphics[width=0.95\linewidth]{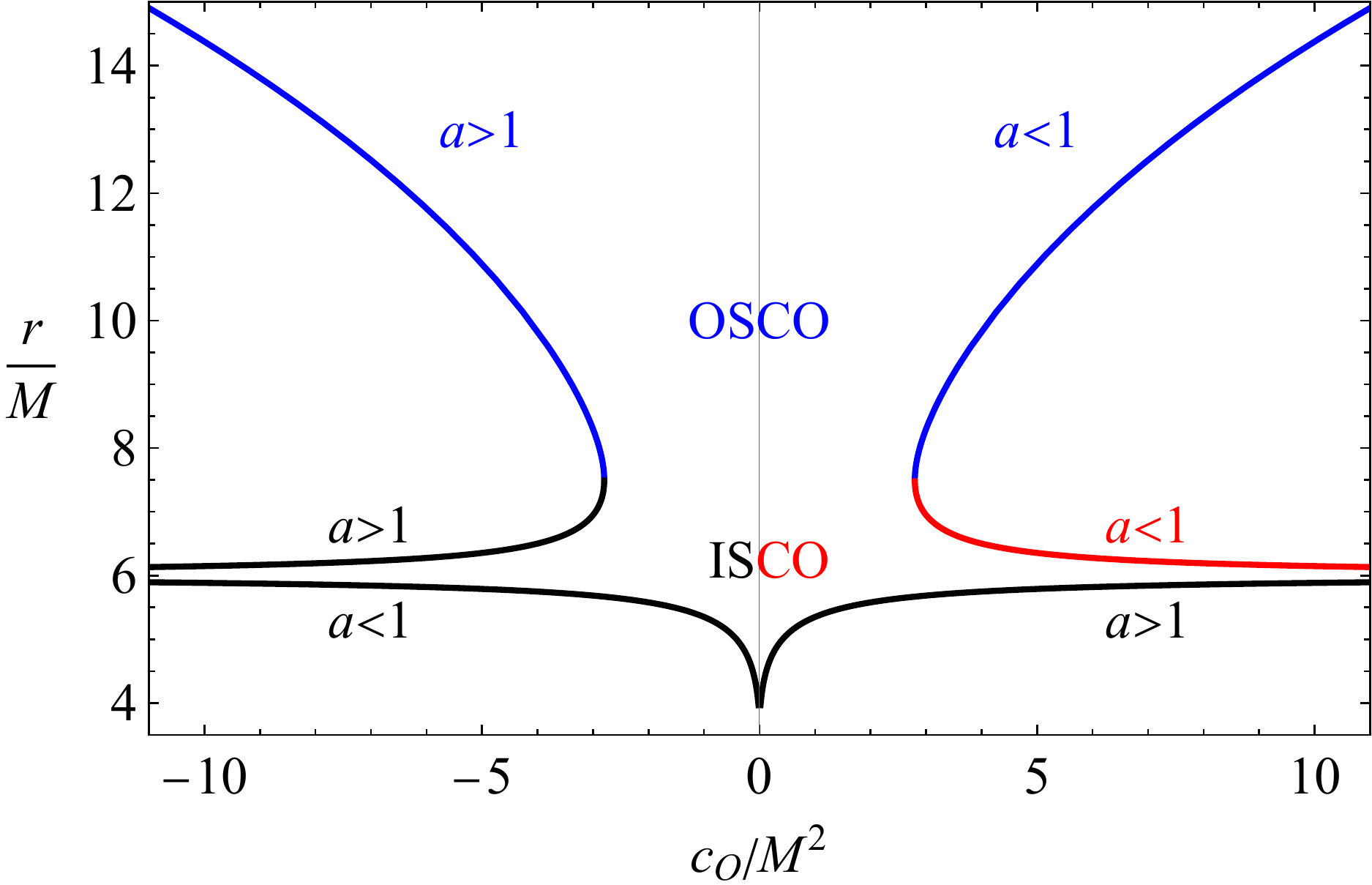}
\caption{The roots of the ISCO equation (\ref{ISCOeq}) as a function of $c_{\cal O}$ for $a>1$ and $a<1$. The black, red and blue curves correspond, respectively, to the roots $r_2$, $r_3$, and $r_4$ given in (\ref{is1}). The root $r_1$ is missing from the plot because it turns out to be negative for the given $a$ and $c_{\cal O}$ ranges. \label{isco}}
\end{figure}

\begin{figure}[ht!]
   \centering
\includegraphics[width=0.97\linewidth]{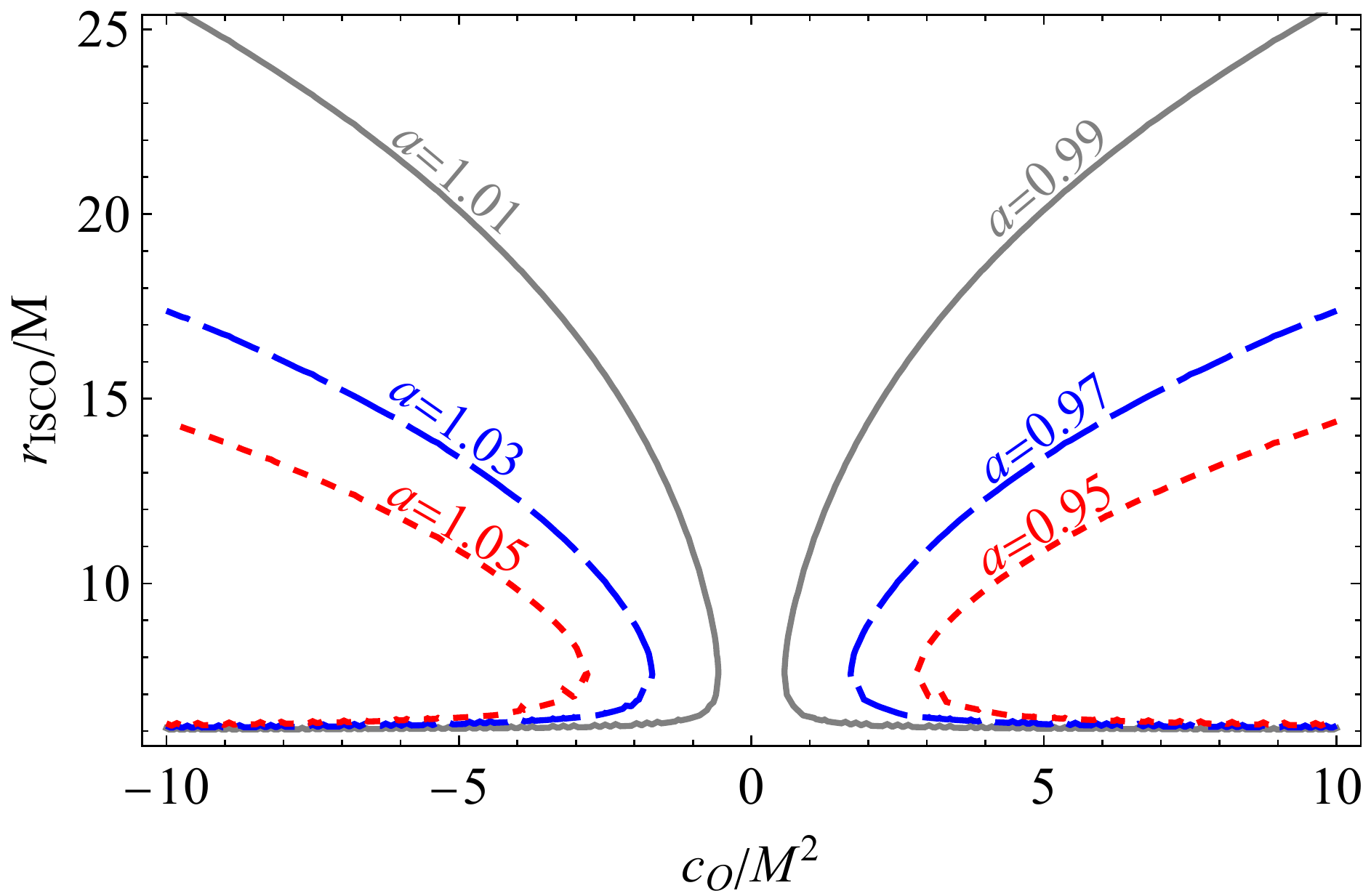}
\caption{The ISCO and OSCO radii as functions of the parameter $c_{\cal O}$ for $a$ values equidistant to unity. \label{isco2}}
\end{figure}

The radius of innermost stable circular orbits (ISCO), denoted as $r_{\rm ISCO}$, is the solution of the equation $\partial_{rr}V_{\rm eff}(r)=0$ along with the conditions (\ref{conditions}). For the effective potential in (\ref{effpotentail}), the ISCO equation takes the form:
\begin{equation}\label{ISCOeq}
 \frac{2M(r-6M)}{r^3(r-3M)}+\frac{a-1}{12\pi c_\text{O}}\frac{4r-15M}{r-3M}=0   \end{equation}
from which it follows that $r_{\rm ISCO} \to 6M$ for 
$a=1$ or $c_\text{O} \to \infty$. Away from this Schwarzschild limit, the ISCO equation (\ref{ISCOeq}) assumes four different solutions $r_1,r_2,r_3$ and $r_4$:
\begin{eqnarray}
\label{is1}
2r_{1,2,3,4}=\frac{15M}{8} + \alpha_1 {\cal P}_1 + \alpha_2 {\cal P}_2
\end{eqnarray}
in which $\alpha_1 = \pm 1$ and $\alpha_2 = \pm 1$ are two independent constants such that 
\begin{eqnarray}
{\cal P}_1^2&=&\frac{1}{12 (1-a)}\sqrt[3]{\frac{{\cal Q}}{2}}+486 \pi  c_{\rm O} M^2 \sqrt[3]{\frac{2}{{\cal Q}}}+\frac{225 M^2}{64}\ , \nonumber\\  {\cal Q} &=&3888 \pi  \Bigg\{225 (a-1)^2 M^4 c_O-16 \pi  (a-1) M^2 c_O^2\nonumber\\ &+&(a-1) M^2 c_O \Bigg[52488 \pi  (a-1) M^2 c_O \nonumber\\ &+&\left(225 (a-1) M^2-16 \pi  c_O\right)^2\Bigg]^{\frac{1}{2}}\Bigg\}\ ,\nonumber\\
{\cal P}_2&=&\left[\frac{1}{4{\cal P}_1}\left(\frac{48\pi c_{\rm O}M}{a-1}-\frac{3375M^2}{64}\right)-\frac{675M^2}{64}-{\cal P}_1^2\right]^{\frac{1}{2}} \ ,\nonumber
\end{eqnarray}
such that the radii  $r_1, \ r_2, \ r_3, \ r_4$ correspond to the  configurations $(\alpha_1,\alpha_2)=(+,+),\, (+,-),\,  (-,+),\, (-,-)$, respectively. A judicious numerical analysis can reveal physics implications of the lengthy analytic solution in (\ref{is1}).  To this end, depicted in Figure~\ref{isco} are the positive real solutions of (\ref{ISCOeq}) for both $a<1$ and $a>1$ regimes. Indeed, black, red and blue lines correspond, respectively, to $r_2$, $r_3$ and $r_4$. Our numerical analysis shows that $r_1$ takes on negative value for all values of $a$ and $c_{\rm O}$. Also, it is observed from the figure and obtained from our numerical analyses that, $r_2$ and $r_3$ correspond to ISCO at $a>1$ \& $c_{\rm O}<0$ and $a<1$ \& $c_{\rm O}>0$, respectively. The root $r_4$, on the other hand, corresponds to the outermost stable circular orbit (OSCO) and coincides with ISCO at critical values of the parameter $c_{\rm O}$. It is clear that ISCO radius increases (decreases) with the increase of the parameter $c_{\rm O}$ at $a>1$ ($a<1$). In other words, ISCO and OSCO come close (go apart) depending on $a$ and $c_{\rm O}$. These features are depicted in Fig.~\ref{isco2} in which values of the $a$ parameter are evenly sampled  around the unity.

\section{Fundamental frequencies} \label{sec4}

In this section, we study the fundamental frequencies of test particles around a black hole in symmergent gravity. In particular, we will explore the effects of the symmergent gravity parameters on the frequencies of the particles in their Keplerian orbits. We will also explore the frequencies of the radial and vertical oscillations of the particles around their stable circular orbits. We will apply our results to the analysis of QPO frequencies.

\subsection{Keplerian frequency}

The angular velocity $\Omega_K=\dot{\phi}/\dot{t}$ of test particles in their circular (Keplerian) orbits around the symmergent black hole can be expressed as, 
\begin{equation}\label{omegaKep}
\Omega_K^2={\frac{a-1}{4 \pi  c_\text{O}}+\frac{ M}{r^3}}
\end{equation}
from which it follows that Keplerian frequency vanishes at the critical distance 
\begin{equation}
    r_{\rm cr}=\left(\frac{4 \pi  c_\text{O} M}{1-a} \right)^{\frac{1}{3}}\ ,
\end{equation}
provided that $a<1$ \& $c_\text{O}>0$ or $a>1$ \& $c_\text{O}<0$. The Keplerian frequency $\Omega_K$ can be converted to ${\rm Hz}$ units (1/seconds) by using the speed of light and the Schwarzschild radius, 
\begin{equation}\label{toHz}
\nu_K = \frac{1}{2\pi}\frac{c^3}{GM} \Omega_K \ 
\end{equation}
which indeed measures in ${\rm Hz}$ with the speed of light  $c=3\cdot 10^8 \rm m/sec$ and the gravitational constant is $G=6.67\cdot 10^{-11}\rm m^3/(kg^2\cdot sec)$.

\begin{figure}[h]
   \centering
\includegraphics[width=0.87\linewidth]{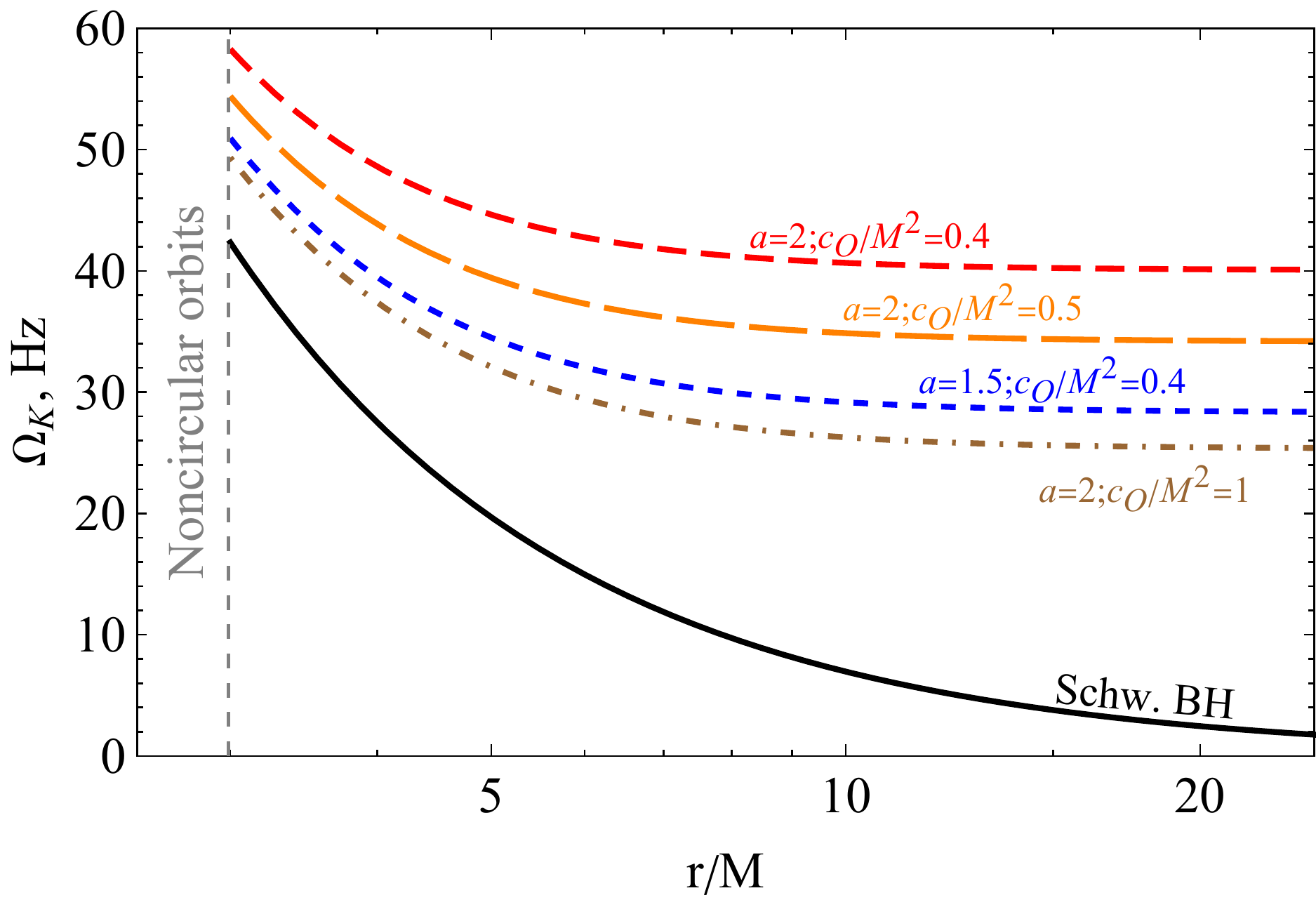}
\includegraphics[width=0.91\linewidth]{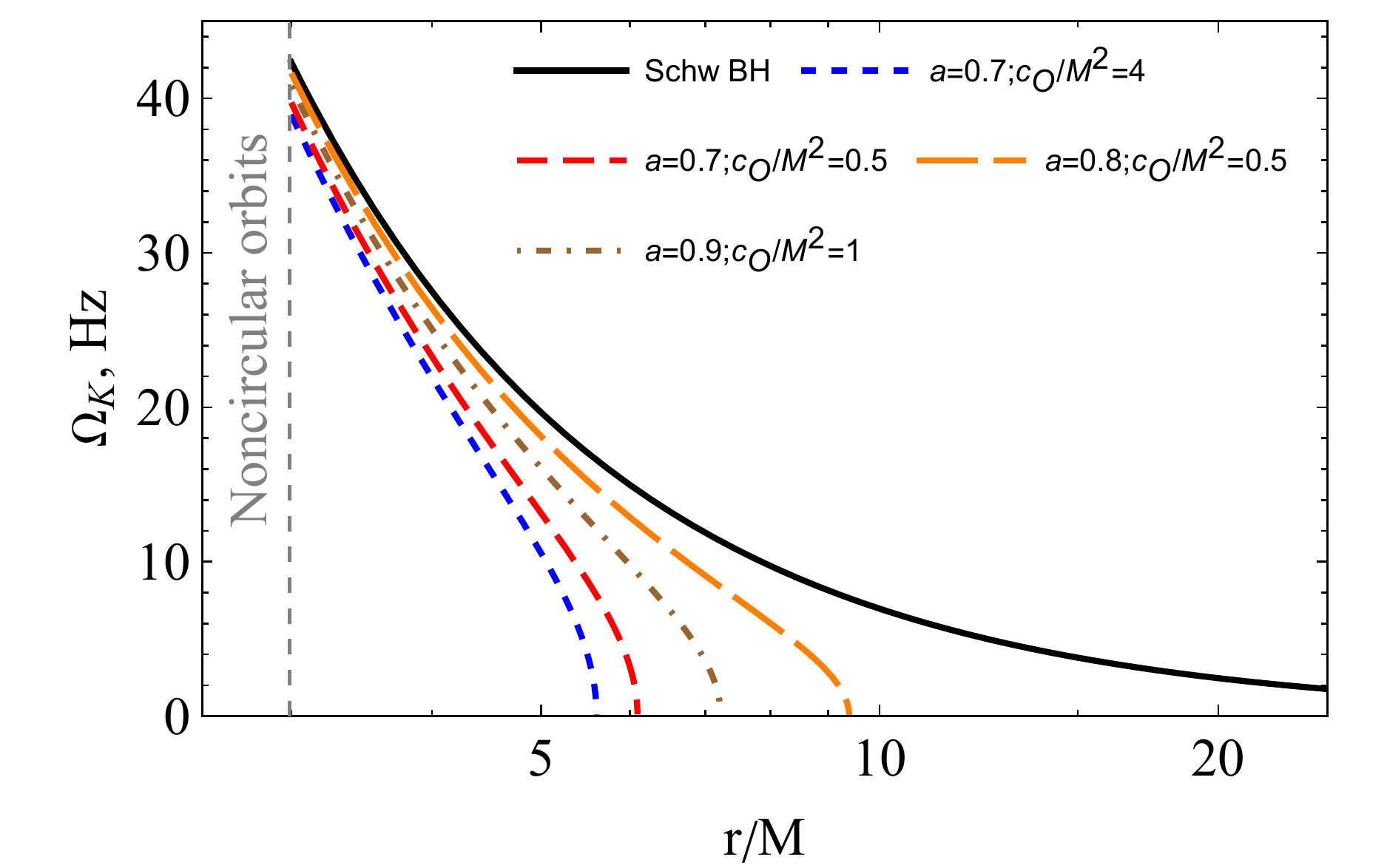}
\caption{The Keplerian frequency $\Omega_K\ (\rm Hz)$ for $a>1$ (top panel) and $a<1$ (bottom panel). It is clear that there doesn't (does) exist critical radii $r_{\rm cr}$  at which $\Omega_K$ vanishes for $a>1$ ($a<1$) for the indicated values of $c_{\rm O}$. \label{Keplerian}}
\end{figure}

Fig.~\ref{Keplerian} shows the radial variation of the frequency of test particles in their Keplerian orbits around the symmergent black hole for different values of the $c_{\rm O}$ and $a$ parameters. The top (bottom) panel corresponds to the case  $a>1$ ($a<1$). In general, as the figures suggest, an increase in $a$ ($c_{\rm O}$) causes an increase (decrease) in the Keplerian frequency. The top panel suggests that there is no critical $r_{\rm cr}$ at which the Keplerian frequency vanishes. The bottom panel, on the other hand, suggests that there are always $r_{\rm cr}$ at which $\Omega_K$ vanishes. For instance, $r_{\rm cr}/M\approx 6$ for $a=0.7$ and $c_{\rm O}/M^2 = 0.5$. The critical radius $r_{\rm cr}$ moves to radial infinity in the Schwarzschild limit. It is clear that test particles fall freely into the central black hole at the critical radii $r_{\rm cr}$.  It is also clear that this critical distance increases as $a$ ($c_{\rm O}$) increases (decreases). 

\subsection{Harmonic oscillations}
Small displacements like  $r_0+\delta r$ and $\pi/2+\delta \theta$ from the stable configuration give cause to oscillations along the radial, angular and vertical axes of the orbits. The small displacements above obey the well-known harmonic oscillator motion \cite{Bardeen68}:  
\begin{eqnarray}
\frac{d^2\delta r}{dt^2}+\Omega_r^2 \delta r=0\ , \qquad \frac{d^2\delta\theta}{dt^2}+\Omega_\theta^2 \delta\theta=0\ ,  
\end{eqnarray}
with the frequencies
\begin{eqnarray}
\Omega_r^2=-\frac{1}{2g_{rr}(u^t)^2}\partial_r^2V_{\rm eff}(r,\theta)\Big |_{\theta=\pi/2}
\end{eqnarray}
for radial oscillations, and 
\begin{eqnarray}
\Omega_\theta^2=-\frac{1}{2g_{\theta\theta}(u^t)^2}\partial_\theta^2V_{\rm eff}(r,\theta)\Big |_{\theta=\pi/2}
\end{eqnarray}
for angular oscillations. In the symmergent gravity effective potential, these harmonic oscillation frequencies take the form, 
\begin{eqnarray}\label{wr}
\frac{\Omega_r^2}{\Omega_K^2} = {\frac{(a-1) r^3 (4 r-15 M)+24 \pi  c_\text{O} M (r-6 M)}{(a-1) r^4+24 \pi  c_\text{O} M r}} 
\end{eqnarray}
and
\begin{eqnarray}
\label{wt}
\Omega_\theta = \Omega_\phi =\Omega_K
\end{eqnarray}
where $\Omega_\phi$ is the angular velocity of the particle measured by an observer at infinity. 
 
\section{Twin-peak QPOs around symmergent black holes} \label{sec5}

In general, QPOs refer to juddering of the light rays from astronomical objects about certain frequencies. As an illustrative case, QPOs in (micro)quasars have come into view in relation to the fundamental frequencies of the particles orbiting the black hole at the center of the (micro)quasar. The main point is that the frequencies of QPOs depend on the spacetime geometry in the black hole environment.  
 
In this section, we explore the upper and lower frequencies of twin-peaked QPOs in the framework of symmergent gravity. Here,
the twin-peak QPOs are picked up on the basis of probing the symmergent gravity parameters. In general, the upper and lower frequencies are described in different ways by the frequencies of the radial, vertical and orbital oscillations:

\begin{itemize}

\item {\bf The relativistic precession (RP) model} has been first proposed in Ref.~\cite{Stella1998ApJL} to explain the physical mechanism behind {\rm kHz} twin-peak QPOs in the frequency range 0.2 to 1.25 {\rm kHz} from neutron stars in LMXRBs. It has been shown that the RP model applies also to black hole candidates in the binary systems of black holes and neutron stars \cite{Stella2001AIPC}. Furthermore, an RP model has been developed in \cite{Ingram2014MNRAS} to obtain mass and spin measurements of black holes at the center of microquasars using the data from the accretion disk power-density spectrum. The upper and lower frequencies in the RP model are given by the fundamental oscillation frequencies $\nu_U=\nu_\phi$ and $\nu_L=\nu_\phi-\nu_r$, respectively. 

\item {\bf The epicyclic resonance (ER) model} is based on the resonances in the axisymmetric oscillation modes in the accretion disc of black holes \cite{Abramowicz2001AA}. It has been shown that the disc oscillation modes are related to the frequencies of (quasi) harmonic oscillations of the circular geodesics of the test particles. Here, we consider two special cases of the ER model: ER2 and ER3. They differ from each other by their oscillation modes. ER2 has the classes of the upper and lower frequencies
$\nu_U=2\nu_\theta-\nu_r\ \&\ \nu_L=\nu_r$ as well as   $\nu_U=\nu_\theta+\nu_r\ \&\ \nu_L=\nu_\theta$. ER3, on the other hand, has only one class of the upper and lower frequencies 
$\nu_U=\nu_\theta+\nu_r\ \&\ \nu_L=\nu_\theta-\nu_r$ \cite{Abramowicz2001AA}.

\item {\bf The warped disc (WD) model} is structured upon the non-axisymmetric modes of the oscillations of the accretion disc around the black holes and neutron stars  \cite{Kato2004PASJ,Kato2008PASJ}. According to the WD model assumptions, the upper and lower frequencies are $\nu_U=2\nu_\phi-\nu_r$,  $\nu_L=2(\nu_\phi-\nu_r)$, and  vertical oscillations gives cause to a thin accretion disc to be warped ~\cite{Kato2004PASJ,Kato2008PASJ}.
\end{itemize} 
Now, using the three models mentioned above,  we analyze possible upper and lower frequencies of the twin-peaked QPOs. These frequencies are generated by the flickering of the test particles in their stable circular orbits around symmergent black holes.

\begin{figure*}[ht!]
   \centering
   \includegraphics[width=0.48\linewidth]{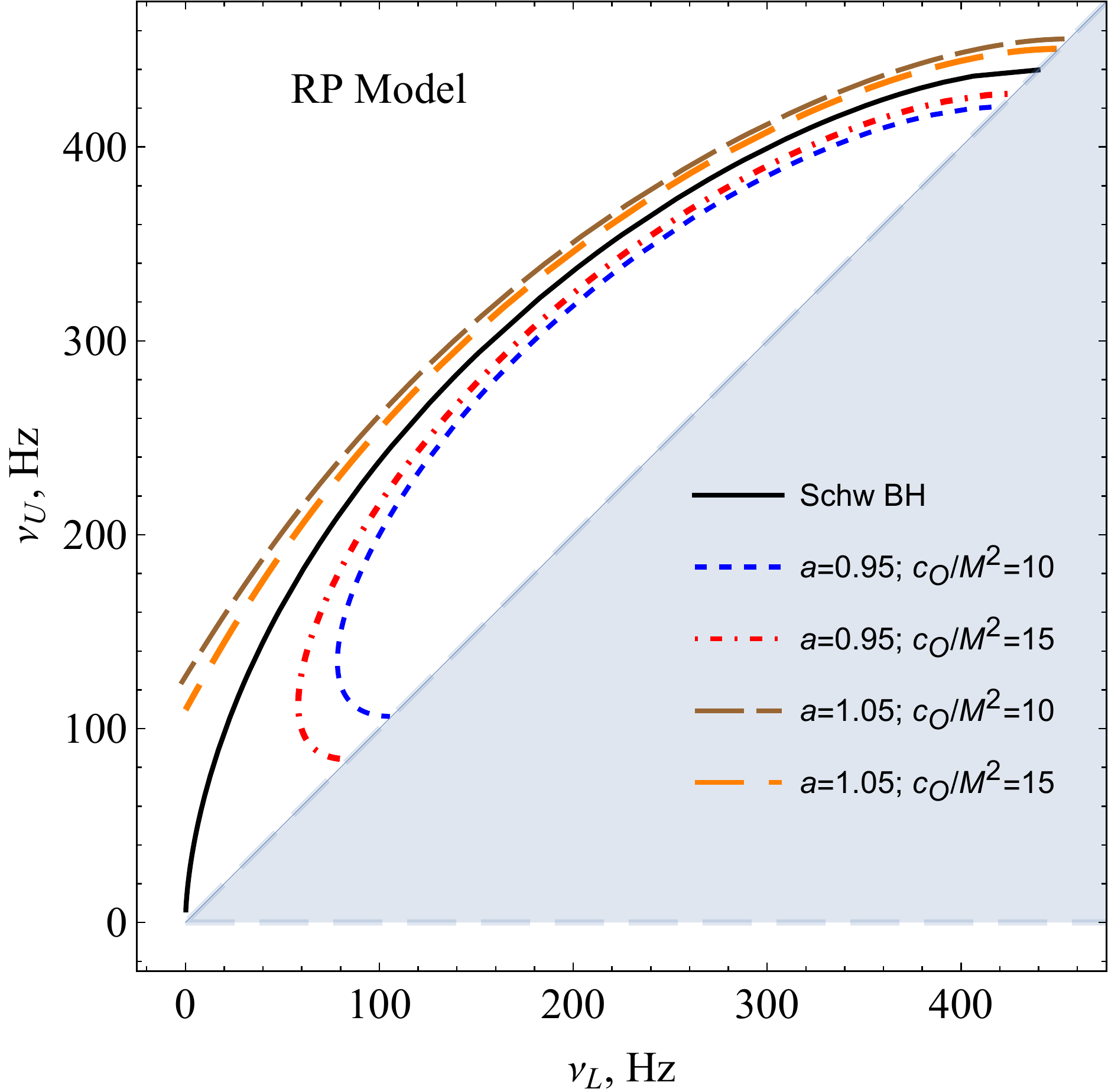}
\includegraphics[width=0.48\linewidth]{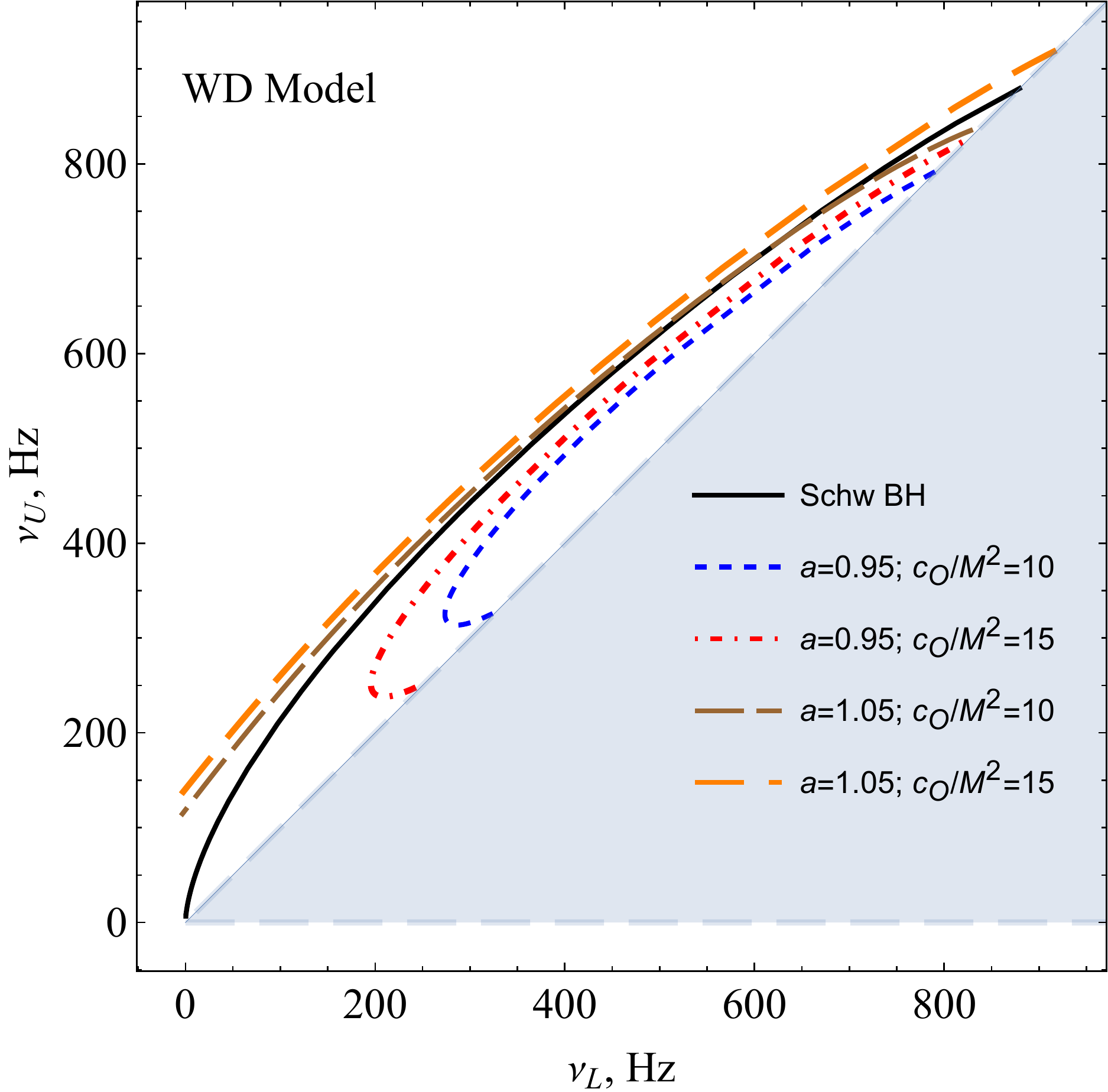}
\includegraphics[width=0.48\linewidth]{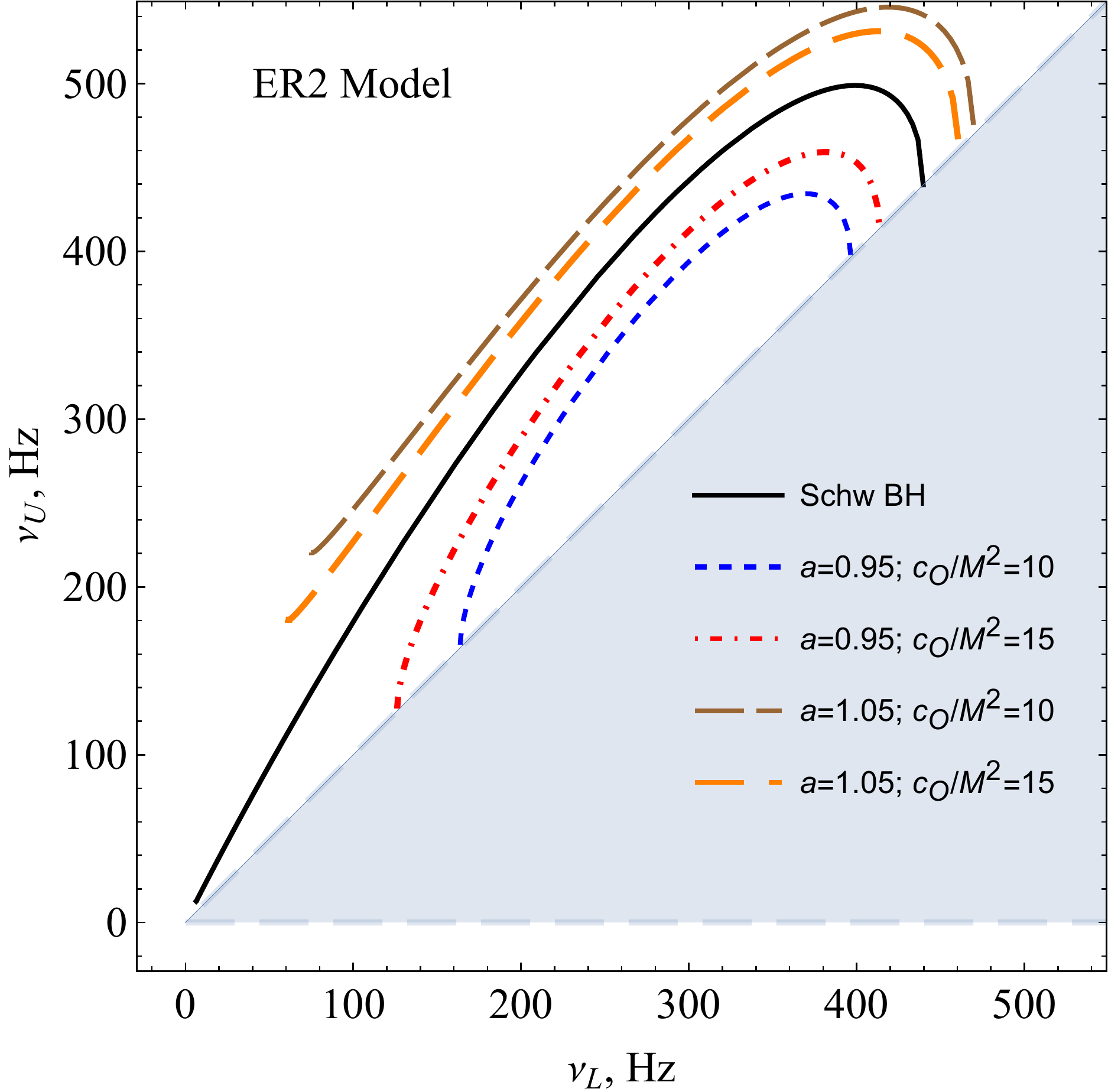}
\includegraphics[width=0.48\linewidth]{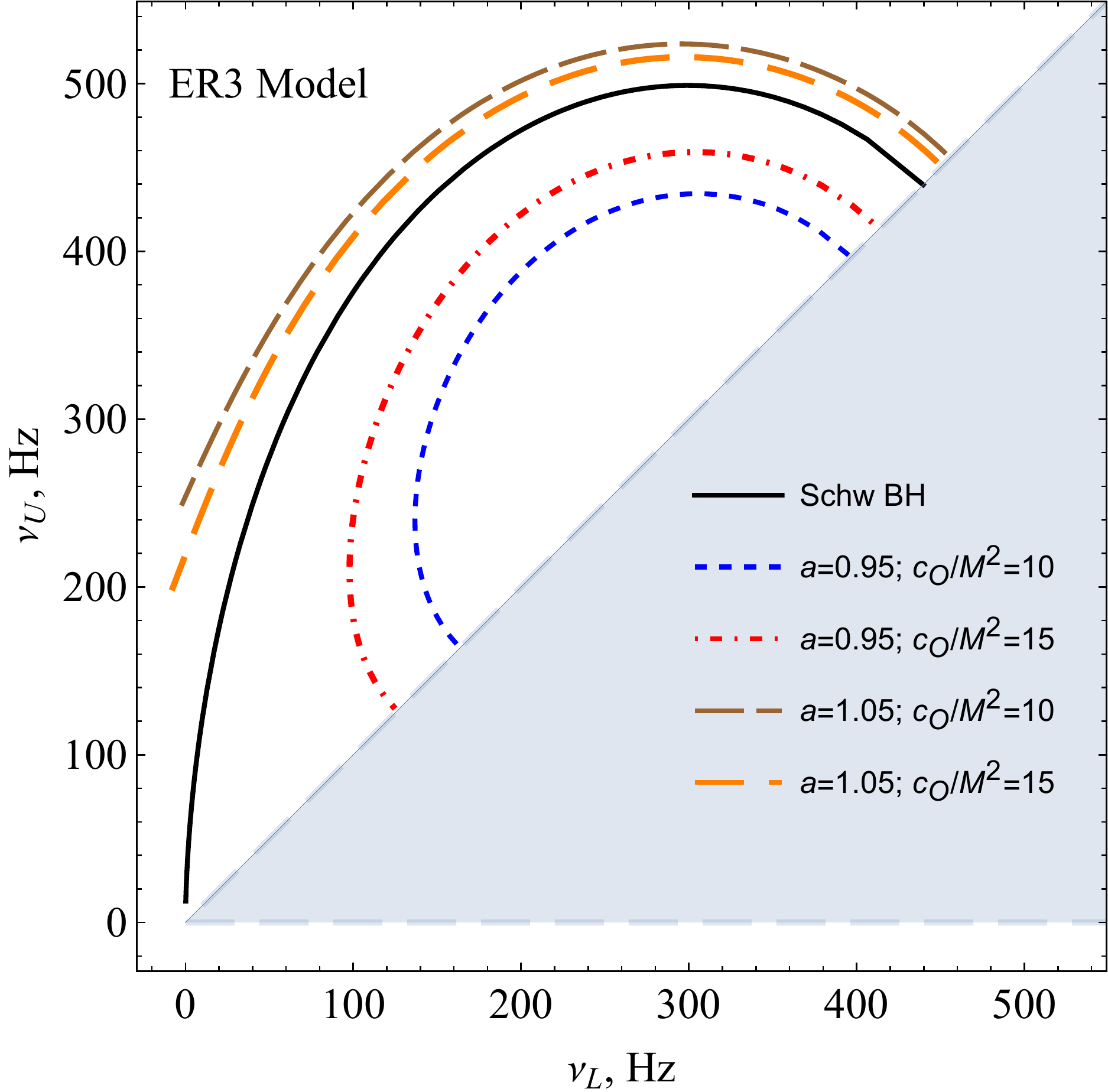}
\caption{The upper ($\nu_U$) and lower ($\nu_L$) frequencies of the twin-peak QPOs around a symmergent black hole of mass $M=5M_\odot$. The upper-left, upper-right, bottom-left and bottom-right panels correspond to the RP, WD, ER2 and ER3 models, respectively. Each panel is a $\nu_L$--$\nu_U$ plane (in ${\rm Hz}$), with various curves resulting from the symmergent gravity parameters $a=0.95\ \&\ 1.05$ and $c_{\rm O}/M^2=10\ \&\ 15$. The $\nu_U$--$\nu_L$ curves are distributed above ($a>1$) and below ($a<1$) the Schwarzschild curve, where $c_{\rm O}/M^2$ values reveal finer structures.  \label{upperlow}}
\end{figure*}

Possible values of the upper and lower frequencies of twin-peak QPOs around a symmergent gravity black hole are plotted in Figure~\ref{upperlow} for the RP, WD, ER2, and ER3 models. The inclined blue-solid $\nu_U=\nu_L$ line is the deadline for twin-peak QPOs since their two peaks coincide for $\nu_U=\nu_L$. The light-blue shaded area is the graveyard for twin-peak QPOs in that no such QPOs can be observed in this domain of $\nu_U$ and $\nu_L$. It is observed from the figure that no low-frequency twin peak QPOs occur. This is due to the presence of a non-zero $a$ parameter because for $a<1$ ratios of the frequencies slightly increase but, for $a>1$, the same ratios decrease compared to the Schwarzschild case ($a=1$). It is also observed from the ER2 and ER3 models that two different QPOs could be observed with similar frequency ratios for upper (lower) frequencies above 450 Hz (250 Hz). Frequency degeneracies as such are observed also at low-frequency twin-peak QPOs around 100 Hz in the RP and WD models (similar low-frequency behavior occurs between 100-200 Hz in the ER3 model).  

\subsection{QPO orbits}

In this part, we shall study the relationships between the parameters $c_{\rm O}$ and $a$ as well as the radii of the QPO orbits in RP, WD, ER2, and ER3 models. To this end, for the ratio of the upper and lower frequencies we use the solutions of the following equations

\begin{eqnarray}
    3\nu_L(r;a, c_{\rm O})=2\nu_U(r; a, c_{\rm O})\ , \\ \quad 4\nu_L(r; a, c_{\rm O})=3\nu_U(r; a, c_{\rm O})\ , \\ \quad 5\nu_L(r; a, c_{\rm O})=4\nu_U(r; a, c_{\rm O})\ ,
\end{eqnarray}
which are too complicated to admit a closed-form analytic solution. Thus, we will solve them  numerically and analyze the effects of symmergent gravity on the QPOs graphically in the frameworks of the RP, WD, ER2 and ER3 models.   

\begin{figure*}[ht!]
   \centering
   \includegraphics[width=0.48\linewidth]{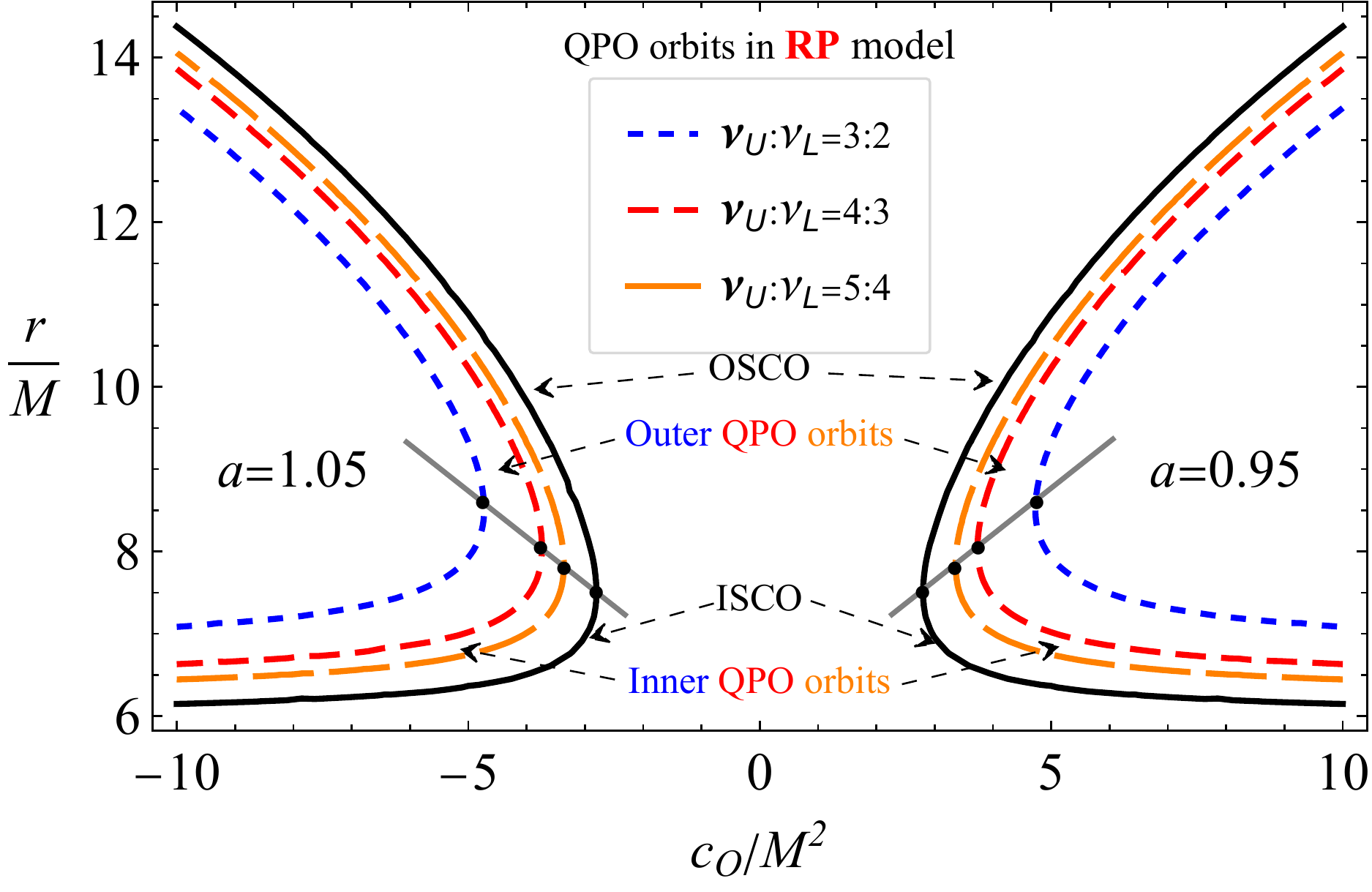}
   \includegraphics[width=0.48\linewidth]{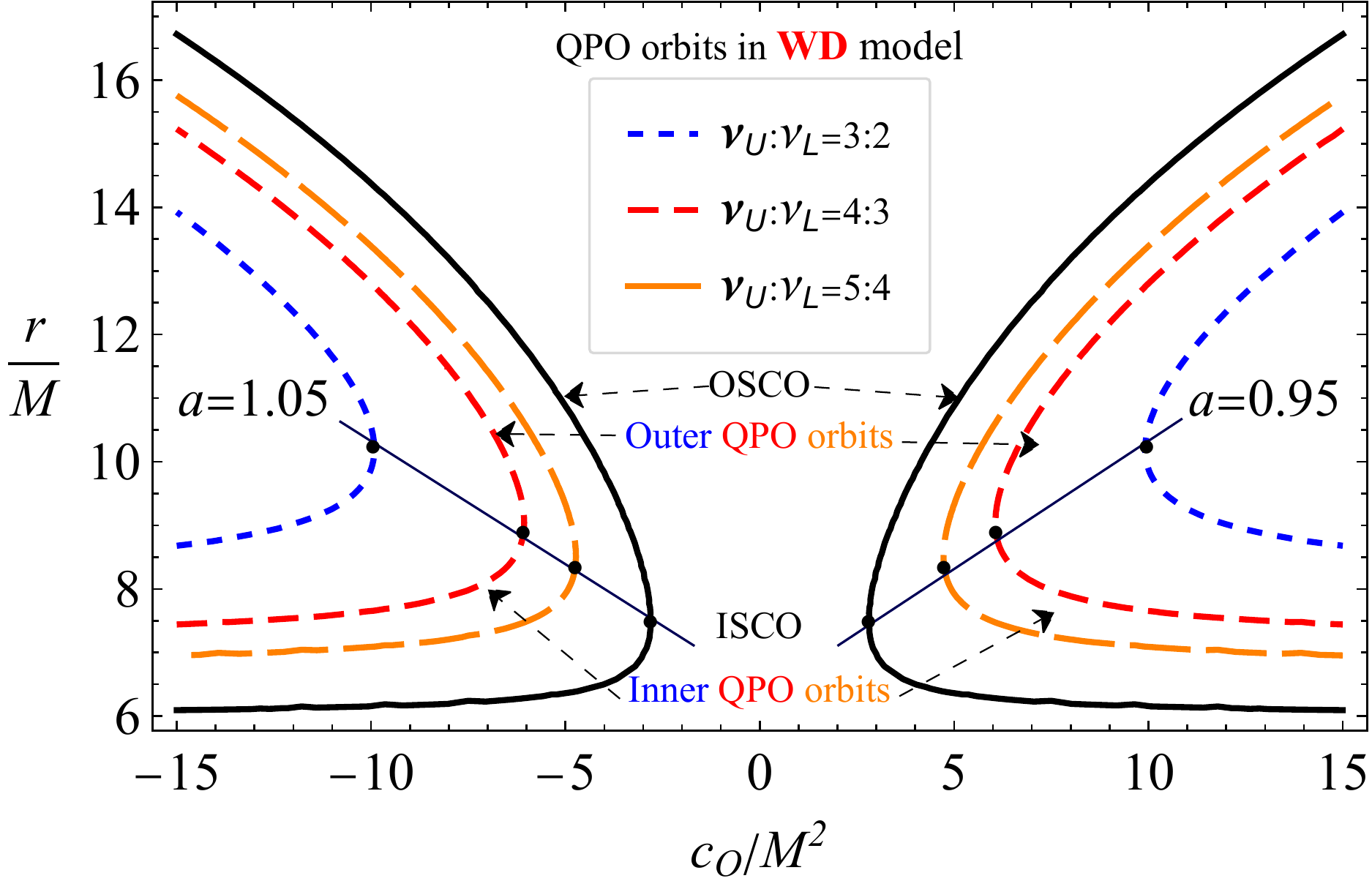}
    \includegraphics[width=0.48\linewidth]{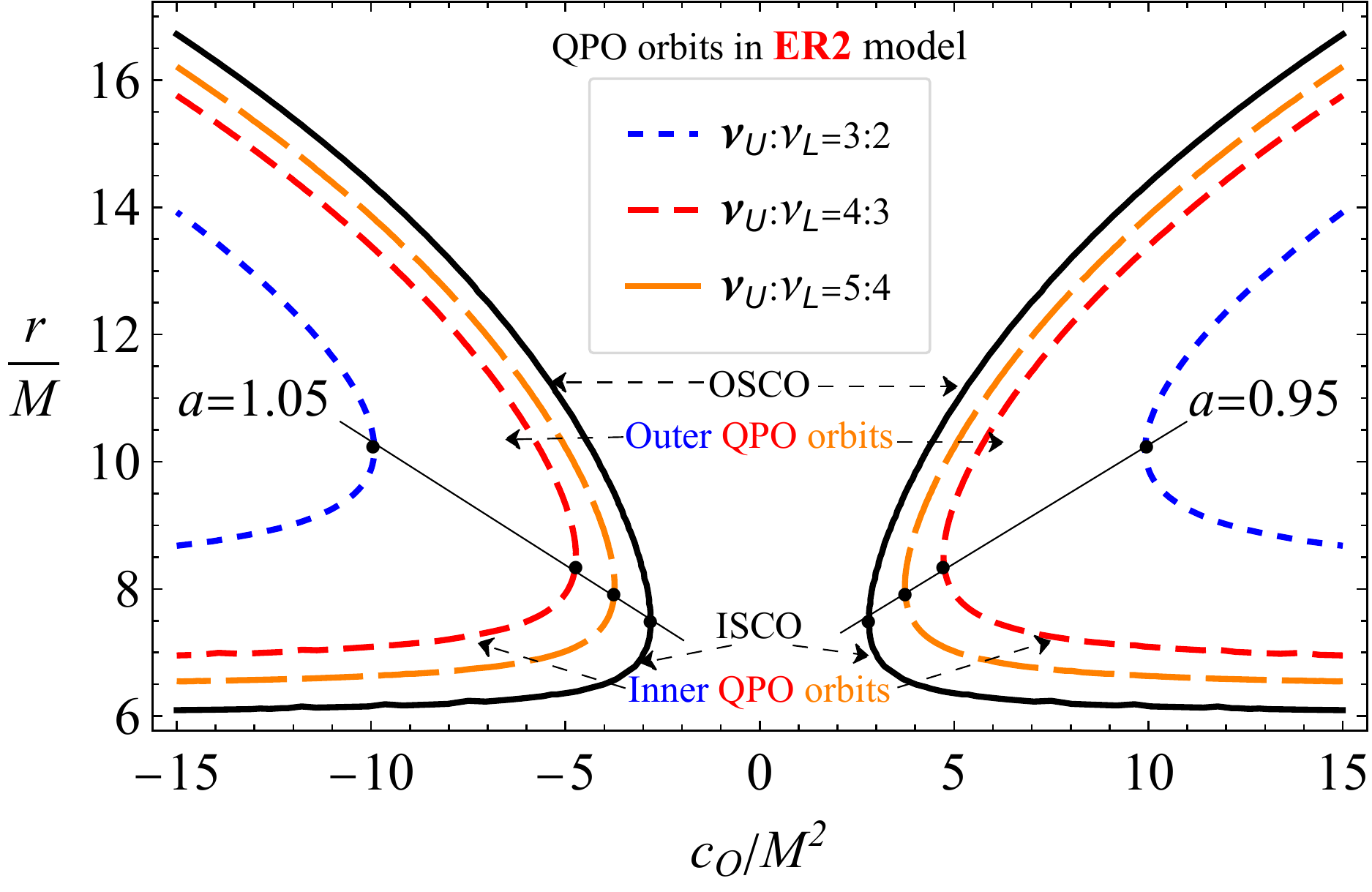}
       \includegraphics[width=0.48\linewidth]{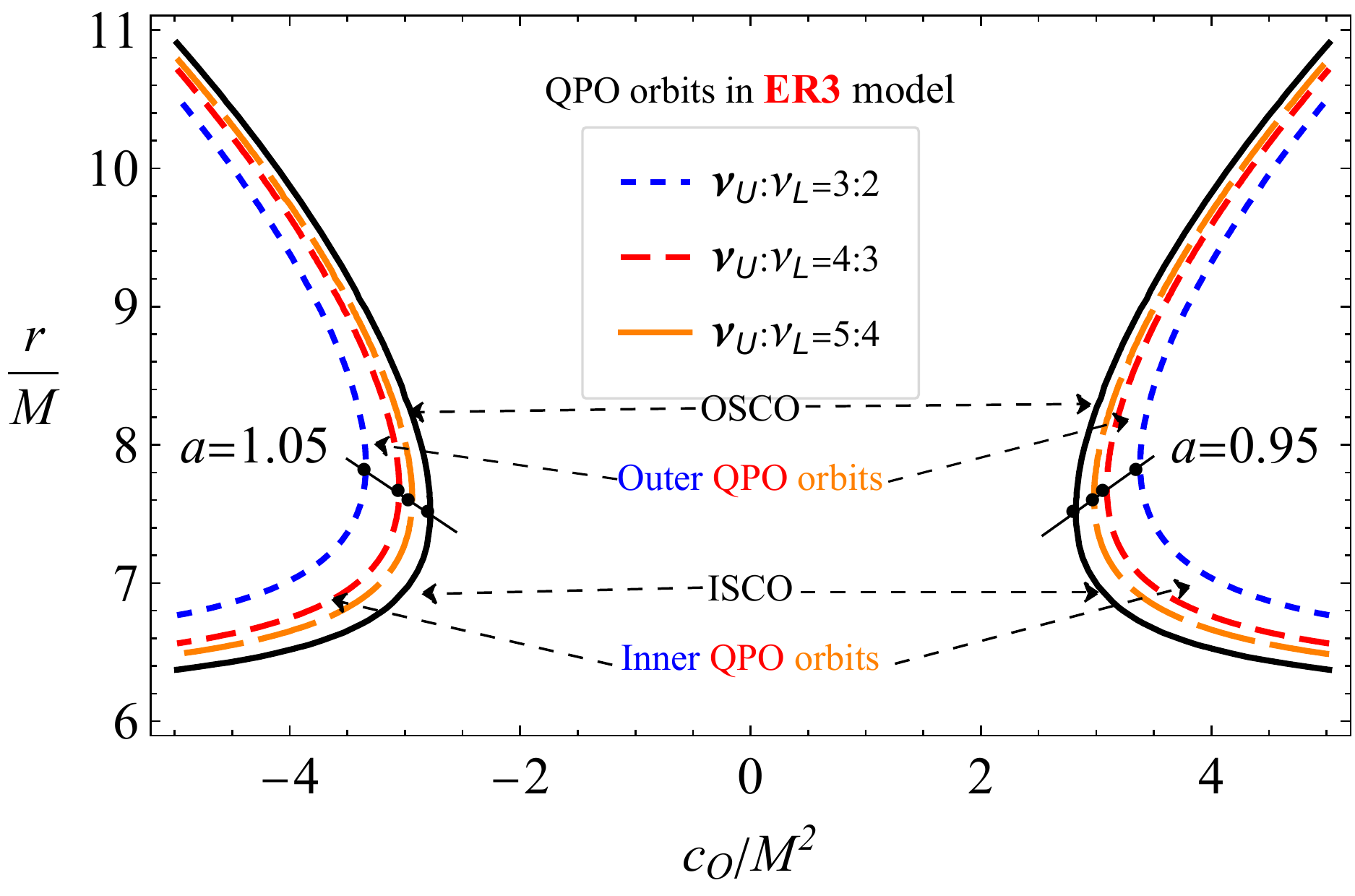}
\caption{The QPO orbits with (O)ISCO in the  RP (upper-left panel), WD (upper-right panel), ER2 (lower-left panel) and ER3 (upper-right panel) models. The QPO orbits lie near (away) from the ISCOs/OSCOs in the RP and ER3 (ER2 and WD) models.
\label{rqpo}}
\end{figure*}

Illustrated in Fig.~\ref{rqpo} are the radii of the QPO orbits as a function of the parameter $c_{\rm O}$ for $a=0.95$ and $a=1.05$ in the QPO models of RP (upper-left panel), WD (upper-right panel), ER2 (lower-left panel) and ER3 (upper-right panel). In each case, we indicate the OSCO and ISCO orbits, as well as the inner and outer QPO orbits. The QPO orbits are located outside the ISCOs and inside the OSCOs. As was mentioned in Fig.~\ref{isco}, the accretion disc is limited by ISCO and OSCO. In Fig.\ref{rqpo}, too, we have found that the outer border in the orbits of QPOs is due to the presence of the symmergent gravity parameters $a$ and $c_{\rm O}$. It follows from the figure that QPO orbits are located near ISCOs/OSCOs in the RP and ER3 models. In the ER2 and WD models, however, QPO orbits are located away from the ISCOs/OSCOs. In general, ISCOs are one of the most important features of the black holes, and their determinations by astrophysical observations are a major issue. To this end, this issue can be resolved by the QPO studies in the RP and ER3 models. Our numerical calculations show that the distance between the QPOs orbits (with the ratio 5:4) and ISCO consists of the ISCO radii less than 5-7 \% in RP and ER3 models when $a=0.95$ and $a=1.05$. It is around the error sizes in astrophysical observations, and one can take therefore the RP and ER3 QPO orbits as ISCOs.

\section{Constraints on symmergent gravity parameters from QPO frequencies} \label{sec6}

In this section, we determine constraints on the symmergent gravity parameters $a$ and $c_{\rm O}$ by using the data from QPOs in the following astrophysical objects: 
\begin{itemize}
\item GRS 1915-105 is observed in upper and lower frequencies $168\pm5$ \& $113\pm3$, respectively, with central black hole mass $12.4^{+2.0}_{-1.8} M_\odot$ \cite{Reid2014ApJ}
    \item GRO J1655-40, powered by central black hole with mass $(5.4\pm0.4)M_\odot$, observed in the high frequencies $441\pm 2$ \& $298\pm 4$ and low frequency $17.3\pm 0.1$ Hz \cite{Stuchlik2015MNRAS,Stuchlik2016AA}.
    \item XTE J1550-564, mass of the black hole at the centre of the microquasar is $(9.1\pm0.61)M_\odot$ and it is detected in $276\pm3$\& $184\pm5$ Hz ~\cite{Stuchlik2013AA,Miller2001ApJ},
    \item H1743+322 microquasar has been found in the frequency band of electromagnetic spectrum $242\pm3$ \& $166\pm5$ Hz, mass of the central black hole lies in the range of between 8 and 14.07 solar mass ~\cite{Stuchlik2015MNRAS,Remillard2006ApJ},
\item a QPO have been detected around SgrA* located at the center of Milky Way galaxy, with the low frequency $1.445 \pm 0.16$ and $0.866\pm0.04$ mHz ~\cite{Vrba2021JCAP,Abramowicz2004ragt}. 
    \end{itemize}
Once the radii of the QPO orbits are determined, it becomes possible to reveal the constraints on the symmergent gravity parameters $a$ and $c_{\rm O}$ by solving the equations 
\begin{equation}
    \nu_{\rm U}(r;a,c_{\rm O})=\nu_{\rm U}^{ob}, \qquad \nu_{\rm L}(r;a,c_{\rm O})=\nu_{\rm L}^{ob},
\end{equation}
for each of the RP, WD, ER2 and ER3 models. In these equations,  $\nu_{\rm U}^{ob}$ and $\nu_{\rm L}^{ob}$ are observational values of the upper and lower frequencies of a given QPO. In order to get relations between the allowed values of the symmergent parameters, we use observational results on the above-mentioned astrophysical objects.

\begin{figure*}[ht!]
   \centering
   \includegraphics[width=0.48\linewidth]{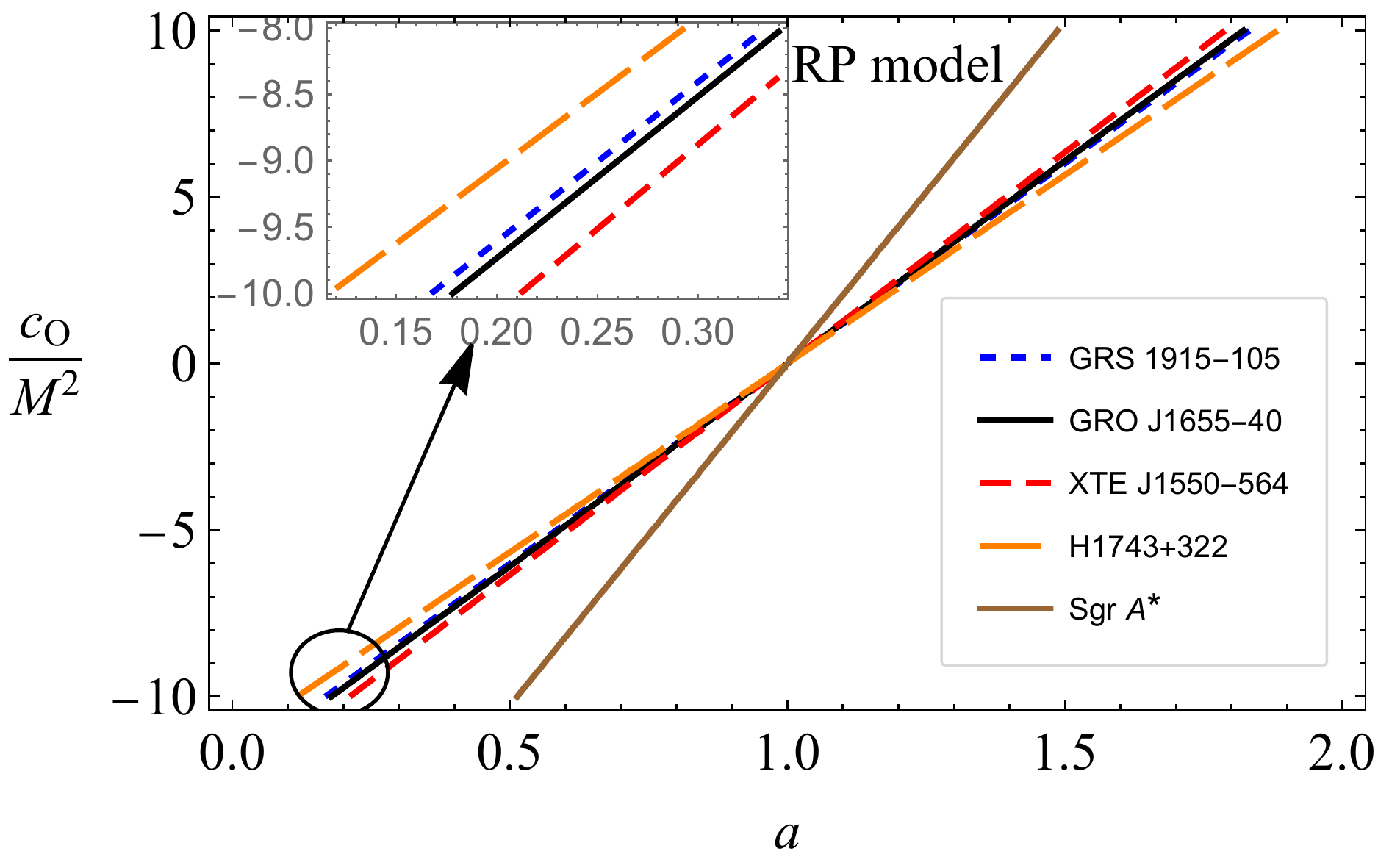}
   \includegraphics[width=0.48\linewidth]{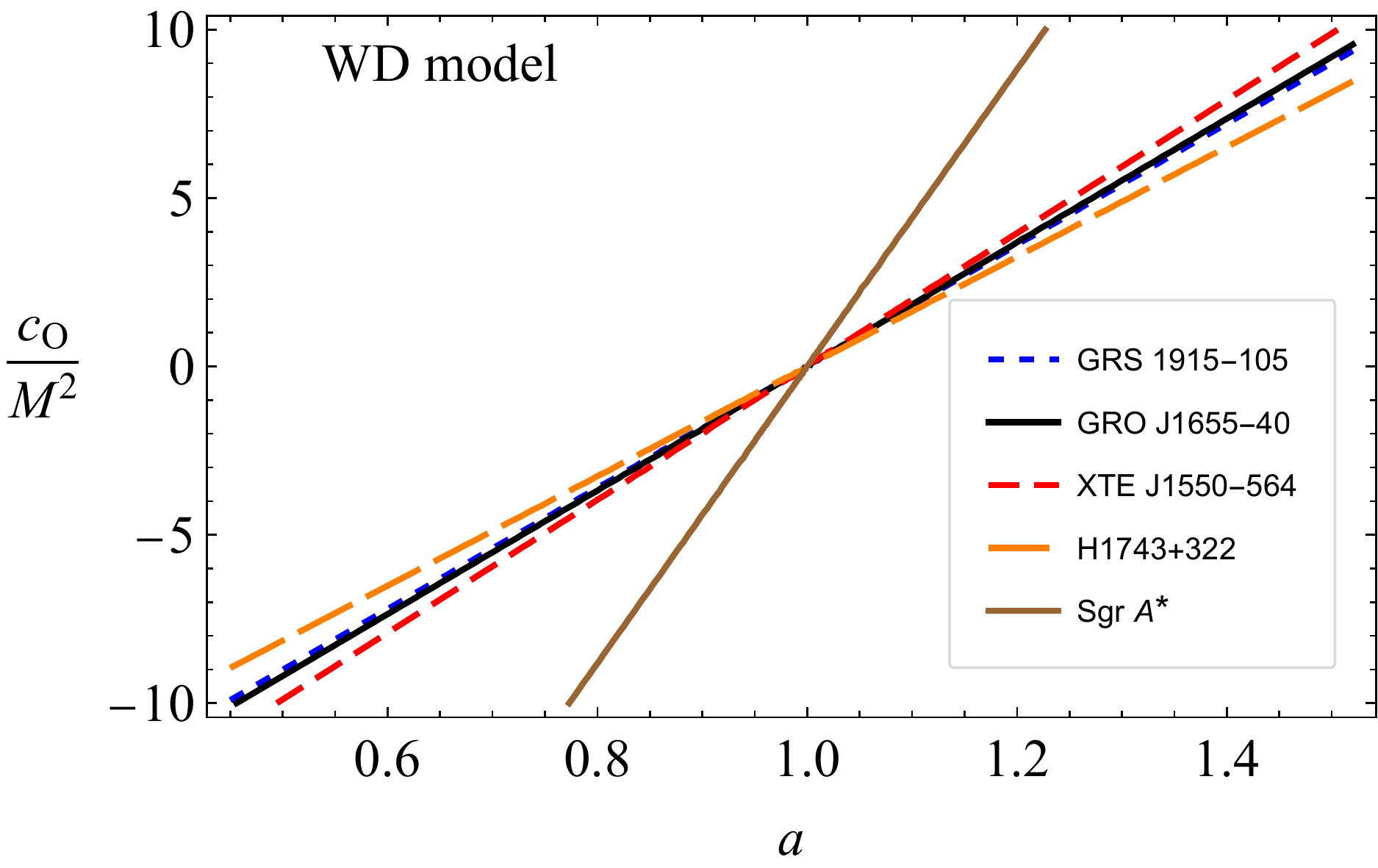}
   \includegraphics[width=0.48\linewidth]{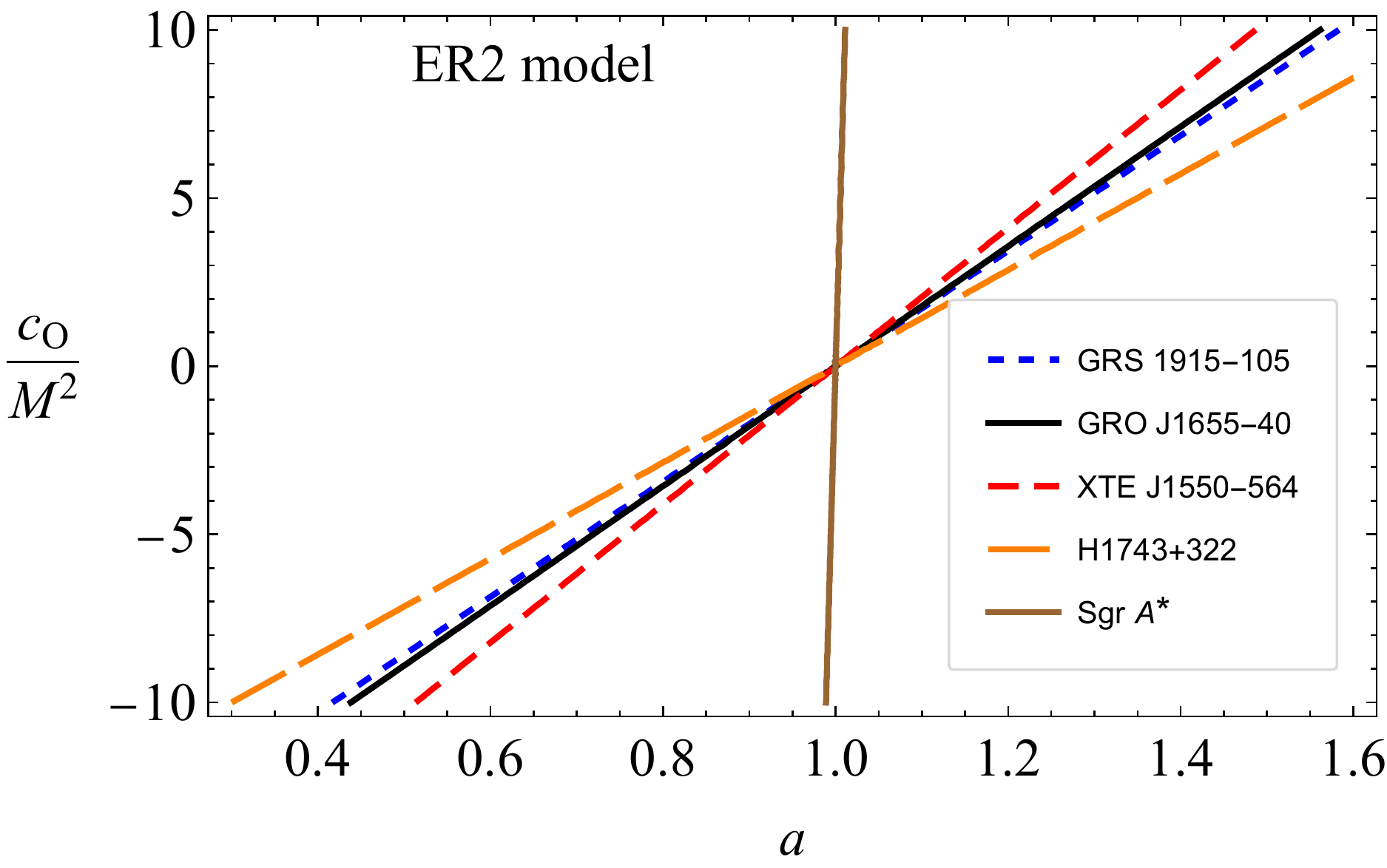}
   \includegraphics[width=0.48\linewidth]{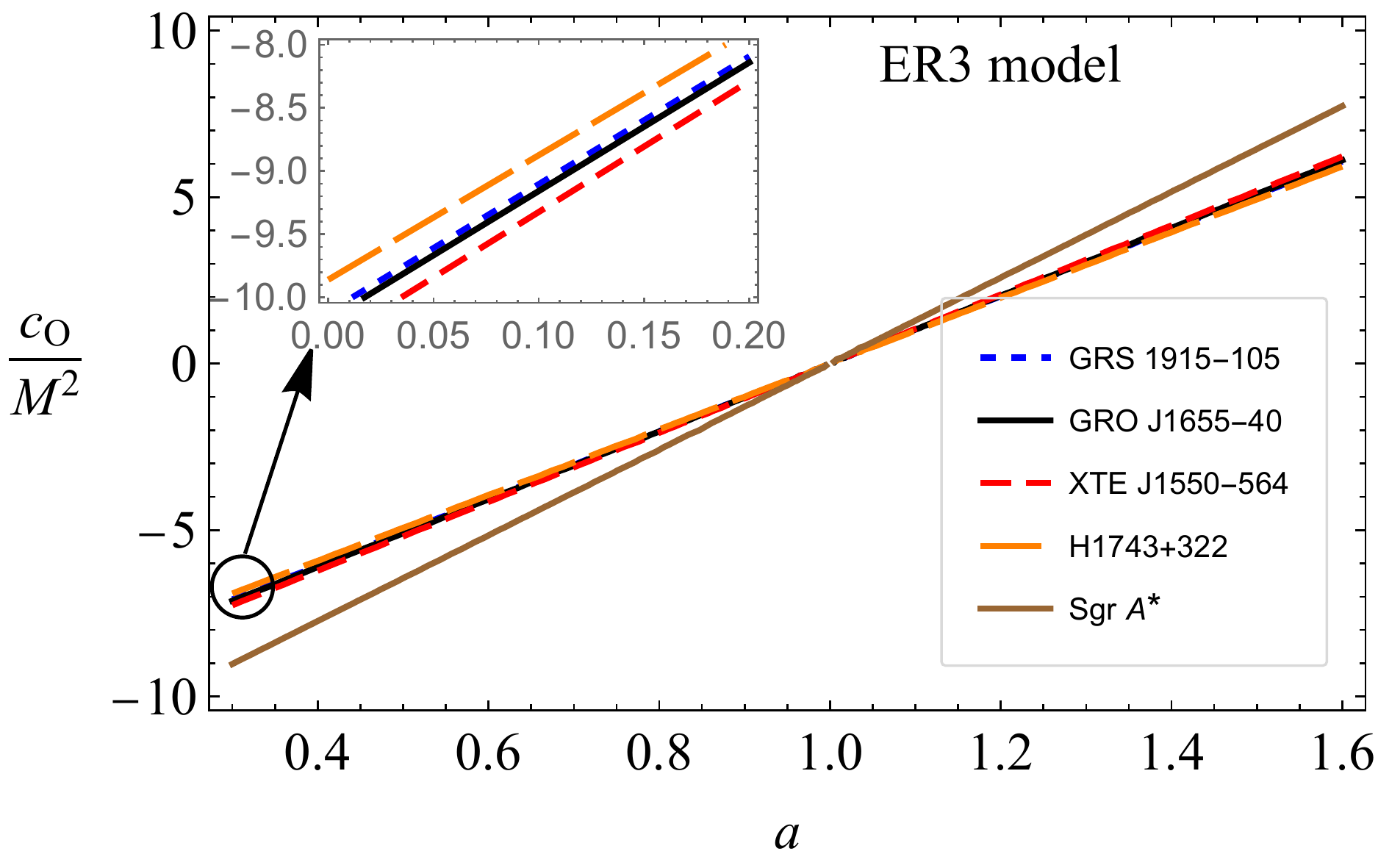}
\caption{The relationship between the symmergent gravity parameters $c_{\rm O}$ and $a$ according to the QPO objects  GRS 1915-105, GRO J1655-40, XTE J1550-654, H1743+322 and Sgr A*. It is clear that the Sgr A* exhibits strong model-dependence and deviates significantly from the microquasar data (ER3 might be an exception, depending on the precision goal). \label{avscGRS}}
\end{figure*}

Figure~\ref{avscGRS} shows relationships between $a$ and $c_{\rm O}$ parameters in symmergent gravity for the twin-peak QPO objects GRS 1915-105, GRO J1655-40, XTE J1550-564, H1743+322, and Sgr A* in the RP (top-left), WD (top-right), ER2 (bottom-left), and ER3 (bottom-right) models. It is clear that $a$ and $c_{\rm O}$ exhibit a linear relationship. It is observed that the data from GRS 1915-105, GRO J1655-40, XTE J1550-564, and H1743+322 exhibit a nearly model-independent behavior.  These microquasars have a frequency ratio of 3:2 and stellar-mass black holes at their centers. The data from Sgr A* exhibits strong model dependence and deviates significantly from the microquasar data. It may be that the existing QPO models (RP, WD, ER2, and ER3) are not sufficiently accurate to cover the Sgr A* data.

\section{Weak deflection angle of light by symmergent black hole for a finite-distance observer and source} \label{sec7}
In this section, we determine the weak deflection angle by a symmergent black hole using the Gauss-Bonnet theorem (GBT) approach developed by 
Ishihara and others \cite{Ishihara:2016vdc,Ishihara:2016sfv}. To this aim, we explore weak deflection angle as a sensitive probe of the loop factor $c_{\rm O}$. The GBT is expressed as \cite{Gibbons:2008rj}
\begin{align}
\iint_{T} K dS + \sum_{a=1}^N \int_{\partial T_a} \kappa_g d\ell +
\sum_{a=1}^N \theta_a = 2\pi ,  \label{GB}
\end{align}
in which $T$ is a two-dimensional orientable surface with boundaries $\partial
T_a$ ($a=1, 2, \cdots, N$) that are differentiable curves, $\theta_a$ ($a=1,
2, \cdots, N$) denote jump angles, $K$ denotes the Gaussian curvature of the
surface $T$, $dS$ is the area element of the surface, $\kappa_g$ means the
geodesic curvature of $\partial T_a$, and $\ell$ is the line element along
the boundary. By applying this fundamental formula to the quadrilateral $%
{}^{\infty}_{R}\Box^{\infty}_{S}$, which consists of the spatial curve for the light ray, two outgoing radial lines from $R$ and  $S$ and a circular arc segment $C_r$ of coordinate radius $r_C$ ($r_C \to \infty$) centered at the lens $L$ which intersects the radial lines through the receiver or the source, Ishihara et al. \cite{Ishihara:2016vdc} obtained the formula
\begin{align}
\alpha &= \Psi_R - \Psi_S + \phi_{RS}  
= - \iint_{{}^{\infty}_{R}\Box^{\infty}_{S}} K dS   \label{alpha-K}
\end{align}
from which it follows that   $\alpha$ is an invariant and remains well-defined even if $L$ is a singularity point since the domain ${}^{\infty}_{R}\Box^{\infty}_{S}$ does not contain the point $L$. It follows that $\alpha=0$ in Euclidean space. This is because $K$ vanishes in Euclidean space, and the area integral of $K$ thus vanishes. Furthermore, \eqref{alpha-K} proves useful for asymptotically flat spacetimes, and cannot be applied to non-asymptotically flat ones that have the metric (lapse) function in Eq.~\eqref{smetric}.

To solve this problem, Ishihara and others \cite{Ishihara:2016vdc} used the main definition $\Psi_R - \Psi_S + \phi_{RS}$ in Eq.~\eqref{alpha-K}, and focused on the calculation of the positional angle $\Psi_S$ ($\Psi_R$) between the radial direction and the light ray at the source (receiver) position, where the relative angle between the sources and the receiver is given by   $\phi_{RS}$.  Rewriting the metric in Eq.~(\ref{metric}) as
\begin{align}
ds^2 &= g_{\mu\nu} dx^{\mu} dx^{\nu}  \nonumber \\
&= -A(r) dt^2 + B(r) dr^2 + C(r) d\Omega^2 ,  \label{ds2-SSS-AB}
\end{align}
with the solid angle $d\Omega^2 \equiv d\theta^2 + \sin^2\theta d\phi^2$ and azimuthal angle $\phi$.  Considering the photon orbital plane at $\theta = \pi/2$, the angle $\Psi$ is found to be \cite{Ishihara:2016vdc}
\begin{equation} \label{e27}
    \sin\Psi=\sqrt{\frac{A(r)}{C(r)}}b,
\end{equation}
where $b$ is the impact parameter of light. Using the lapse function \eqref{smetric},  one finds
\begin{align}
    \Psi&=\arcsin(bu)-\frac{bMu^{2}}{\sqrt{1-b^{2}u^{2}}}+\frac{a-1}{48\pi c_{O}}\frac{b}{u\sqrt{1-b^{2}u^{2}}} \nonumber\\
    &- \frac{a-1}{48\pi c_{O}}\frac{bM\left(2b^{2}u^{2}-1\right)}{\left(1-b^{2}u^{2}\right)^{3/2}}.
\end{align}
as the explicit expression of $\Psi$, with the variable $u=1/r$.  Taking into account the finite distance between the lensing object and the source and the receiver,  one gets 
\begin{equation}
    \Psi_{\text{R}}-\Psi_{\text{S}} = \Psi|_{u\rightarrow u_{\text{R}}} - \pi + \Psi|_{u\rightarrow u_{\text{S}}}
\end{equation}
where $u_R$ ($u_S$) is the inverse-position of the receiver (source). The methodology used by Ishihara and others \cite{Ishihara:2016vdc} determines the source-receiver angle $\phi_{RS}$ as,
\begin{equation}
\label{phiRS}
   \phi_{RS} =\int_{u_{R}}^{u_{o}}\frac{1}{\sqrt{F(u)}}du+\int_{u_{S}}^{u_{o}}\frac{1}{\sqrt{F(u)}}du,
\end{equation}
where the orbit equation $F(u)$ is given by 
\begin{equation}
    F(u) \equiv \left(\frac{du}{d\phi}\right)^2 = \frac{u^4C(u)(C(u)-A(u)b^2)}{A(u)B(u)b^2}
\end{equation}
after setting $ds^2=0$ in \eqref{ds2-SSS-AB}. The inverse-position $u_o$ in (\ref{phiRS}) is the inverse of the closest approach, which can be determined by an iterative solution of $F(u)$. In fact, $u_o=\sin\phi/b$ turns out to be sufficient so that the azimuthal angle is found to be
\begin{align}
    \phi&=\arcsin(bu)+\frac{M}{b}\frac{b^2u^2-2}{\sqrt{-b^2u^2+1}}+\frac{a-1}{48\pi c_{O}}\frac{b^{3}u}{\sqrt{1-b^{2}u^{2}}} \nonumber\\
    &-\frac{a-1}{48\pi c_{O}}\frac{bM\left(3b^{2}u^{2}-2\right)}{\left(1-b^{2}u^{2}\right)^{3/2}},
\end{align}
with which one finds 
\begin{equation}
    \phi_{\text{RS}}=\phi_{\text{R}}-\phi_{\text{S}}.
\end{equation}
such that $\phi_{\text{R}}=\pi-\phi|_{u\rightarrow u_{\text{R}}}$ and   $\phi_{\text{S}}=\phi|_{u\rightarrow u_{\text{S}}}$ are the azimuthal angles of the receiver and the source, respectively \cite{Takizawa:2020egm,Ono:2019hkw}. 
Having determined all three angles in Eq.~(\ref{alpha-K}), one arrives at the angle formula
\begin{align}
\hat{\alpha} = & \frac{2M}{b}\left(\sqrt{1 - b^2 u^2_S} + \sqrt{1 - b^2
u^2_R}\right)  \nonumber \\
& + \frac{(a-1)b}{48\pi c_{O}}\left(\frac{\sqrt{1 - b^2 u^2_S}}{u_S} + \frac{\sqrt{1
- b^2 u^2_R}}{u_R}\right)  \nonumber \\
& - \frac{(a-1)bM}{48\pi c_{O}}\left(\frac{1}{\sqrt{1 - b^2 u^2_S}} + \frac{1}{%
\sqrt{1 - b^2 u^2_R}}\right)  \label{alpha-Kottler}
\end{align}
whose second line is seen to diverge as $r\to \infty$. This apparent divergence implies that the symmergent black hole does not allow $r \to \infty$ ($u \to 0$) limit. In spite of this, it is possible to regularize $\hat{\alpha}$ by choosing $u$ small yet finite (in accordance with the realistic situations) so that the formula \eqref{alpha-Kottler} takes its small-$u$ apporximate form
\begin{equation} \label{alpha-Kottler2}
    \alpha = \frac{4M}{b} + \frac{(a-1)b}{48\pi c_{O}}\left(\frac{1}{u_S} + \frac{1}{u_R}\right)-\frac{(a-1)bM}{24\pi c_{O}}.
\end{equation}

\begin{figure*}[ht!]
   \centering
\includegraphics[width=0.45\linewidth]{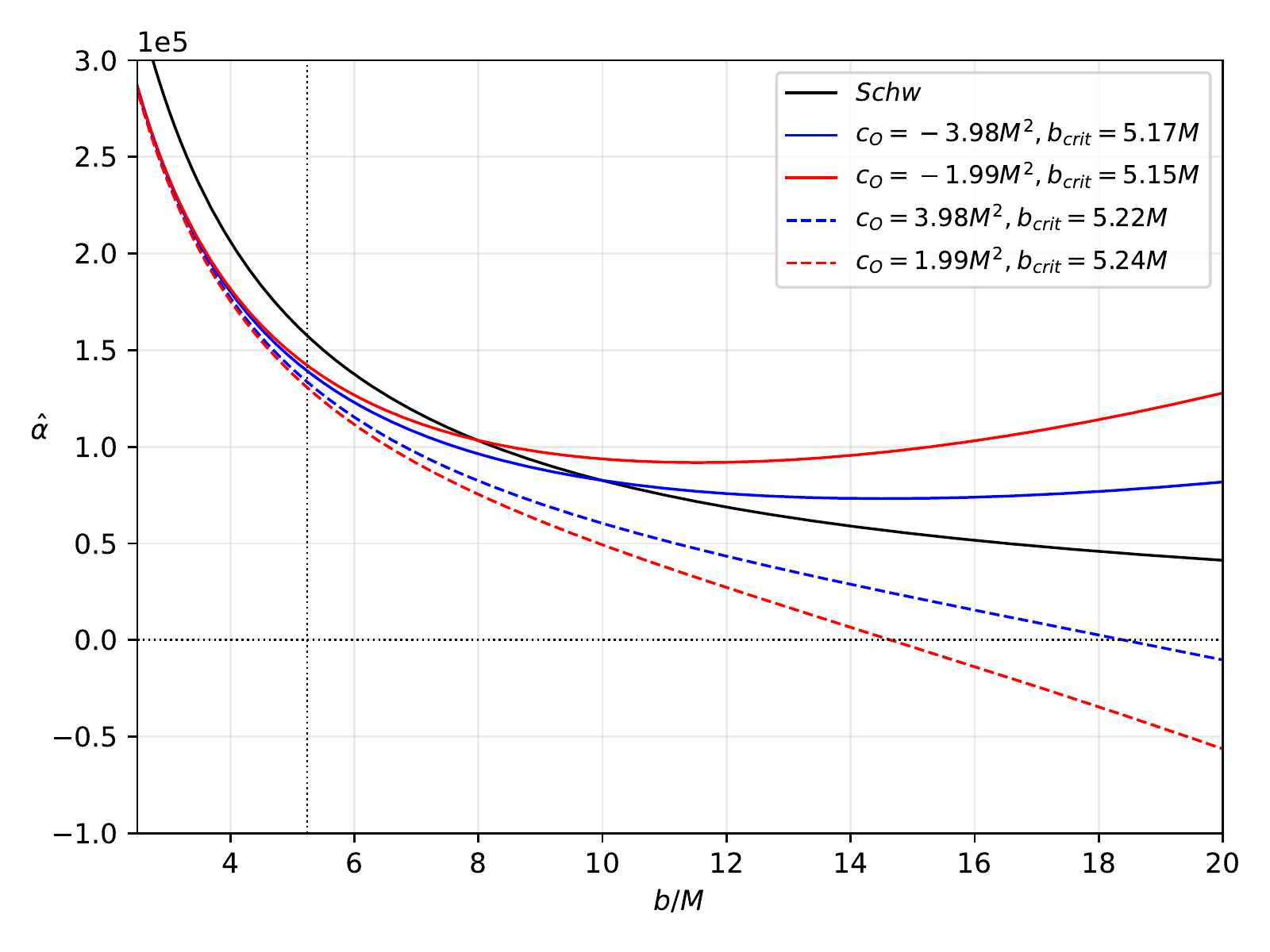}
\includegraphics[width=0.45\linewidth]{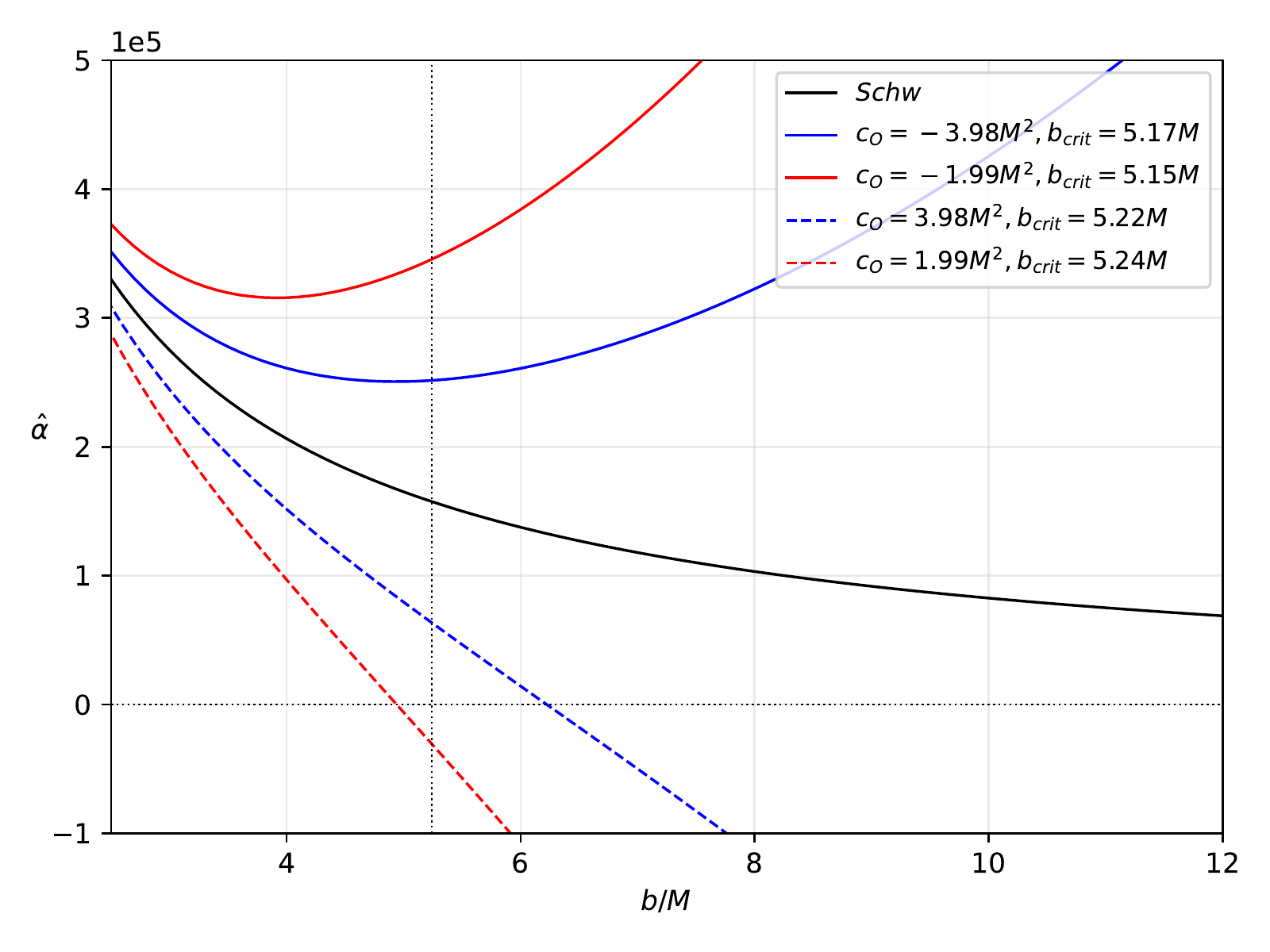}
\caption{The weak deflection angle ($\mu$as) as a function of the impact parameter $b$ for different values of the symmergent gravity parameters. The two panels clearly show the difference between the nearby-case where $u=0.5/b$ (left-panel depicting equation \eqref{alpha-Kottler}) and the faraway-case where $u=0.02/b$ (right-panel depicting equation \eqref{alpha-Kottler2}). In the plots, we set $a=0.9$, took $u_\text{S}=u_\text{R}=u$, and included the highest critical impact parameter indicated by the vertical dotted line ($b_{\rm crit}/M \simeq 5.1$). \label{wda1}}
\end{figure*}

The deflection angle ${\hat \alpha}$ is depicted in Figure \ref{wda1}. The left-panel shows the exact formula \eqref{alpha-Kottler} as a function of the impact parameter $b$ for nearby observers ($u_{\rm R}=u_{\rm S}=0.5/b$), with the parameter values $a=0.9$ and $c_{\rm O}/M^2=\pm 1.99$ and $\pm 3.98$. The right-panel of Figure \ref{wda1}, on the other hand, shows the small-$u$ limit in \eqref{alpha-Kottler2} as a function of the impact parameter $b$ for faraway observers ($u_R=u_S=0.02/b$), with the parameter values $a=0.9$ and $c_{\rm O}/M^2=\pm 1.99$ and $\pm 3.98$. In both panels, the Schwarzschild  case (black curve) decreases monotonically as $b/M \to \infty$. It is observed that  $c_{\rm O}>0$ ($c_{\rm O}<0$) mimics the dS (AdS) behaviour. This comes to mean that $n_{\rm B}>n_{\rm F}$ ($n_{\rm B}<n_{\rm F}$) leads to dS (AdS) behaviours. Interestingly, for an observer near  the black hole, when $b/M \to b_{\rm crit}/M$, $\hat{\alpha}$ remains below the Schwarzschild deflection angle such that the AdS case ($c_{\rm O}<0 \Rightarrow n_{\rm B}<n_{\rm F}$) stays slightly higher. At high values of $b/M$, the AdS (dS) case leads to an increasing (decreasing and vanishing) $\hat{\alpha}$. Meanwhile, an observer far from the black hole perceives a higher value for $\hat{\alpha}$ around $b_\text{crit}/M$ for the AdS case. Clearly, $\hat{\alpha}$ does not exist at all in the dS case, even for low values of $b/M$. These two panels reveal the effects of $a$ and $c_\text{O}$ on the deflection angle. Here, we remark that negative deflection angle in the dS ($c_{\rm O}>0 \Rightarrow n_{\rm B}>n_{\rm F}$) case comes to mean that at such values of $b/M$ the photons are repelled. There are certain cases where repulsive deflection can also happen \cite{Nakashi2019}. Nonetheless, one notes that negative deflection angle does not produce the Einstein ring, as we shall see in the next discussion.

\begin{figure}[h]
   \centering
    \includegraphics[width=\linewidth]{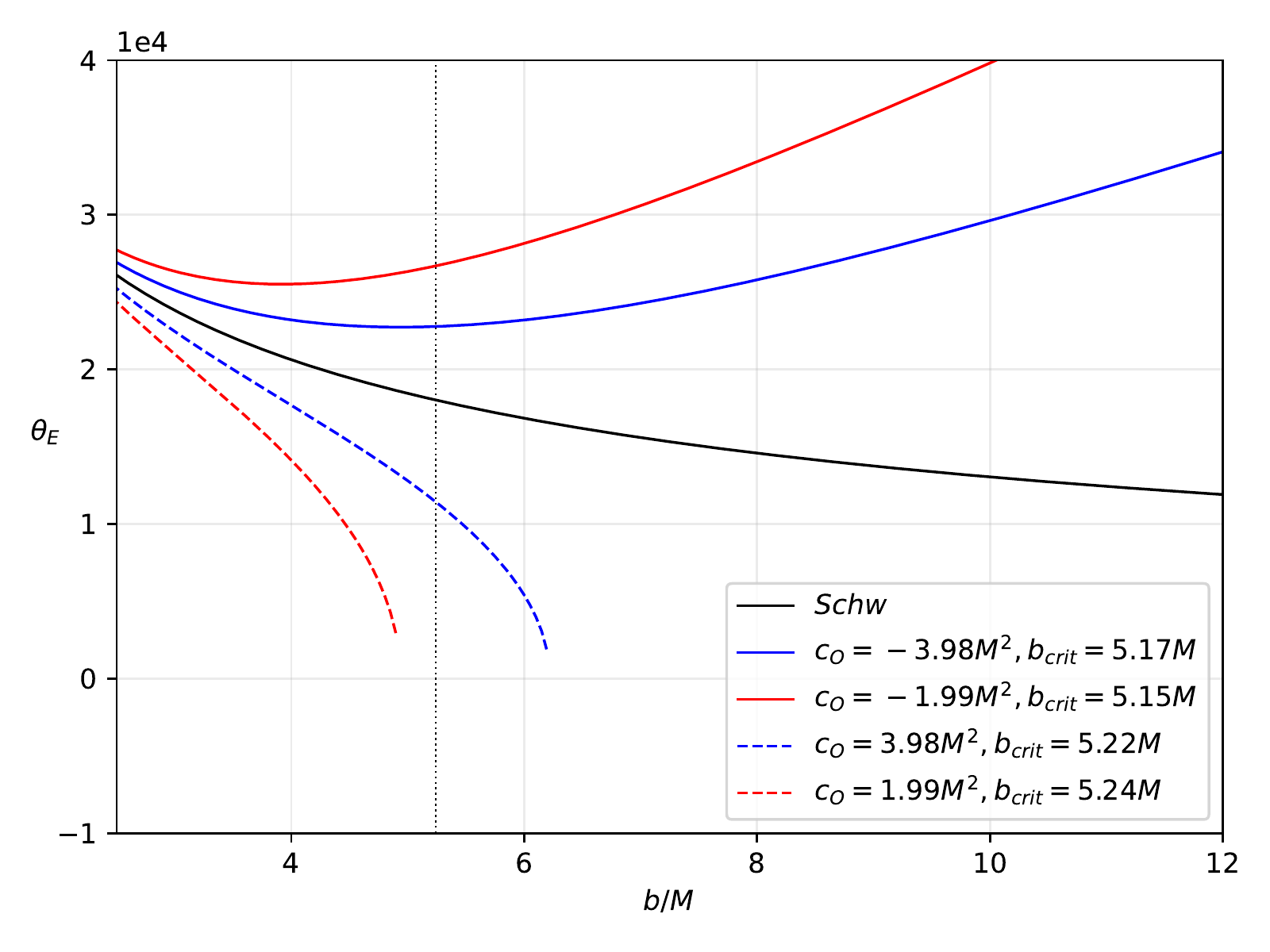}
    \caption{The Einstein ring ($\mu$as) as a function of the impact parameter for $a=0.9$, and $c_{\rm O}/M^2=\pm 1.99$ and $\pm 3.98$. This figure corresponds to the right-panel of Figure \ref{wda1}. In this plott we set $a=0.9$, took $d_\text{S}=d_\text{R}=50b$, and included the highest critical impact parameter indicated by the vertical dotted line.}
    \label{erings}
\end{figure}

The weak deflection angle has direct application in a phenomenon called Einstein ring. Here, we will determine the effects of the symmergent gravity parameters on the Einstein ring. To begin with, let us define the distance of the source and the receiver with respect to the lensing object as $d_\text{S}$ and, $d_\text{R}$ respectively. Through the thin lens equation, we then have $d_\text{RS}=d_\text{S}+d_\text{R}$ so that the position of the weak field images is given by \cite{Bozza2008}
\begin{equation}
    d_\text{RS}\tan\beta=\frac{d_\text{R}\sin\theta-d_\text{S}\sin(\hat{\alpha}-\theta)}{\cos(\hat{\alpha}-\theta)}.
\end{equation}
It is well known that an Einstein ring is formed when $\beta=0$, and the above equation leads then to the angular radius  \cite{Virbhadra:1999nm,Virbhadra:2002ju,Bozza:2001xd,Bozza:2002zj,Hasse:2001by,Perlick:2003vg}
\begin{equation}
\label{approx-wda}
    \theta_\text{E}\approx \frac{d_\text{S}}{d_\text{RS}}\hat{\alpha}.
\end{equation}
In addition, since the Einstein ring is assumed to be small it is safe to take relation $b=d_\text{R}\sin\theta \sim d_\text{R}\theta$. Then,  the weak deflection angle (\ref{approx-wda}) takes the explicit form 
\begin{equation} \label{eqring}
    \theta_\text{E}\approx \sqrt{\frac{d_\text{S}}{d_\text{RS}d_\text{R}}\left[4M+\frac{(a-1)(d_\text{S}+d_\text{R})}{48\pi}\epsilon-\frac{(a-1)M}{24\pi}\epsilon\right]}
\end{equation}
after introducing $\epsilon=b^2/c_\text{O}$. This angular radius is depicted in Figure \ref{erings} as a function of the impact parameter for $a=0.9$  and $c_{\rm O}/M^2=\pm 1.99$ and $\pm 3.98$. This plot corresponds to the right-panel in Figure \ref{wda1}. The sensitivity of the Einstein ring to the model parameters ensures that it can be used to probe the symmergent gravity framework. 

Let us now consider Sgr A*. Our location is
$d_\text{R}\approx 8.3$ kpc  from the galactic center. In the Schwarzschild limit, the angular radius is $\theta_\text{E}^\text{Schw}= 1.186$ arc sec. Using equation \eqref{eqring}, one finds that small deviations from  $\theta_\text{E}^\text{Schw}$ at $a=0.90$ requires $c_\text{O} \sim 4\times 10^{31}$, which means that Nature contains much more bosonic degrees of freedom than the fermionic ones. 

\section{Perihelion shift} \label{sec8}
\begin{figure}[h]
   \centering
    \includegraphics[scale=0.6]{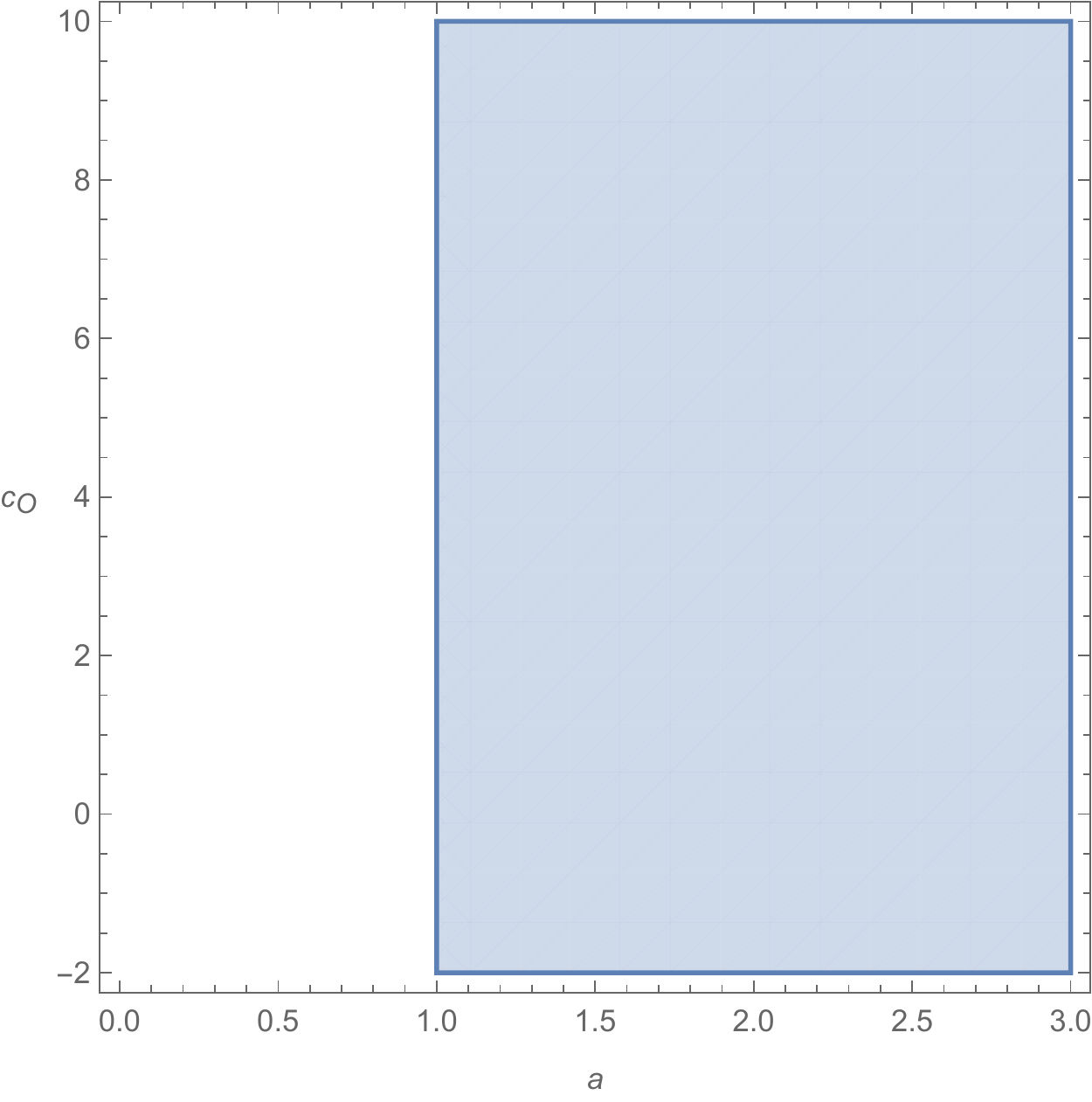}
    \caption{The perihelion shift bound on $c_{\rm O}$ in equation \eqref{eq18b}. The shaded area is the allowed region by the Mercury's perihelion shift \label{cobound}.}
\end{figure}

The perihelion shift of planetary motion around a symmergent black hole can qualify as a sensitive probe of the model parameters. To calculate it, we start with the effective potential for radial motion (in geometrical units $G=1$ and $c=1$)
\begin{equation}
V_{eff}(r)=\frac{1}{2} +\frac{(a-1)}{8\pi
c_\text{O}}\frac{{\cal L}^{2}}{6}-\frac{ M}{r}+\frac{{\cal L}^{2}}{2r^{2}}-\frac{{M} {\cal L}^{2}}{%
r^{3}}+\frac{ (a-1)}{8\pi c_\text{O}}\frac{r^{2}}{6}
\end{equation}%
in which ${\cal L}$ is the specific orbital angular momentum, $R$ is the semimajor axis of the elliptical orbit, and $e$ is the orbit's eccentricity \cite{Arakida:2012ya}. To determine the orbit, we use the variable $u=1/r$ so that $\frac{dr}{d\lambda }=-{\cal L}\frac{du}{d\phi }$. Then, for non-circular orbits and massive particle's the equation of radial motion becomes
\begin{equation}
\frac{d^{2}u}{d\phi ^{2}}+u=\frac{M}{{\cal L}^{2}}+3{M}u^{2}+\frac{1}{%
3{\cal L}^{2}u^{3}}\frac{(a-1)}{8\pi c_\text{O}} \label{eq11}
\end{equation}%
whose first term at the right-hand side suffices for determining the contribution of the quasi-Newtonian orbit:
\begin{equation}
u_{0}=\frac{1}{r}=\frac{M}{{\cal L}^{2}}\left[ 1+e \cos (\phi -\phi _\text{o})%
\right] ,  \label{eq12}
\end{equation}%
in which $\phi _{0}$ is a constant of integration. The most direct way to determine the eccentricity $e$ is to apply successive approximations about the quasi-Newtonian orbit in (\ref{eq12}). This is because the last two terms in equation \eqref{eq11} are small in comparison to the Newtonian contribution and, using the standard perturbation techniques, one can easily obtain the first order correction to be $u\cong u_{0}+u_{1}$ ($u_{1} << u_{0}$), where \cite{Islam:1983rxp,Freire:2001hmq}. The perturbation $u_1(\phi)$ (orbit of small eccentricity) obeys the equation
\begin{equation}
\frac{d^{2}u_{1}}{d\phi ^{2}}+u_{1}\cong \frac{6{M}^{3}}{{\cal L}^{4}}\left( 1-%
\frac{{\cal L}^{8}}{6{M}^{6}}\frac{(a-1)}{8\pi c_\text{O}}\right) e \cos(\phi -\phi_\text{o}),  \label{eq13}
\end{equation}%
with the solution
\begin{equation}
u_{1}=\frac{3{M}^{3}}{{\cal L}^{4}}\left( 1-\frac{{\cal L}^{8}}{6{M}^{6}}\frac{(a-1)%
}{8\pi c_\text{O}}\right) e \phi \sin(\phi -\phi_\text{o}).  \label{eq14}
\end{equation}
Having this correction at hand, one obtains
\begin{eqnarray}
u\cong \frac{{M}}{{\cal L}^{2}}\left\{ 1+e\left[ \cos (\phi -\phi _\text{o})+ H\right] \right\} ,  \label{eq15}
\end{eqnarray}%
with 
\begin{equation}
H=\frac{3{M}^{2}}{{\cal L}^{2}}\left( 1-\frac{{\cal L}^{8}}{6{M}^{6}}\frac{(a-1)}{8\pi c_\text{O}}%
\right) \phi \sin(\phi -\phi_\text{o})
\end{equation}
so that the radial coordinates takes the form
\begin{equation}
u=\frac{1}{r}\cong \frac{{M}}{{\cal L}^{2}}\left\{ 1+e\cos (\phi -\phi_\text{o}-\Delta \phi_\text{o})]\right\} \label{eq17}
\end{equation}
after introducing the angular shift
\begin{equation}
\Delta \phi _{0}=3\left( \frac{M}{{\cal L}}\right) ^{2}\left( 1-\frac{{\cal L}^{8}%
}{6{M}^{6}}\frac{(a-1)}{8\pi c_\text{O}}\right) \phi .  \label{eq16}
\end{equation}%
As a result of the radial coordinate 
\eqref{eq17},  precession per revolution is given by
\begin{equation}
\Delta \phi ={6\pi }\left( \frac{{M}}{{\cal L}}\right) ^{2}\left( 1-\frac{%
{\cal L}^{8}}{6{M}^{6}}\frac{(a-1)}{8\pi c_\text{O}}\right),  \label{eq18}
\end{equation}
and using ${\cal L}=\sqrt{G M R\left(1-e^{2}\right)}$ this precession takes the form  
\begin{equation}
\Delta\phi =\frac{6\pi M }{(1-e^2)R}-\Delta \phi_\text{symmergent} \label{eq199}
\end{equation}
where
\begin{equation}
\Delta \phi_\text{symmergent}={\frac {\left( 1-e^{2} \right) ^{3}{R}^{3} \left( a-1
 \right) }{8 M{c_\text{O}}}}.
\end{equation}
It is clear that this symmergent perihelion shift  vanishes as $a \to 1$ or $c_\text{O} \to \infty$. The perihelion shift of the planet Mercury is obtained with an accuracy better than $5x10^{-3}$ \cite{Islam:1983rxp,Freire:2001hmq,will} so that, using the total shift in (\ref{eq199}), the symmergent gravity parameter $c_\text{O}$ is found to obey the bound 
\begin{equation}
\Bigg|\frac{1}{G c_\text{O}}\Bigg|<\frac{8\pi 10^{-42}}{(a-1)}\ \text{cm}^{-2}. \label{eq18b}
\end{equation}%
which is illustrated in Figure \ref{cobound} in the $c_{\rm O}$ versus $a$ plane.

\begin{figure}[h]
   \centering
\includegraphics[scale=0.75]{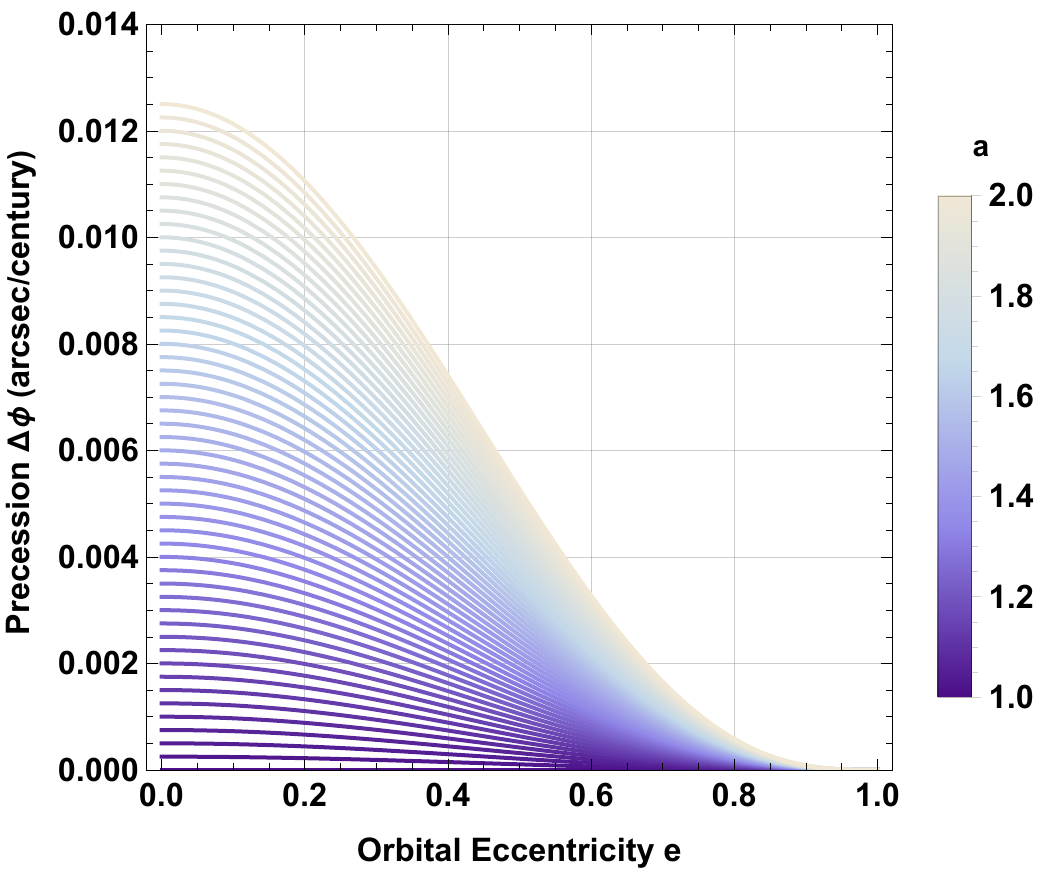}
\includegraphics[scale=0.75]{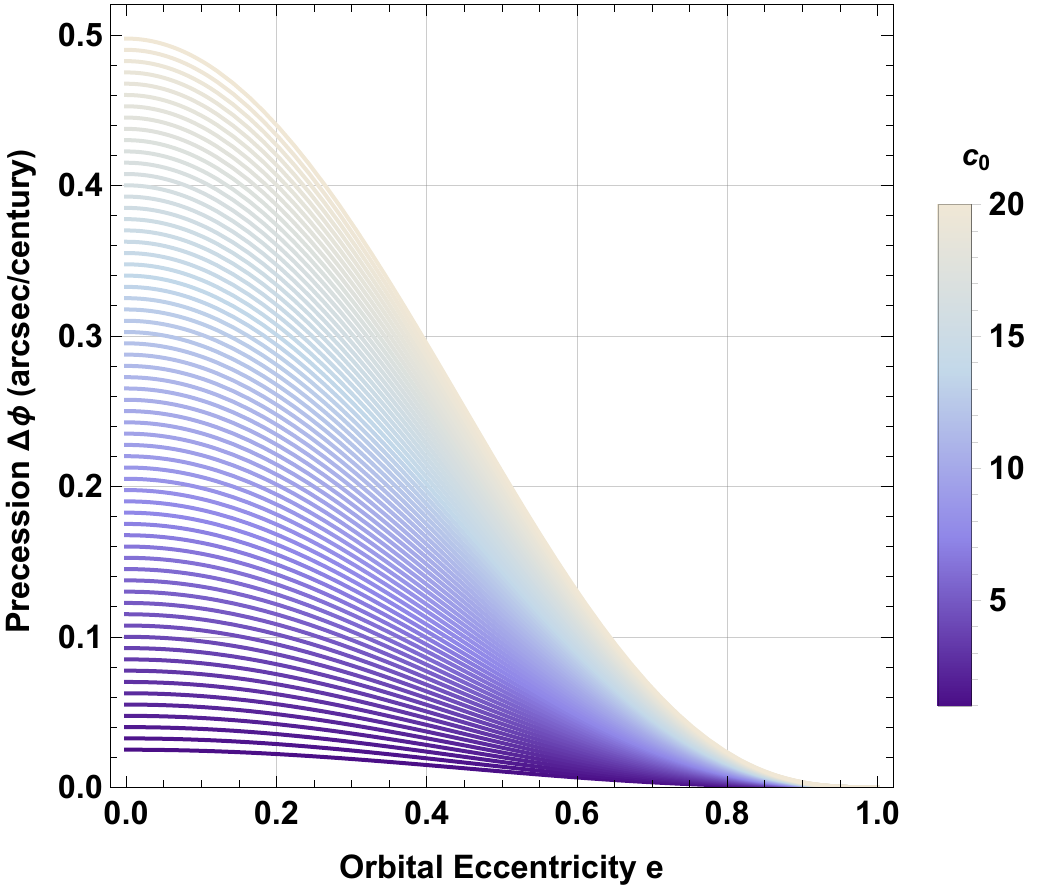}
\caption{The precession $\Delta \phi$ as a function of the orbital eccentricity $e$ for a finely-smapled values of $a$ and $c_{\rm O}$, after setting $G=M=R=c=1$. In general, larger the $a$ and $c_{\rm O}$ larger the precession for small eccentricities. \label{eccentricityplot}}
\end{figure}

\section{Photonsphere and symmergent black hole shadow as observed by a stationary and a co-moving observer} \label{sec9}
In this section, we study bounds on the symmergent gravity parameters from the photonsphere radius and black hole shadow. We do this first for a stationary observer and then for an observer co-moving with the cosmic expansion.

\subsection{Shadow radius for a stationary observer}
Here, we will explore the effects of the loop factor $c_\text{O}$ on  photonsphere and shadow radius of a black hole for a stationary observer. We will do this exploration by following the formalism developed by Perlick and others  \cite{Perlick:2018}. To this end, we begin with the Lagrangian,
\begin{equation}
{\mathbb{L}}(x , \dot{x} ) =
 \frac{1}{2} \left(- A(r) \dot{t}{}^2
+ B(r)\dot{r}{}^2 + C(r) \dot{\varphi}{}^2 \right)
\end{equation}
from which we derive the geodesics along the equatorial plane. This Lagrangian gives rise to two constants of motion (energy $E$ and angular momentum $L$)
\begin{equation}
E = A(r)   \dot{t}  , \qquad
L  =  C(r)  \dot{\varphi},
\end{equation}
and an additional relation
\begin{equation}\label{}
- A(r) \dot{t}{}^2  + 
B(r)\dot{r}{}^2  +  C(r) \dot{\varphi}{}^2
 =  0 
\end{equation}
for null geodesics (massless particles). This latter relation leads to the orbit equation for null light rays
\begin{equation}\label{eq:orbit}
\left( \frac{dr}{d\varphi} \right) ^2  =  r^4 \left(  
\frac{E^2 }{ L^2} - \frac{a-1}{24\pi c_\text{O}}  -  
\frac{1}{r^2}  + \frac{2M}{r^3} \right).
\end{equation}
after using the conserved quantities $E$ and $L$. Under the conditions $dr/d \varphi =0$ and $d^2r/d \varphi ^2 =0$, the radius of the photonsphere 
\begin{equation} \label{erph}
    r_\text{ph} = 3M
\end{equation}
turns out to be independent of the parameters $a$ and $c_{\rm O}$. This, of course, does not mean that the shadow itself is independent of the symmergent gravity parameters. Indeed, as we shall see later, escaping light rays are succumbed to the astrophysical environment in a way sensitive to  $a$ and $c_\text{O}$.

In the presence of small perturbations in their circular motions at $r_{\text{ph}}=3M$,  the light rays can either spiral to the black hole or escape to infinity. The impact parameter for light rays is given by $b = L/E$, and its critical value $b_\text{crit}$ can be determined via the condition $dr/d \varphi =0$ in the orbit equation. Thus, using \eqref{erph}, it is found to be
\begin{equation}
    b_\text{crit}^2 = \frac{27M^2}{\frac{3(a-1)}{8\pi c_\text{O}}M^2+1}.
\end{equation}

Let us consider a static observer with coordinates ($t_\text{o}, r_\text{o}, \theta_\text{o}=\pi/2, \phi=0$). Light which escaped from the photonsphere then approaches and leaves the observer at some angle $\theta$. To this end, the shadow angle is given by
\begin{equation}
\tan \theta_{\text{sh}}  =  
\frac{b_\text{crit}^2}{C(r_\text{o})/A(r_\text{o})}\frac{d \varphi}{dr}
 \Bigg|_{r = r_\text{o}}.
\end{equation}
By a simple trigonometric identity and with the use of the orbit equation \eqref{eq:orbit}, the exact expression for the radius of the shadow can then be derived as
\begin{align} \label{eqshad}
    R_{\text{sh}} = r_\text{ph}\left[ \frac{2\left(1-\frac{2M}{r_\text{o}}+\frac{a-1}{24\pi c_\text{O}}r_\text{o}^2 \right)}{1-\frac{M}{r_\text{ph}}+\frac{a-1}{36\pi c_\text{O}} r_\text{ph}^2} \right]^{1/2}
\end{align}
where we have written $R_{\text{sh}}=r_\text{o}\sin\theta_\text{sh}$. Here, if $r_\text{o}=3M$, we get $R_{\text{sh}}=3M$, which implies that $\theta_{\text{sh}}=\pi/2$ and half of the sky is dark. Such a result is in accordance with the results of \cite{Perlick:2018}. We need to take note again that the overall contribution of the symmergent parameters mimics the effects of the cosmological constant. Indeed, looking at \eqref{eqshad} with $a<1$, we see that  $c_{\rm O}<0$ ($n_{\rm B}<n_{\rm F}$) mimics the AdS case while $c_{\rm O}<0$ ($n_{\rm B}<n_{\rm F}$) does the dS case.

\begin{figure}[h]
   \centering
    \includegraphics[width=\linewidth]{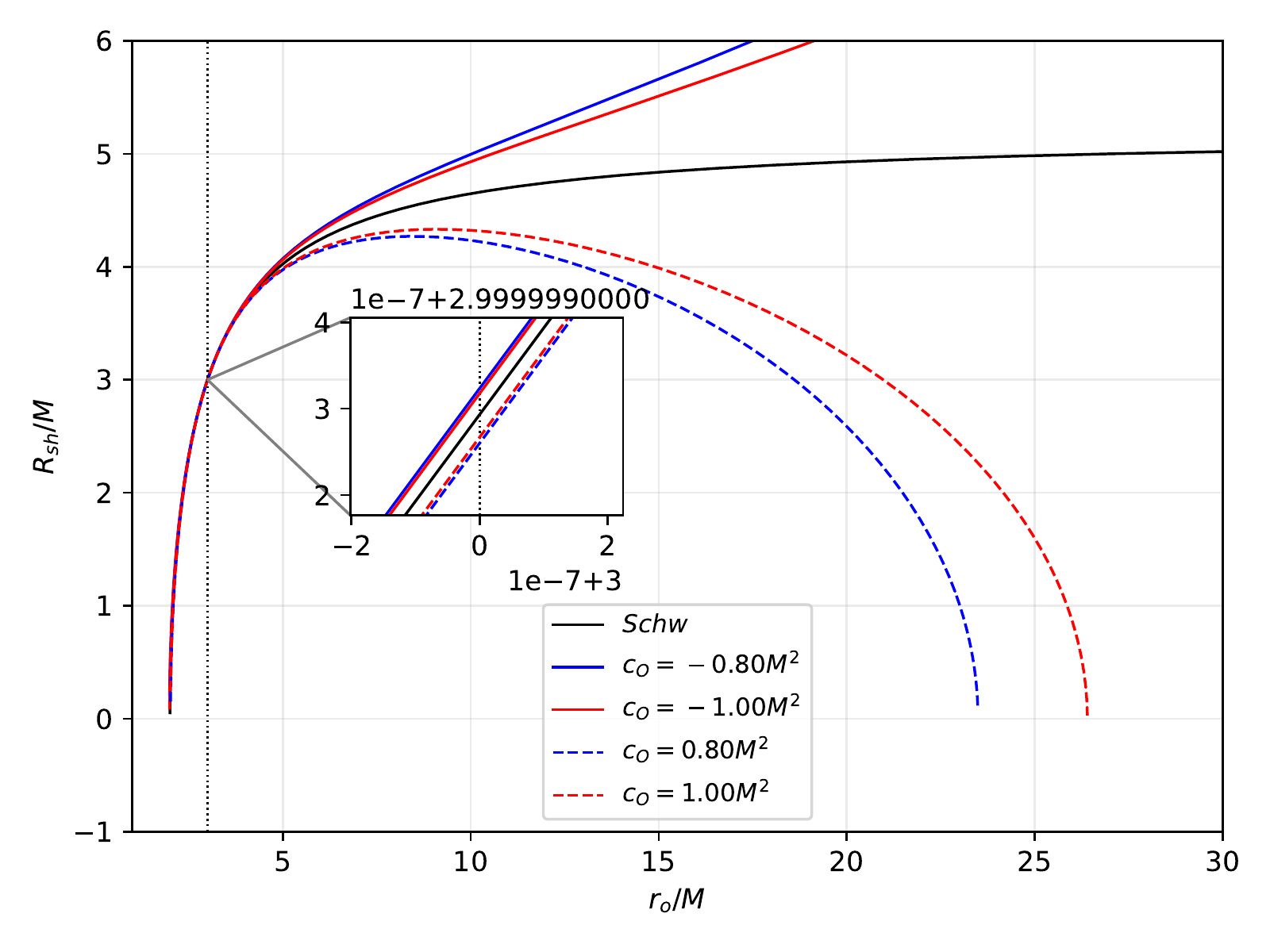}
    \caption{The shadow radius $R_{sh}$ as a function of the observer's position $r_\text{o}$ for $a=0.9$. The solid lines correspond to $c_{\rm O}<0$ ($n_\text{B}<n_\text{F}$). The dashed lines, on the other hand, correspond to $c_{\rm O}>0$ ($n_\text{B}>n_\text{F}$). The vertical dotted line is $r_\text{ph}=3m$. \label{rph}}
\end{figure}

Depicted in Figure \ref{rph} is the variation of the shadow radius  in equation \eqref{eqshad} with the position $r_o$ of the observer for $a=0.9$ and $c_{\rm O}=\pm 0.8$ and $\pm 1$. The plot also contains the Schwarzschild case (black curve), for comparison. As expected, the Schwarzschild case approaches asymptotically to $3\sqrt{3}M$ at large $r_\text{o}/M$. One again notes that the symmergent gravity contribution mimics the effects of the cosmological constant, and significant effects are possible only at very large distances that are comparable to the cosmological horizon. Even at the location of Earth, its effects remain negligible. However, we can scale the parameters such that symmergent gravity effects get pronounced. In Figure \ref{rph}, we can clearly see that the effect is unobservable weak at small distances (like, for example, $r_\text{o}=3M$). At large distances, however, shadow radii in symmergent gravity deviate significantly from the Schwarzschild curve, such that the dS  ($c_{\rm O}>0 \Rightarrow n_{\rm B}>n_{\rm F}$) and AdS ($c_{\rm O}<0 \Rightarrow n_{\rm B}<n_{\rm F}$) are clearly distinguished. The dS type collapses the shadow radius at very large distances, and it implies that black holes near the cosmological horizon may have very small shadow radii. The AdS case, however, is blind to such critical radii as its shadow radius increases indefinitely.

\begin{figure}
   \centering
    \includegraphics[width=\linewidth]{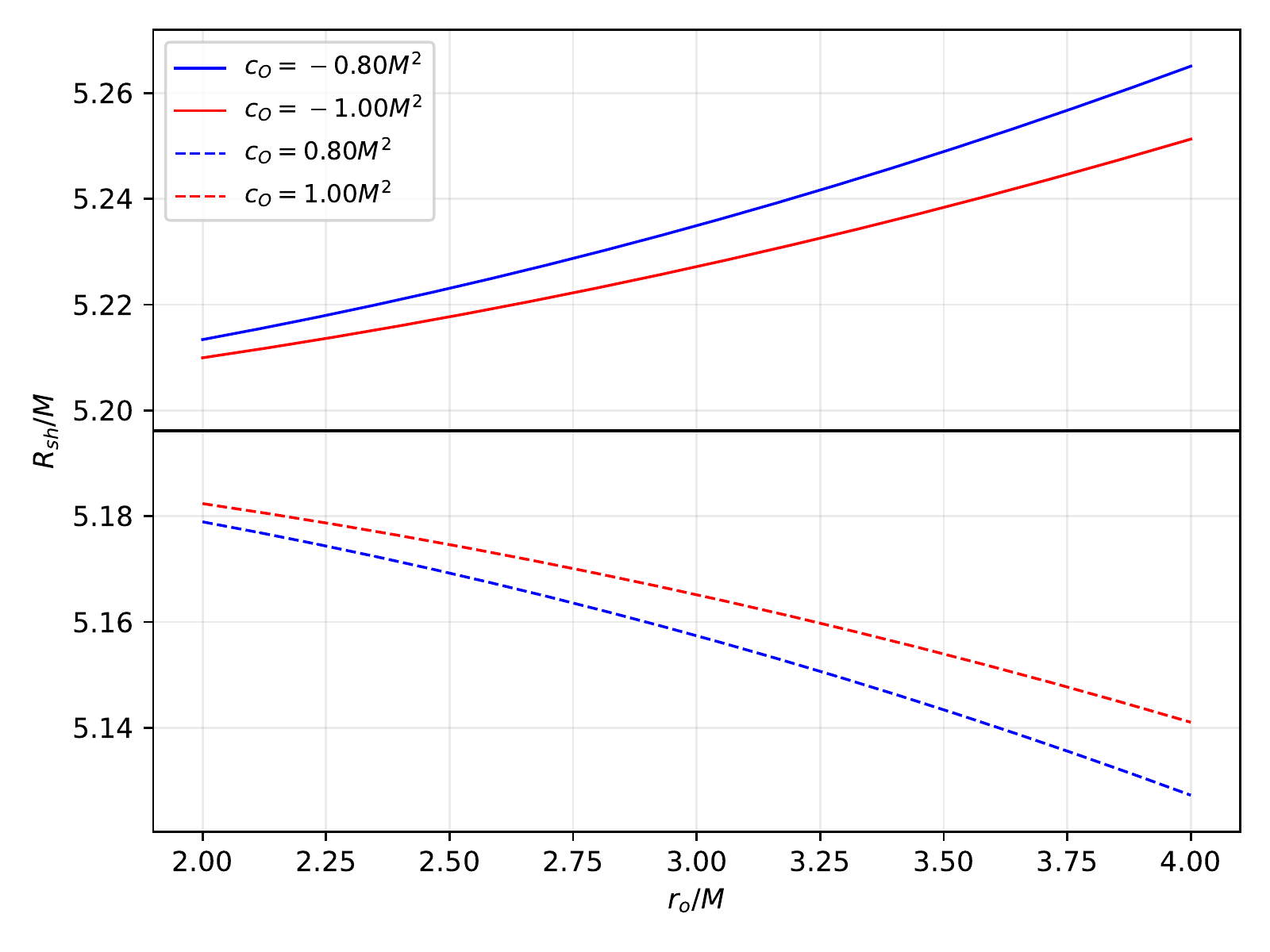}
    \caption{The shadow radius $R_{sh}$ as a function of the observer position $r_\text{o}$ for an observer nearby the black hole. In the plot, with $a=0.9$,   solid lines correspond to $c_{\rm O}<0$ ($n_\text{B}<n_\text{F}$). The dashed lines, on the other hand, correspond to $c_{\rm O}>0$ ($n_\text{B}>n_\text{F}$).
    The horizontal solid line (black) is the shadow radius for the Schwarzschild case. \label{rph2}}
\end{figure}

The shadow radius of the black hole Sgr A* can be approximated as $3\sqrt{3}M$. It is so because $r_\text{o}>>M$, and $\frac{a-1}{24\pi c_\text{O}}<<M$ 
considering our position $r_o$ relative to Sgr  A*.  Keeping in mind that we are still way too far from the cosmological horizon, these conditions allow us to carry out a Taylor expansion in \eqref{eqshad} to find
\begin{equation} \label{shadapprox}
    R_{\text{sh}} \approx 3\sqrt{3}M+\frac{\sqrt{3}}{16}\frac{a-1}{\pi c_\text{O}}Mr_\text{o}^2+\mathcal{O}(M^2r_\text{o}^{-1},M^2r_\text{o},M^3,M^3r_\text{o}^{-2}),
\end{equation}
which makes sense only in the inset plot in Figure \ref{rph} because the dS structure is still allowed (not applicable to regions exceeding the scaled cosmological horizon). Thus, with this scaling, Earth's position falls somewhere within the inset plot, where the effects of the symmergent gravity can be disentangled. All these details are explicitly demonstrated in Figure \ref{rph2}.
We can estimate further, for a given $a$, value of $c_\text{O}$ that gives an appreciable deviation to the shadow radius, and it turns out to be 
\begin{equation} \label{est1}
    c_\text{O}^\text{sh}=\frac{(1-a)r_\text{o}^2}{48 \pi},
\end{equation}
which relates $c_{\rm O}$ to the observer's position $r_o$.

\subsection{Shadow radius for an observer co-moving with the cosmic expansion}
\begin{figure}[h]
   \centering
    \includegraphics[width=\linewidth]{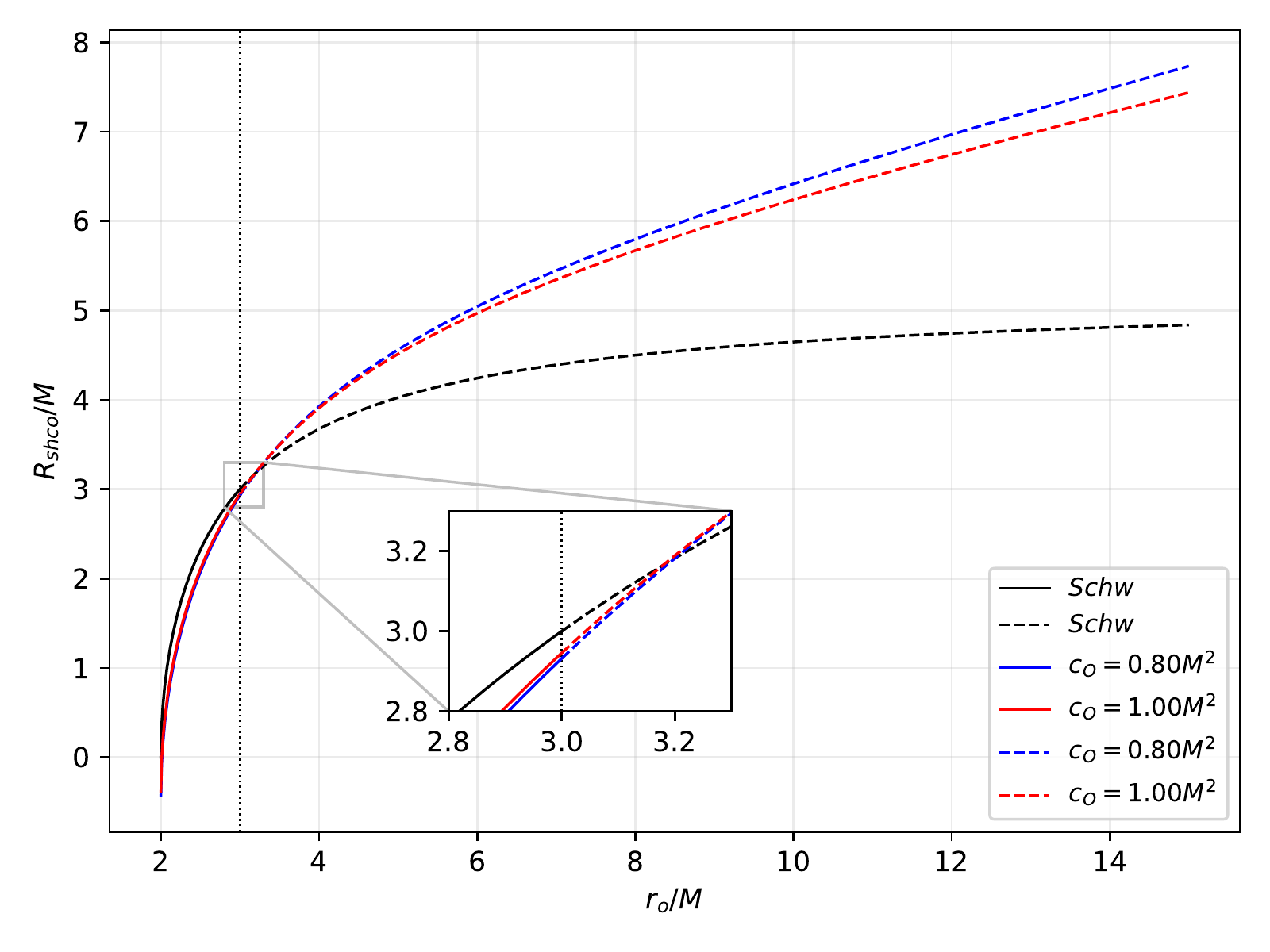}
    \caption{The shadow radius $R_{shco}$ as a function of the co-moving observer position $r_\text{o}$ for $a=0.9$. The black (solid/dashed) line corresponds to the Schwarzschild case. The blue/red solid (dashed) lines represent the upper (lower) sign in equation \eqref{shacomov}. \label{figshacomov}}
\end{figure}

Here, we study black hole shadow as observed by a co-moving observer. For this purpose, we follow the general procedure laid out by Perlick and others \cite{Perlick:2018}. Also, we incorporate  dS type symmergent black hole by using the lapse function
\begin{equation}
    f(r) = 1 - \frac{2M}{r} - \frac{(1-a) r^2}{24\pi c_\text{O}} 
\end{equation}
in the analyses that follow. The first step of the analysis is to transform the metric to a twiddled coordinate system (or the McVittie coordinates) via the following transformations,
\begin{align}
dr &= e^{H_\text{o} \tilde{t}} \left( 1- \frac{M^2}{4 \tilde{r}{}^2}  e^{-2H_\text{o}\tilde{t}} \right)
\left( d \tilde{r}+\tilde{r}  H_\text{o}  d \tilde{t} \right)  , \nonumber\\
dt &=\dfrac{
\left( 1 - \dfrac{M}{2 \tilde{r}} e^{-H_\text{o} \tilde{t}} \right)^2  d \tilde{t}
+ H_\text{o}  \tilde{r} e^{2 H_\text{o} \tilde{t}} 
\left( 1+ \dfrac{M}{2 \tilde{r}} e^{-H_\text{o} \tilde{t}} \right)^6 d \tilde{r}
}{
\left( 1 - \dfrac{M}{2 \tilde{r}} e^{-H_\text{o} \tilde{t}} \right)^2
- H_\text{o}^2 \tilde{r}{}^2 e^{2 H_\text{o} \tilde{t}}
\left( 1 + \dfrac{M}{2 \tilde{r}} e^{-H_\text{o} \tilde{t}} \right)^6
}\ ,
\end{align}
in which $H_\text{o} = \sqrt{(1-a)/24\pi c_\text{O}}$ is the Hubble constant measured by the observer. This transformation puts the metric into a new form,

\begin{align}
    \tilde{g}{}_{\mu \nu} d \tilde{x}{}^{\mu} d \tilde{x}{}^{\nu} &=
- \left( 1-\dfrac{M}{2 \tilde{r}}  e^{-H_\text{o}\tilde{t}} \right)^2
 \left( 1+\dfrac{M}{2 \tilde{r}}  e^{-H_\text{o}\tilde{t}} \right)^{-2}
c^2 d \tilde{t}{}^2 \nonumber\\
&+  e^{2 H_\text{o} \tilde{t}}
\left( 1 + \dfrac{M}{2 \tilde{r}} e^{-H_\text{o} \tilde{t}}\right)^4
 \left( d \tilde{r}{}^2 + \tilde{r}{}^2 d \Omega ^2 \right)\ .
\end{align}
Using the standard aberration formula, the shadow angular radius formula in Ref.~\cite{Perlick:2018} can be rectified and, as a result,  the shadow radius takes the form
\begin{equation}
    R_\text{shco}^{2} = 
\frac{\left( 1 - v^2 \right)
}{
\left( 1 \pm  v \sqrt{1-R_\text{sh}^2/r_\text{o}^2} \right) ^2} R_\text{sh}^2\ ,
\end{equation}
for a co-moving observer.  The velocity $v$ of the observer 
\begin{equation}
    \tilde{g}{}_{\mu \nu} u_\text{shco}^{\mu}  u_\text{shco}^{\nu} = -(1-v^2)^{-1/2}
\end{equation}
is set by the twiddled metric. In fact, using the normalization condition
\begin{align}
    g_{\mu \nu}u_\text{sh}^{\mu}  u_\text{sh}^{\nu} &= -1
\end{align}
one gets the velocity, 

\begin{equation}
    v = \sqrt{\frac{(1-a)}{2\pi c_\text{O}}} r_\text{o} \left(1-\frac{2M}{r_\text{o}}\right)^{-1/2}\ .
\end{equation}
It is clear from the velocity formula that if $a<1$, $c_\text{O}$ is not allowed to be negative (at least for the parameters in this study). At any rate, this condition corresponds to a dS ($c_{\rm O}>0 \Rightarrow n_{\rm B}>n_{\rm F}$) type symmergent black hole, which is consistent with the fact that we are embedded in a dS-type Universe. After substituting these formulae, we finally get the exact expression for shadow radius as observed by a co-moving observer
\begin{align} \label{shacomov}
    R_\text{shco}&=3\sqrt{3}M\sqrt{\left(1-\frac{2M}{r_\text{o}}\right)\left[1-\frac{9(1-a)M^{2}}{8\pi c_\text{O}}\right]} \nonumber\\
    &\mp 3\sqrt{3}r_\text{o}M\sqrt{\frac{1-a}{24\pi c}}\sqrt{1-
    \frac{27M^{2}}{r_\text{o}^{2}}\left(1-\frac{2M}{r_\text{o}}\right)}\ .
\end{align}
In this formula, upper (lower) sign is used for the domain $r_\text{h1} < r_\text{o} < r_\text{h2}$  ($r_\text{h1} < r_\text{o} < \infty$). These two domains are illustrated in Figure \ref{figshacomov}.
Comparing Figure \ref{figshacomov} to Figure \ref{rph}, one observes that a co-moving observer can have the ability to measure a considerable deviation from the shadow radius under the effect of dS symmergent spacetime in regions near $r_\text{o} = 3M$. The reason for this is that the deviation due to a static observer is vanishingly small. Needless to say, this region corresponds to locations in the Universe  far from the cosmological horizon. It is highly interesting that theoretical considerations do not admit the AdS ($c_{\rm O}<0 \Rightarrow n_{\rm B}<n_{\rm F}$) symmergent case, and this is in perfect agreement with the already observed dS - Universe \cite{Planck:2018vyg}. One nevertheless keeps in mind that if the black hole being observed is embedded in space with extreme expansion, a co-moving observer (knowing that he/she lives in the dS type Universe) will perceive a larger shadow radius  (very large deviation).

\begin{figure}[h]
   \centering
\includegraphics[width=\linewidth]{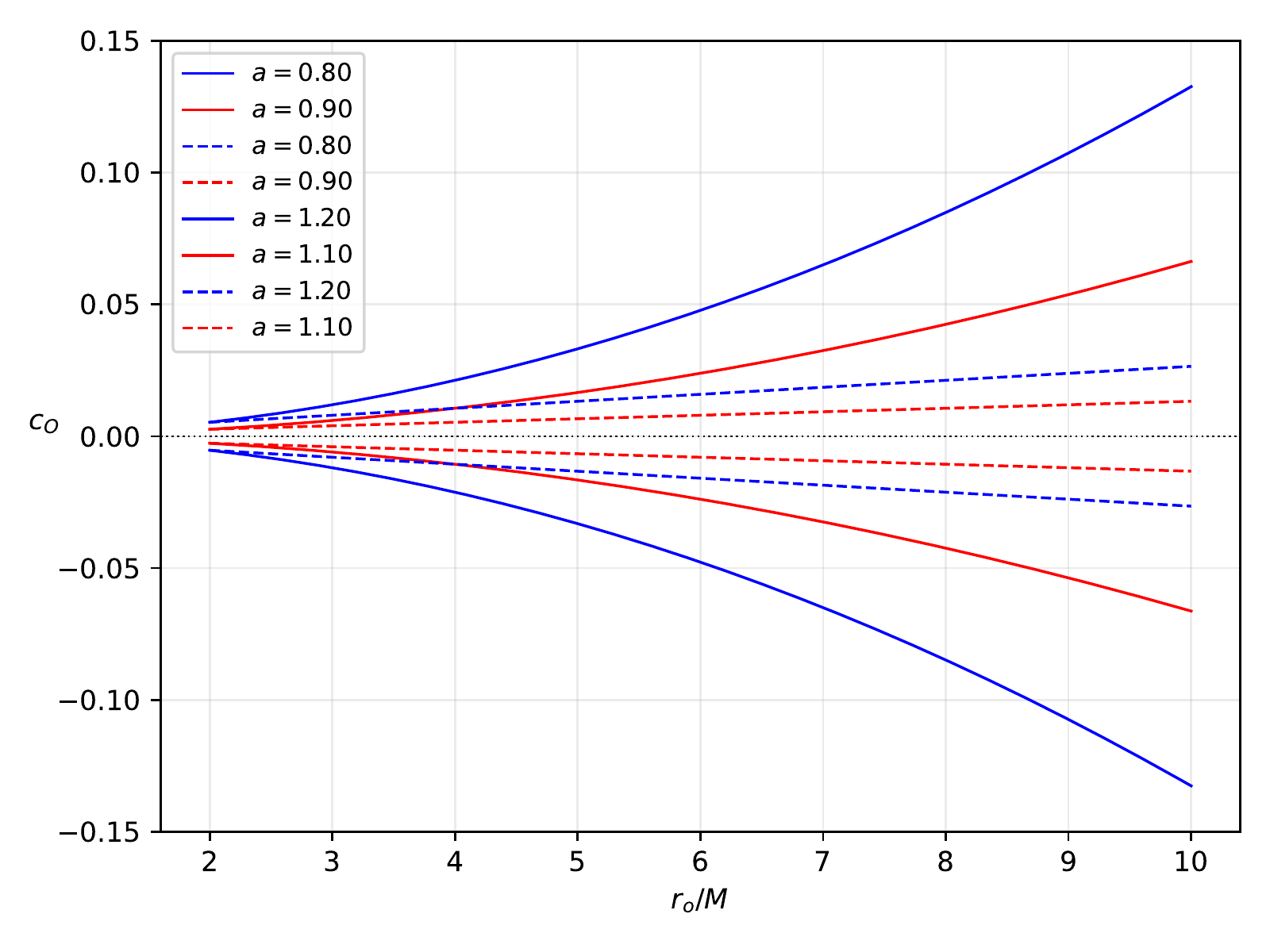}
\caption{The dependence of $c_\text{O}$ on the observer's position $r_\text{o}$ for different values of $a$. Here, solid and dashed lines correspond to stationary and co-moving observers, respectively. \label{c_o}}
\end{figure}

In similarity to the shadow radius for a static observer, the conditions $r_\text{o}>>m$ and $\frac{a-1}{24\pi c_\text{O}}<<m$ applies if Earth is co-moving with the cosmological expansion. In this case, one can perform a Taylor expansion of equation \eqref{shacomov} to get, 
\begin{equation} \label{shacomovapprox}
    R_\text{shco}\approx 3\sqrt{3}M\left[1+\frac{(1-a)}{24\pi c_\text{O}}r_\text{o}\right]\ ,
\end{equation}
which is in agreement with equation (40) of the reference \cite{Perlick:2018}. Then, 
similar to \eqref{est1}, we can have an estimate of $c_\text{O}$ such that it gives cause to a significant deviation from the shadow radius of the black hole under observation, and we end up with
\begin{align} \label{est2}
    c_\text{O}^\text{shco}&=\frac{(1-a)r_\text{o}}{24 \pi}\ .
\end{align}
To contrast observations of stationary and co-moving observers, we plot  the equations \eqref{est1} and \eqref{est2} in Figure \ref{c_o}. It follows from the figure that the requisite $n_\text{B}-n_\text{F}$ (boson-fermion number difference) for an observer to perceive a considerable deviation on the shadow radius depends on the state of the observer. Indeed, a co-moving observer needs less $n_\text{B}-n_\text{F}$ to perceive a considerable deviation than a stationary observer.

\section{Conclusions}

In the present work, we have performed a comprehensive and systematic study of the symmergent gravity in black hole environments.

We started with the black hole solution in the symmergent gravity, with a detailed analysis of the properties of their event horizons. We have shown that the outer and cosmologic-like horizons exist in the cases $c_{\rm O}<0$ \& $a>1$ and $c_{\rm O}>0$ \& $a<1$, and an increase in $a$ at $c_{\rm O}<0$ causes these horizons to move away from each other, namely the outer (cosmological) horizon moves to larger (smaller) radii. These horizon properties enabled the determination of the allowed ranges of the symmergent gravity parameters $a$ and $c_{\rm O}$.

Having revealed the horizon properties, we turned to an investigation of the dynamics of test particles around the symmergent black hole. We have found that there are four solutions for ISCO radius together with the OSCO. In fact, ISCOs and OSCOs exist in the cases $c_{\rm O}<0$ \& $a>1$ and $c_{\rm O}>0$ \& $a<1$. 

We have also studied the effects of symmergent gravity parameters on the frequencies of Keplerian orbits, as well as the radial oscillations of test particles along their circular stable orbits. In our analyses, we employed the known QPO models of  RP, WD, ER2, and ER3 models. We have shown that low-frequency QPOs switch off in the symmergent gravity and the abundance of the high-frequency twin-peak QPOs increase (decreases) at $a>1$ ($a<1$). We have utilized the QPOs of the microquasars GRS 1915-105, GRO J1655-40, XTE J1550-564, and H1743+322 as well as the supermassive black hole Sgr A* to constrain the symmergent gravity parameters, and found a nearly model-independent relationship between $a$ and $c_{\rm O}$.    

We have further studied the effects of symmergent gravity parameters on the weak deflection angle as well as the shadow of the black hole. We have obtained the general expression of weak deflection angle using appropriate models in a way involving finite distances of the source and the receiver. We have determined variations of the weak deflection angle with the symmergent gravity parameters $a$ and $c_\text{O}$. Our results, despite the scaling of symmergent gravity effects via the selected values of $a$ and $c_\text{O}$, suggest that it is possible to disentangle symmergent effects via observable deviations, especially in the Einstein ring formation. We also studied how the dS and AdS type symmergent geometries are sensitive to the said deviations. Considering the center of our galaxy, a small deviation from the Schwarzschild case requires $c_\text{O} \sim 4 \times 10^{31}$, which means that Nature contains much more bosonic degrees of freedom than the fermionic ones. (One keeps in mind that, by the nature of the symmergent gravity, the extra particles do not need to couple to the known particles).  Concerning the measurements of the deflection angle, we believe that it could be possible to resolve symmergent gravity effects with the advent of the space technologies, such as the improved ESA GAIA mission and the Event horizon telescope, and the more powerful VLBI RadioAstron.

For the symmergent black hole shadow, we derived an exact analytical formula for the shadow radius, both for static and co-moving observers. Our analysis suggests that an observer,  co-moving with the Universe's expansion, can measure a larger deviation from the shadow radius than a static observer. A co-moving observer can also measure a larger shadow radius for black holes embedded in the extreme expansion of spacetime. We contrasted determinations of the parameter $c_\text{O}$ for static and co-moving observers and found that the latter observes less difference between $n_\text{B}$ and $n_\text{F}$ under the effect of cosmic expansion.

Our analysis in this work can be extended to other physical phenomena. One immediate extension would be the generalization of the metric to spinning symmergent black holes. Another extension would be examining the effects of matter and radiation-dominated Universes on shadow radius as perceived by a co-moving observer. Yet another extension would be the analysis of a varying Hubble's constant near the cosmic horizon. Concerning the astrophysical objects, our analysis here can be extended to neutron stars, magnetars, and as such to determine the allowed ranges of the symmergent gravity parameters. It would be fair to say that there is a wealth of phenomena that could be analyzed in the framework of symmergent gravity. 



\begin{thebibliography}{146}
\expandafter\ifx\csname natexlab\endcsname\relax\def\natexlab#1{#1}\fi
\expandafter\ifx\csname bibnamefont\endcsname\relax
  \def\bibnamefont#1{#1}\fi
\expandafter\ifx\csname bibfnamefont\endcsname\relax
  \def\bibfnamefont#1{#1}\fi
\expandafter\ifx\csname citenamefont\endcsname\relax
  \def\citenamefont#1{#1}\fi
\expandafter\ifx\csname url\endcsname\relax
  \def\url#1{\texttt{#1}}\fi
\expandafter\ifx\csname urlprefix\endcsname\relax\def\urlprefix{URL }\fi
\providecommand{\bibinfo}[2]{#2}
\providecommand{\eprint}[2][]{\url{#2}}
\bibliographystyle{apsrev}
\bibitem[{\citenamefont{Abbott et~al.}(2016)}]{LIGOScientific:2016aoc}
\bibinfo{author}{\bibfnamefont{B.~P.} \bibnamefont{Abbott}}
  \bibnamefont{et~al.} (\bibinfo{collaboration}{LIGO Scientific, Virgo}),
  \bibinfo{journal}{Phys. Rev. Lett.} \textbf{\bibinfo{volume}{116}},
  \bibinfo{pages}{061102} (\bibinfo{year}{2016}), \eprint{1602.03837}.

\bibitem[{\citenamefont{Akiyama et~al.}(2019)}]{Event2021a}
\bibinfo{author}{\bibfnamefont{K.}~\bibnamefont{Akiyama}} \bibnamefont{et~al.}
  (\bibinfo{collaboration}{Event Horizon Telescope}),
  \bibinfo{journal}{Astrophys. J. Lett.} \textbf{\bibinfo{volume}{875}},
  \bibinfo{pages}{L1} (\bibinfo{year}{2019}), \eprint{1906.11238}.

\bibitem[{\citenamefont{Akiyama et~al.}(2022)}]{Event2021b}
\bibinfo{author}{\bibfnamefont{K.}~\bibnamefont{Akiyama}} \bibnamefont{et~al.}
  (\bibinfo{collaboration}{Event Horizon Telescope}),
  \bibinfo{journal}{Astrophys. J. Lett.} \textbf{\bibinfo{volume}{930}},
  \bibinfo{pages}{L12} (\bibinfo{year}{2022}).

\bibitem[{\citenamefont{Sakharov}(1967)}]{sakharov}
\bibinfo{author}{\bibfnamefont{A.~D.} \bibnamefont{Sakharov}},
  \bibinfo{journal}{Dokl. Akad. Nauk Ser. Fiz.} \textbf{\bibinfo{volume}{177}},
  \bibinfo{pages}{70} (\bibinfo{year}{1967}).

\bibitem[{\citenamefont{Visser}(2002)}]{visser}
\bibinfo{author}{\bibfnamefont{M.}~\bibnamefont{Visser}},
  \bibinfo{journal}{Mod. Phys. Lett. A} \textbf{\bibinfo{volume}{17}},
  \bibinfo{pages}{977} (\bibinfo{year}{2002}), \eprint{gr-qc/0204062}.

\bibitem[{\citenamefont{Verlinde}(2017)}]{verlinde}
\bibinfo{author}{\bibfnamefont{E.~P.} \bibnamefont{Verlinde}},
  \bibinfo{journal}{SciPost Phys.} \textbf{\bibinfo{volume}{2}},
  \bibinfo{pages}{016} (\bibinfo{year}{2017}), \eprint{1611.02269}.

\bibitem[{\citenamefont{Macias and Camacho}(2008)}]{incompatible}
\bibinfo{author}{\bibfnamefont{A.}~\bibnamefont{Macias}} \bibnamefont{and}
  \bibinfo{author}{\bibfnamefont{A.}~\bibnamefont{Camacho}},
  \bibinfo{journal}{Phys. Lett. B} \textbf{\bibinfo{volume}{663}},
  \bibinfo{pages}{99} (\bibinfo{year}{2008}).

\bibitem[{\citenamefont{Wald}(2018)}]{wald}
\bibinfo{author}{\bibfnamefont{R.~M.} \bibnamefont{Wald}},
  \bibinfo{journal}{Einstein Stud.} \textbf{\bibinfo{volume}{14}},
  \bibinfo{pages}{439} (\bibinfo{year}{2018}), \eprint{0907.0416}.

\bibitem[{\citenamefont{Dyson}(2013)}]{dyson}
\bibinfo{author}{\bibfnamefont{F.}~\bibnamefont{Dyson}}, \bibinfo{journal}{Int.
  J. Mod. Phys. A} \textbf{\bibinfo{volume}{28}}, \bibinfo{pages}{1330041}
  (\bibinfo{year}{2013}).

\bibitem[{\citenamefont{'t~Hooft and Veltman}(1974)}]{thooft}
\bibinfo{author}{\bibfnamefont{G.}~\bibnamefont{'t~Hooft}} \bibnamefont{and}
  \bibinfo{author}{\bibfnamefont{M.~J.~G.} \bibnamefont{Veltman}},
  \bibinfo{journal}{Ann. Inst. H. Poincare Phys. Theor. A}
  \textbf{\bibinfo{volume}{20}}, \bibinfo{pages}{69} (\bibinfo{year}{1974}).

\bibitem[{\citenamefont{Demir}(2021{\natexlab{a}})}]{demir1}
\bibinfo{author}{\bibfnamefont{D.}~\bibnamefont{Demir}}, \bibinfo{journal}{Gen.
  Rel. Grav.} \textbf{\bibinfo{volume}{53}}, \bibinfo{pages}{22}
  (\bibinfo{year}{2021}{\natexlab{a}}), \eprint{2101.12391}.

\bibitem[{\citenamefont{Weinberg}(1979)}]{weinberg}
\bibinfo{author}{\bibfnamefont{S.}~\bibnamefont{Weinberg}},
  \bibinfo{journal}{Physica A} \textbf{\bibinfo{volume}{96}},
  \bibinfo{pages}{327} (\bibinfo{year}{1979}).

\bibitem[{\citenamefont{Burgess}(2020)}]{eff-action2}
\bibinfo{author}{\bibfnamefont{C.~P.} \bibnamefont{Burgess}},
  \emph{\bibinfo{title}{{Introduction to Effective Field Theory}}}
  (\bibinfo{publisher}{Cambridge University Press}, \bibinfo{year}{2020}), ISBN
  \bibinfo{isbn}{978-1-139-04804-0, 978-0-521-19547-8}.

\bibitem[{\citenamefont{Demir}(2019)}]{demir2}
\bibinfo{author}{\bibfnamefont{D.}~\bibnamefont{Demir}}, \bibinfo{journal}{Adv.
  High Energy Phys.} \textbf{\bibinfo{volume}{2019}}, \bibinfo{pages}{4652048}
  (\bibinfo{year}{2019}), \eprint{1901.07244}.

\bibitem[{\citenamefont{Demir}(2016)}]{demir3}
\bibinfo{author}{\bibfnamefont{D.~A.} \bibnamefont{Demir}},
  \bibinfo{journal}{Adv. High Energy Phys.} \textbf{\bibinfo{volume}{2016}},
  \bibinfo{pages}{6727805} (\bibinfo{year}{2016}), \eprint{1605.00377}.

\bibitem[{\citenamefont{Birrell and Davies}(1984)}]{birrel}
\bibinfo{author}{\bibfnamefont{N.~D.} \bibnamefont{Birrell}} \bibnamefont{and}
  \bibinfo{author}{\bibfnamefont{P.~C.~W.} \bibnamefont{Davies}},
  \emph{\bibinfo{title}{{Quantum Fields in Curved Space}}}, Cambridge
  Monographs on Mathematical Physics (\bibinfo{publisher}{Cambridge Univ.
  Press}, \bibinfo{address}{Cambridge, UK}, \bibinfo{year}{1984}), ISBN
  \bibinfo{isbn}{978-0-521-27858-4, 978-0-521-27858-4}.

\bibitem[{\citenamefont{Demir}(2021{\natexlab{b}})}]{demir4}
\bibinfo{author}{\bibfnamefont{D.}~\bibnamefont{Demir}},
  \bibinfo{journal}{Galaxies} \textbf{\bibinfo{volume}{9}}, \bibinfo{pages}{33}
  (\bibinfo{year}{2021}{\natexlab{b}}), \eprint{2105.04277}.

\bibitem[{\citenamefont{\c{C}imdiker}(2020)}]{irfan}
\bibinfo{author}{\bibfnamefont{I.~I.} \bibnamefont{\c{C}imdiker}},
  \bibinfo{journal}{Phys. Dark Univ.} \textbf{\bibinfo{volume}{30}},
  \bibinfo{pages}{100736} (\bibinfo{year}{2020}).

\bibitem[{\citenamefont{\c{C}imdiker et~al.}(2021)\citenamefont{\c{C}imdiker,
  Demir, and \"Ovg\"un}}]{symmergent-bh}
\bibinfo{author}{\bibfnamefont{I.}~\bibnamefont{\c{C}imdiker}},
  \bibinfo{author}{\bibfnamefont{D.}~\bibnamefont{Demir}}, \bibnamefont{and}
  \bibinfo{author}{\bibfnamefont{A.}~\bibnamefont{\"Ovg\"un}},
  \bibinfo{journal}{Phys. Dark Univ.} \textbf{\bibinfo{volume}{34}},
  \bibinfo{pages}{100900} (\bibinfo{year}{2021}), \eprint{2110.11904}.

\bibitem[{\citenamefont{Bozza}(2003)}]{Bozza:2002af}
\bibinfo{author}{\bibfnamefont{V.}~\bibnamefont{Bozza}},
  \bibinfo{journal}{Phys. Rev. D} \textbf{\bibinfo{volume}{67}},
  \bibinfo{pages}{103006} (\bibinfo{year}{2003}), \eprint{gr-qc/0210109}.

\bibitem[{\citenamefont{Virbhadra and Ellis}(2000)}]{Virbhadra:1999nm}
\bibinfo{author}{\bibfnamefont{K.~S.} \bibnamefont{Virbhadra}}
  \bibnamefont{and} \bibinfo{author}{\bibfnamefont{G.~F.~R.}
  \bibnamefont{Ellis}}, \bibinfo{journal}{Phys. Rev. D}
  \textbf{\bibinfo{volume}{62}}, \bibinfo{pages}{084003}
  (\bibinfo{year}{2000}), \eprint{astro-ph/9904193}.

\bibitem[{\citenamefont{Virbhadra and Ellis}(2002)}]{Virbhadra:2002ju}
\bibinfo{author}{\bibfnamefont{K.~S.} \bibnamefont{Virbhadra}}
  \bibnamefont{and} \bibinfo{author}{\bibfnamefont{G.~F.~R.}
  \bibnamefont{Ellis}}, \bibinfo{journal}{Phys. Rev. D}
  \textbf{\bibinfo{volume}{65}}, \bibinfo{pages}{103004}
  (\bibinfo{year}{2002}).

\bibitem[{\citenamefont{Bozza et~al.}(2001)\citenamefont{Bozza, Capozziello,
  Iovane, and Scarpetta}}]{Bozza:2001xd}
\bibinfo{author}{\bibfnamefont{V.}~\bibnamefont{Bozza}},
  \bibinfo{author}{\bibfnamefont{S.}~\bibnamefont{Capozziello}},
  \bibinfo{author}{\bibfnamefont{G.}~\bibnamefont{Iovane}}, \bibnamefont{and}
  \bibinfo{author}{\bibfnamefont{G.}~\bibnamefont{Scarpetta}},
  \bibinfo{journal}{Gen. Rel. Grav.} \textbf{\bibinfo{volume}{33}},
  \bibinfo{pages}{1535} (\bibinfo{year}{2001}), \eprint{gr-qc/0102068}.

\bibitem[{\citenamefont{Bozza}(2002)}]{Bozza:2002zj}
\bibinfo{author}{\bibfnamefont{V.}~\bibnamefont{Bozza}},
  \bibinfo{journal}{Phys. Rev. D} \textbf{\bibinfo{volume}{66}},
  \bibinfo{pages}{103001} (\bibinfo{year}{2002}), \eprint{gr-qc/0208075}.

\bibitem[{\citenamefont{Hasse and Perlick}(2002)}]{Hasse:2001by}
\bibinfo{author}{\bibfnamefont{W.}~\bibnamefont{Hasse}} \bibnamefont{and}
  \bibinfo{author}{\bibfnamefont{V.}~\bibnamefont{Perlick}},
  \bibinfo{journal}{Gen. Rel. Grav.} \textbf{\bibinfo{volume}{34}},
  \bibinfo{pages}{415} (\bibinfo{year}{2002}), \eprint{gr-qc/0108002}.

\bibitem[{\citenamefont{Perlick}(2004)}]{Perlick:2003vg}
\bibinfo{author}{\bibfnamefont{V.}~\bibnamefont{Perlick}},
  \bibinfo{journal}{Phys. Rev. D} \textbf{\bibinfo{volume}{69}},
  \bibinfo{pages}{064017} (\bibinfo{year}{2004}), \eprint{gr-qc/0307072}.

\bibitem[{\citenamefont{He et~al.}(2020)\citenamefont{He, Zhou, Feng, Mu, Wang,
  Li, Pan, and Lin}}]{He:2020eah}
\bibinfo{author}{\bibfnamefont{G.}~\bibnamefont{He}},
  \bibinfo{author}{\bibfnamefont{X.}~\bibnamefont{Zhou}},
  \bibinfo{author}{\bibfnamefont{Z.}~\bibnamefont{Feng}},
  \bibinfo{author}{\bibfnamefont{X.}~\bibnamefont{Mu}},
  \bibinfo{author}{\bibfnamefont{H.}~\bibnamefont{Wang}},
  \bibinfo{author}{\bibfnamefont{W.}~\bibnamefont{Li}},
  \bibinfo{author}{\bibfnamefont{C.}~\bibnamefont{Pan}}, \bibnamefont{and}
  \bibinfo{author}{\bibfnamefont{W.}~\bibnamefont{Lin}}, \bibinfo{journal}{Eur.
  Phys. J. C} \textbf{\bibinfo{volume}{80}}, \bibinfo{pages}{835}
  (\bibinfo{year}{2020}).

\bibitem[{\citenamefont{Bozza}(2008)}]{Bozza2008}
\bibinfo{author}{\bibfnamefont{V.}~\bibnamefont{Bozza}},
  \bibinfo{journal}{Phys. Rev. D} \textbf{\bibinfo{volume}{78}},
  \bibinfo{pages}{103005} (\bibinfo{year}{2008}).

\bibitem[{\citenamefont{Gibbons and Werner}(2008)}]{Gibbons:2008rj}
\bibinfo{author}{\bibfnamefont{G.~W.} \bibnamefont{Gibbons}} \bibnamefont{and}
  \bibinfo{author}{\bibfnamefont{M.~C.} \bibnamefont{Werner}},
  \bibinfo{journal}{Class. Quant. Grav.} \textbf{\bibinfo{volume}{25}},
  \bibinfo{pages}{235009} (\bibinfo{year}{2008}), \eprint{0807.0854}.

\bibitem[{\citenamefont{\"Ovg\"un}(2018)}]{Ovgun:2018fnk}
\bibinfo{author}{\bibfnamefont{A.}~\bibnamefont{\"Ovg\"un}},
  \bibinfo{journal}{Phys. Rev. D} \textbf{\bibinfo{volume}{98}},
  \bibinfo{pages}{044033} (\bibinfo{year}{2018}), \eprint{1805.06296}.

\bibitem[{\citenamefont{\"Ovg\"un}(2019{\natexlab{a}})}]{Ovgun:2019wej}
\bibinfo{author}{\bibfnamefont{A.}~\bibnamefont{\"Ovg\"un}},
  \bibinfo{journal}{Phys. Rev. D} \textbf{\bibinfo{volume}{99}},
  \bibinfo{pages}{104075} (\bibinfo{year}{2019}{\natexlab{a}}),
  \eprint{1902.04411}.

\bibitem[{\citenamefont{\"Ovg\"un}(2019{\natexlab{b}})}]{Ovgun:2018oxk}
\bibinfo{author}{\bibfnamefont{A.}~\bibnamefont{\"Ovg\"un}},
  \bibinfo{journal}{Universe} \textbf{\bibinfo{volume}{5}},
  \bibinfo{pages}{115} (\bibinfo{year}{2019}{\natexlab{b}}),
  \eprint{1806.05549}.

\bibitem[{\citenamefont{Javed et~al.}(2019{\natexlab{a}})\citenamefont{Javed,
  Abbas, and \"Ovg\"un}}]{Javed:2019kon}
\bibinfo{author}{\bibfnamefont{W.}~\bibnamefont{Javed}},
  \bibinfo{author}{\bibfnamefont{J.}~\bibnamefont{Abbas}}, \bibnamefont{and}
  \bibinfo{author}{\bibfnamefont{A.}~\bibnamefont{\"Ovg\"un}},
  \bibinfo{journal}{Eur. Phys. J. C} \textbf{\bibinfo{volume}{79}},
  \bibinfo{pages}{694} (\bibinfo{year}{2019}{\natexlab{a}}),
  \eprint{1908.09632}.

\bibitem[{\citenamefont{Javed et~al.}(2019{\natexlab{b}})\citenamefont{Javed,
  Abbas, and \"Ovg\"un}}]{Javed:2019rrg}
\bibinfo{author}{\bibfnamefont{W.}~\bibnamefont{Javed}},
  \bibinfo{author}{\bibfnamefont{j.}~\bibnamefont{Abbas}}, \bibnamefont{and}
  \bibinfo{author}{\bibfnamefont{A.}~\bibnamefont{\"Ovg\"un}},
  \bibinfo{journal}{Phys. Rev. D} \textbf{\bibinfo{volume}{100}},
  \bibinfo{pages}{044052} (\bibinfo{year}{2019}{\natexlab{b}}),
  \eprint{1908.05241}.

\bibitem[{\citenamefont{Javed et~al.}(2019{\natexlab{c}})\citenamefont{Javed,
  Babar, and \"Ovg\"un}}]{Javed:2019ynm}
\bibinfo{author}{\bibfnamefont{W.}~\bibnamefont{Javed}},
  \bibinfo{author}{\bibfnamefont{R.}~\bibnamefont{Babar}}, \bibnamefont{and}
  \bibinfo{author}{\bibfnamefont{A.}~\bibnamefont{\"Ovg\"un}},
  \bibinfo{journal}{Phys. Rev. D} \textbf{\bibinfo{volume}{100}},
  \bibinfo{pages}{104032} (\bibinfo{year}{2019}{\natexlab{c}}),
  \eprint{1910.11697}.

\bibitem[{\citenamefont{Javed et~al.}(2020{\natexlab{a}})\citenamefont{Javed,
  Hamza, and \"Ovg\"un}}]{Javed:2020lsg}
\bibinfo{author}{\bibfnamefont{W.}~\bibnamefont{Javed}},
  \bibinfo{author}{\bibfnamefont{A.}~\bibnamefont{Hamza}}, \bibnamefont{and}
  \bibinfo{author}{\bibfnamefont{A.}~\bibnamefont{\"Ovg\"un}},
  \bibinfo{journal}{Phys. Rev. D} \textbf{\bibinfo{volume}{101}},
  \bibinfo{pages}{103521} (\bibinfo{year}{2020}{\natexlab{a}}),
  \eprint{2005.09464}.

\bibitem[{\citenamefont{Javed et~al.}(2019{\natexlab{d}})\citenamefont{Javed,
  Babar, and \"Ovg\"un}}]{Javed:2019qyg}
\bibinfo{author}{\bibfnamefont{W.}~\bibnamefont{Javed}},
  \bibinfo{author}{\bibfnamefont{R.}~\bibnamefont{Babar}}, \bibnamefont{and}
  \bibinfo{author}{\bibfnamefont{A.}~\bibnamefont{\"Ovg\"un}},
  \bibinfo{journal}{Phys. Rev. D} \textbf{\bibinfo{volume}{99}},
  \bibinfo{pages}{084012} (\bibinfo{year}{2019}{\natexlab{d}}),
  \eprint{1903.11657}.

\bibitem[{\citenamefont{\"Ovg\"un et~al.}(2019)\citenamefont{\"Ovg\"un,
  Sakall\i{}, and Saavedra}}]{Ovgun:2018fte}
\bibinfo{author}{\bibfnamefont{A.}~\bibnamefont{\"Ovg\"un}},
  \bibinfo{author}{\bibfnamefont{I.}~\bibnamefont{Sakall\i{}}},
  \bibnamefont{and} \bibinfo{author}{\bibfnamefont{J.}~\bibnamefont{Saavedra}},
  \bibinfo{journal}{Annals Phys.} \textbf{\bibinfo{volume}{411}},
  \bibinfo{pages}{167978} (\bibinfo{year}{2019}), \eprint{1806.06453}.

\bibitem[{\citenamefont{Javed et~al.}(2020{\natexlab{b}})\citenamefont{Javed,
  Abbas, and \"Ovg\"un}}]{Javed:2019jag}
\bibinfo{author}{\bibfnamefont{W.}~\bibnamefont{Javed}},
  \bibinfo{author}{\bibfnamefont{J.}~\bibnamefont{Abbas}}, \bibnamefont{and}
  \bibinfo{author}{\bibfnamefont{A.}~\bibnamefont{\"Ovg\"un}},
  \bibinfo{journal}{Annals Phys.} \textbf{\bibinfo{volume}{418}},
  \bibinfo{pages}{168183} (\bibinfo{year}{2020}{\natexlab{b}}),
  \eprint{2007.16027}.

\bibitem[{\citenamefont{Werner}(2012)}]{Werner_2012}
\bibinfo{author}{\bibfnamefont{M.~C.} \bibnamefont{Werner}},
  \bibinfo{journal}{Gen. Rel. Grav.} \textbf{\bibinfo{volume}{44}},
  \bibinfo{pages}{3047} (\bibinfo{year}{2012}), \eprint{1205.3876}.

\bibitem[{\citenamefont{Ishihara et~al.}(2016)\citenamefont{Ishihara, Suzuki,
  Ono, Kitamura, and Asada}}]{Ishihara:2016vdc}
\bibinfo{author}{\bibfnamefont{A.}~\bibnamefont{Ishihara}},
  \bibinfo{author}{\bibfnamefont{Y.}~\bibnamefont{Suzuki}},
  \bibinfo{author}{\bibfnamefont{T.}~\bibnamefont{Ono}},
  \bibinfo{author}{\bibfnamefont{T.}~\bibnamefont{Kitamura}}, \bibnamefont{and}
  \bibinfo{author}{\bibfnamefont{H.}~\bibnamefont{Asada}},
  \bibinfo{journal}{Phys. Rev. D} \textbf{\bibinfo{volume}{94}},
  \bibinfo{pages}{084015} (\bibinfo{year}{2016}), \eprint{1604.08308}.

\bibitem[{\citenamefont{Ishihara et~al.}(2017)\citenamefont{Ishihara, Suzuki,
  Ono, and Asada}}]{Ishihara:2016sfv}
\bibinfo{author}{\bibfnamefont{A.}~\bibnamefont{Ishihara}},
  \bibinfo{author}{\bibfnamefont{Y.}~\bibnamefont{Suzuki}},
  \bibinfo{author}{\bibfnamefont{T.}~\bibnamefont{Ono}}, \bibnamefont{and}
  \bibinfo{author}{\bibfnamefont{H.}~\bibnamefont{Asada}},
  \bibinfo{journal}{Phys. Rev. D} \textbf{\bibinfo{volume}{95}},
  \bibinfo{pages}{044017} (\bibinfo{year}{2017}), \eprint{1612.04044}.

\bibitem[{\citenamefont{Ono et~al.}(2017)\citenamefont{Ono, Ishihara, and
  Asada}}]{Ono:2017pie}
\bibinfo{author}{\bibfnamefont{T.}~\bibnamefont{Ono}},
  \bibinfo{author}{\bibfnamefont{A.}~\bibnamefont{Ishihara}}, \bibnamefont{and}
  \bibinfo{author}{\bibfnamefont{H.}~\bibnamefont{Asada}},
  \bibinfo{journal}{Phys. Rev. D} \textbf{\bibinfo{volume}{96}},
  \bibinfo{pages}{104037} (\bibinfo{year}{2017}), \eprint{1704.05615}.

\bibitem[{\citenamefont{Pantig and
  Rodulfo}(2020{\natexlab{a}})}]{Pantig:2020odu}
\bibinfo{author}{\bibfnamefont{R.~C.} \bibnamefont{Pantig}} \bibnamefont{and}
  \bibinfo{author}{\bibfnamefont{E.~T.} \bibnamefont{Rodulfo}},
  \bibinfo{journal}{Chin. J. Phys.} \textbf{\bibinfo{volume}{66}},
  \bibinfo{pages}{691} (\bibinfo{year}{2020}{\natexlab{a}}),
  \eprint{2003.00764}.

\bibitem[{\citenamefont{Pantig et~al.}(2022)\citenamefont{Pantig, Yu, Rodulfo,
  and \"Ovg\"un}}]{Pantig2022}
\bibinfo{author}{\bibfnamefont{R.~C.} \bibnamefont{Pantig}},
  \bibinfo{author}{\bibfnamefont{P.~K.} \bibnamefont{Yu}},
  \bibinfo{author}{\bibfnamefont{E.~T.} \bibnamefont{Rodulfo}},
  \bibnamefont{and}
  \bibinfo{author}{\bibfnamefont{A.}~\bibnamefont{\"Ovg\"un}},
  \bibinfo{journal}{Annals of Physics} \textbf{\bibinfo{volume}{436}},
  \bibinfo{pages}{168722} (\bibinfo{year}{2022}), \eprint{2104.04304}.

\bibitem[{\citenamefont{Pantig and
  \"Ovg\"un}(2022{\natexlab{a}})}]{Pantig2022a}
\bibinfo{author}{\bibfnamefont{R.~C.} \bibnamefont{Pantig}} \bibnamefont{and}
  \bibinfo{author}{\bibfnamefont{A.}~\bibnamefont{\"Ovg\"un}},
  \bibinfo{journal}{Eur. Phys. J. C} \textbf{\bibinfo{volume}{82}},
  \bibinfo{pages}{391} (\bibinfo{year}{2022}{\natexlab{a}}),
  \eprint{2201.03365}.

\bibitem[{\citenamefont{Pantig and
  \"Ovg\"un}(2022{\natexlab{b}})}]{Pantig2022b}
\bibinfo{author}{\bibfnamefont{R.~C.} \bibnamefont{Pantig}} \bibnamefont{and}
  \bibinfo{author}{\bibfnamefont{A.}~\bibnamefont{\"Ovg\"un}}
  (\bibinfo{year}{2022}{\natexlab{b}}), \eprint{2202.07404},
  \urlprefix\url{https://arxiv.org/abs/2202.07404}.

\bibitem[{\citenamefont{Li and \"Ovg\"un}(2020)}]{Li:2020dln}
\bibinfo{author}{\bibfnamefont{Z.}~\bibnamefont{Li}} \bibnamefont{and}
  \bibinfo{author}{\bibfnamefont{A.}~\bibnamefont{\"Ovg\"un}},
  \bibinfo{journal}{Phys. Rev. D} \textbf{\bibinfo{volume}{101}},
  \bibinfo{pages}{024040} (\bibinfo{year}{2020}), \eprint{2001.02074}.

\bibitem[{\citenamefont{Li et~al.}(2020{\natexlab{a}})\citenamefont{Li, Zhang,
  and \"Ovg\"un}}]{Li:2020wvn}
\bibinfo{author}{\bibfnamefont{Z.}~\bibnamefont{Li}},
  \bibinfo{author}{\bibfnamefont{G.}~\bibnamefont{Zhang}}, \bibnamefont{and}
  \bibinfo{author}{\bibfnamefont{A.}~\bibnamefont{\"Ovg\"un}},
  \bibinfo{journal}{Phys. Rev. D} \textbf{\bibinfo{volume}{101}},
  \bibinfo{pages}{124058} (\bibinfo{year}{2020}{\natexlab{a}}),
  \eprint{2006.13047}.

\bibitem[{\citenamefont{Synge}(1966)}]{Synge1966}
\bibinfo{author}{\bibfnamefont{J.~L.} \bibnamefont{Synge}},
  \bibinfo{journal}{Mon. Not. R. Astron. Soc.} \textbf{\bibinfo{volume}{131}},
  \bibinfo{pages}{463} (\bibinfo{year}{1966}), ISSN \bibinfo{issn}{0035-8711}.

\bibitem[{\citenamefont{Luminet}(1979)}]{Luminet1979}
\bibinfo{author}{\bibfnamefont{J.~P.} \bibnamefont{Luminet}},
  \bibinfo{journal}{Astron. Astrophys.} \textbf{\bibinfo{volume}{75}},
  \bibinfo{pages}{228} (\bibinfo{year}{1979}).

\bibitem[{\citenamefont{Tsupko et~al.}(2020)\citenamefont{Tsupko, Fan, and
  Bisnovatyi-Kogan}}]{Tsupko_2020}
\bibinfo{author}{\bibfnamefont{O.~Y.} \bibnamefont{Tsupko}},
  \bibinfo{author}{\bibfnamefont{Z.}~\bibnamefont{Fan}}, \bibnamefont{and}
  \bibinfo{author}{\bibfnamefont{G.~S.} \bibnamefont{Bisnovatyi-Kogan}},
  \bibinfo{journal}{Class. Quant. Grav.} \textbf{\bibinfo{volume}{37}},
  \bibinfo{pages}{065016} (\bibinfo{year}{2020}), \eprint{1905.10509}.

\bibitem[{\citenamefont{Hioki and Maeda}(2009)}]{Hioki2009}
\bibinfo{author}{\bibfnamefont{K.}~\bibnamefont{Hioki}} \bibnamefont{and}
  \bibinfo{author}{\bibfnamefont{K.-i.} \bibnamefont{Maeda}},
  \bibinfo{journal}{Phys. Rev. D} \textbf{\bibinfo{volume}{80}},
  \bibinfo{pages}{024042} (\bibinfo{year}{2009}), \eprint{0904.3575}.

\bibitem[{\citenamefont{Dymnikova and Kraav}(2019)}]{Dymnikova2019}
\bibinfo{author}{\bibfnamefont{I.}~\bibnamefont{Dymnikova}} \bibnamefont{and}
  \bibinfo{author}{\bibfnamefont{K.}~\bibnamefont{Kraav}},
  \bibinfo{journal}{Universe} \textbf{\bibinfo{volume}{5}}, \bibinfo{pages}{1}
  (\bibinfo{year}{2019}), ISSN \bibinfo{issn}{22181997}.

\bibitem[{\citenamefont{Wei et~al.}(2019)\citenamefont{Wei, Zou, Liu, and
  Mann}}]{Wei2019}
\bibinfo{author}{\bibfnamefont{S.-W.} \bibnamefont{Wei}},
  \bibinfo{author}{\bibfnamefont{Y.-C.} \bibnamefont{Zou}},
  \bibinfo{author}{\bibfnamefont{Y.-X.} \bibnamefont{Liu}}, \bibnamefont{and}
  \bibinfo{author}{\bibfnamefont{R.~B.} \bibnamefont{Mann}},
  \bibinfo{journal}{JCAP} \textbf{\bibinfo{volume}{08}}, \bibinfo{pages}{030}
  (\bibinfo{year}{2019}), \eprint{1904.07710}.

\bibitem[{\citenamefont{Xu et~al.}(2018)\citenamefont{Xu, Hou, and
  Wang}}]{Xu2018a}
\bibinfo{author}{\bibfnamefont{Z.}~\bibnamefont{Xu}},
  \bibinfo{author}{\bibfnamefont{X.}~\bibnamefont{Hou}}, \bibnamefont{and}
  \bibinfo{author}{\bibfnamefont{J.}~\bibnamefont{Wang}},
  \bibinfo{journal}{JCAP} \textbf{\bibinfo{volume}{10}}, \bibinfo{pages}{046}
  (\bibinfo{year}{2018}), \eprint{1806.09415}.

\bibitem[{\citenamefont{Hou et~al.}(2018)\citenamefont{Hou, Xu, and
  Wang}}]{Hou:2018avu}
\bibinfo{author}{\bibfnamefont{X.}~\bibnamefont{Hou}},
  \bibinfo{author}{\bibfnamefont{Z.}~\bibnamefont{Xu}}, \bibnamefont{and}
  \bibinfo{author}{\bibfnamefont{J.}~\bibnamefont{Wang}},
  \bibinfo{journal}{JCAP} \textbf{\bibinfo{volume}{12}}, \bibinfo{pages}{040}
  (\bibinfo{year}{2018}), \eprint{1810.06381}.

\bibitem[{\citenamefont{Bambi et~al.}(2019)\citenamefont{Bambi, Freese,
  Vagnozzi, and Visinelli}}]{Bambi2019}
\bibinfo{author}{\bibfnamefont{C.}~\bibnamefont{Bambi}},
  \bibinfo{author}{\bibfnamefont{K.}~\bibnamefont{Freese}},
  \bibinfo{author}{\bibfnamefont{S.}~\bibnamefont{Vagnozzi}}, \bibnamefont{and}
  \bibinfo{author}{\bibfnamefont{L.}~\bibnamefont{Visinelli}},
  \bibinfo{journal}{Phys. Rev. D} \textbf{\bibinfo{volume}{100}},
  \bibinfo{pages}{044057} (\bibinfo{year}{2019}), \eprint{1904.12983}.

\bibitem[{\citenamefont{Konoplya}(2019)}]{Konoplya2019}
\bibinfo{author}{\bibfnamefont{R.~A.} \bibnamefont{Konoplya}},
  \bibinfo{journal}{Phys. Lett. B} \textbf{\bibinfo{volume}{795}},
  \bibinfo{pages}{1} (\bibinfo{year}{2019}), \eprint{1905.00064}.

\bibitem[{\citenamefont{Pantig}(2021)}]{Pantig2021}
\bibinfo{author}{\bibfnamefont{R.~C.} \bibnamefont{Pantig}}, in
  \emph{\bibinfo{booktitle}{Proceedings of the Samahang Pisika ng Pilipinas}}
  (\bibinfo{year}{2021}), vol.~\bibinfo{volume}{39}, pp.
  \bibinfo{pages}{SPP--2021--1C--03}.

\bibitem[{\citenamefont{Okyay and \"Ovg\"un}(2022)}]{Okyay:2021nnh}
\bibinfo{author}{\bibfnamefont{M.}~\bibnamefont{Okyay}} \bibnamefont{and}
  \bibinfo{author}{\bibfnamefont{A.}~\bibnamefont{\"Ovg\"un}},
  \bibinfo{journal}{JCAP} \textbf{\bibinfo{volume}{01}}, \bibinfo{pages}{009}
  (\bibinfo{year}{2022}), \eprint{2108.07766}.

\bibitem[{\citenamefont{Uniyal et~al.}(2022)\citenamefont{Uniyal, Pantig, and
  \"Ovg\"un}}]{Uniyal:2022vdu}
\bibinfo{author}{\bibfnamefont{A.}~\bibnamefont{Uniyal}},
  \bibinfo{author}{\bibfnamefont{R.~C.} \bibnamefont{Pantig}},
  \bibnamefont{and} \bibinfo{author}{\bibfnamefont{A.}~\bibnamefont{\"Ovg\"un}}
  (\bibinfo{year}{2022}), \eprint{2205.11072}.

\bibitem[{\citenamefont{Kuang and \"Ovg\"un}(2022)}]{Kuang:2022xjp}
\bibinfo{author}{\bibfnamefont{X.-M.} \bibnamefont{Kuang}} \bibnamefont{and}
  \bibinfo{author}{\bibfnamefont{A.}~\bibnamefont{\"Ovg\"un}}
  (\bibinfo{year}{2022}), \eprint{2205.11003}.

\bibitem[{\citenamefont{Kumar et~al.}(2020)\citenamefont{Kumar, Ghosh, and
  Wang}}]{Kumar:2020hgm}
\bibinfo{author}{\bibfnamefont{R.}~\bibnamefont{Kumar}},
  \bibinfo{author}{\bibfnamefont{S.~G.} \bibnamefont{Ghosh}}, \bibnamefont{and}
  \bibinfo{author}{\bibfnamefont{A.}~\bibnamefont{Wang}},
  \bibinfo{journal}{Phys. Rev. D} \textbf{\bibinfo{volume}{101}},
  \bibinfo{pages}{104001} (\bibinfo{year}{2020}), \eprint{2001.00460}.

\bibitem[{\citenamefont{Belhaj et~al.}(2020)\citenamefont{Belhaj, Benali,
  El~Balali, El~Moumni, and Ennadifi}}]{Belhaj:2020rdb}
\bibinfo{author}{\bibfnamefont{A.}~\bibnamefont{Belhaj}},
  \bibinfo{author}{\bibfnamefont{M.}~\bibnamefont{Benali}},
  \bibinfo{author}{\bibfnamefont{A.}~\bibnamefont{El~Balali}},
  \bibinfo{author}{\bibfnamefont{H.}~\bibnamefont{El~Moumni}},
  \bibnamefont{and} \bibinfo{author}{\bibfnamefont{S.~E.}
  \bibnamefont{Ennadifi}}, \bibinfo{journal}{Class. Quant. Grav.}
  \textbf{\bibinfo{volume}{37}}, \bibinfo{pages}{215004}
  (\bibinfo{year}{2020}), \eprint{2006.01078}.

\bibitem[{\citenamefont{Li et~al.}(2020{\natexlab{b}})\citenamefont{Li, Guo,
  and Chen}}]{Li2020}
\bibinfo{author}{\bibfnamefont{P.-C.} \bibnamefont{Li}},
  \bibinfo{author}{\bibfnamefont{M.}~\bibnamefont{Guo}}, \bibnamefont{and}
  \bibinfo{author}{\bibfnamefont{B.}~\bibnamefont{Chen}},
  \bibinfo{journal}{Phys. Rev. D} \textbf{\bibinfo{volume}{101}},
  \bibinfo{pages}{084041} (\bibinfo{year}{2020}{\natexlab{b}}),
  \eprint{2001.04231}.

\bibitem[{\citenamefont{\"Ovg\"un and Sakall\i{}}(2020)}]{Ovgun:2020gjz}
\bibinfo{author}{\bibfnamefont{A.}~\bibnamefont{\"Ovg\"un}} \bibnamefont{and}
  \bibinfo{author}{\bibfnamefont{I.}~\bibnamefont{Sakall\i{}}},
  \bibinfo{journal}{Class. Quant. Grav.} \textbf{\bibinfo{volume}{37}},
  \bibinfo{pages}{225003} (\bibinfo{year}{2020}), \eprint{2005.00982}.

\bibitem[{\citenamefont{\"Ovg\"un et~al.}(2020)\citenamefont{\"Ovg\"un,
  Sakall\i{}, Saavedra, and Leiva}}]{Ovgun:2019jdo}
\bibinfo{author}{\bibfnamefont{A.}~\bibnamefont{\"Ovg\"un}},
  \bibinfo{author}{\bibfnamefont{I.}~\bibnamefont{Sakall\i{}}},
  \bibinfo{author}{\bibfnamefont{J.}~\bibnamefont{Saavedra}}, \bibnamefont{and}
  \bibinfo{author}{\bibfnamefont{C.}~\bibnamefont{Leiva}},
  \bibinfo{journal}{Mod. Phys. Lett. A} \textbf{\bibinfo{volume}{35}},
  \bibinfo{pages}{2050163} (\bibinfo{year}{2020}), \eprint{1906.05954}.

\bibitem[{\citenamefont{\"Ovg\"un et~al.}(2018)\citenamefont{\"Ovg\"un,
  Sakall\i{}, and Saavedra}}]{Ovgun:2018tua}
\bibinfo{author}{\bibfnamefont{A.}~\bibnamefont{\"Ovg\"un}},
  \bibinfo{author}{\bibfnamefont{I.}~\bibnamefont{Sakall\i{}}},
  \bibnamefont{and} \bibinfo{author}{\bibfnamefont{J.}~\bibnamefont{Saavedra}},
  \bibinfo{journal}{JCAP} \textbf{\bibinfo{volume}{10}}, \bibinfo{pages}{041}
  (\bibinfo{year}{2018}), \eprint{1807.00388}.

\bibitem[{\citenamefont{Ling et~al.}(2021)\citenamefont{Ling, Guo, Liu, Kuang,
  and Wang}}]{Ling:2021vgk}
\bibinfo{author}{\bibfnamefont{R.}~\bibnamefont{Ling}},
  \bibinfo{author}{\bibfnamefont{H.}~\bibnamefont{Guo}},
  \bibinfo{author}{\bibfnamefont{H.}~\bibnamefont{Liu}},
  \bibinfo{author}{\bibfnamefont{X.-M.} \bibnamefont{Kuang}}, \bibnamefont{and}
  \bibinfo{author}{\bibfnamefont{B.}~\bibnamefont{Wang}},
  \bibinfo{journal}{Phys. Rev. D} \textbf{\bibinfo{volume}{104}},
  \bibinfo{pages}{104003} (\bibinfo{year}{2021}), \eprint{2107.05171}.

\bibitem[{\citenamefont{Belhaj et~al.}(2021)\citenamefont{Belhaj, Belmahi,
  Benali, El~Hadri, El~Moumni, and Torrente-Lujan}}]{Belhaj:2020okh}
\bibinfo{author}{\bibfnamefont{A.}~\bibnamefont{Belhaj}},
  \bibinfo{author}{\bibfnamefont{H.}~\bibnamefont{Belmahi}},
  \bibinfo{author}{\bibfnamefont{M.}~\bibnamefont{Benali}},
  \bibinfo{author}{\bibfnamefont{W.}~\bibnamefont{El~Hadri}},
  \bibinfo{author}{\bibfnamefont{H.}~\bibnamefont{El~Moumni}},
  \bibnamefont{and}
  \bibinfo{author}{\bibfnamefont{E.}~\bibnamefont{Torrente-Lujan}},
  \bibinfo{journal}{Phys. Lett. B} \textbf{\bibinfo{volume}{812}},
  \bibinfo{pages}{136025} (\bibinfo{year}{2021}), \eprint{2008.13478}.

\bibitem[{\citenamefont{Abdikamalov et~al.}(2019)\citenamefont{Abdikamalov,
  Abdujabbarov, Ayzenberg, Malafarina, Bambi, and
  Ahmedov}}]{Abdikamalov:2019ztb}
\bibinfo{author}{\bibfnamefont{A.~B.} \bibnamefont{Abdikamalov}},
  \bibinfo{author}{\bibfnamefont{A.~A.} \bibnamefont{Abdujabbarov}},
  \bibinfo{author}{\bibfnamefont{D.}~\bibnamefont{Ayzenberg}},
  \bibinfo{author}{\bibfnamefont{D.}~\bibnamefont{Malafarina}},
  \bibinfo{author}{\bibfnamefont{C.}~\bibnamefont{Bambi}}, \bibnamefont{and}
  \bibinfo{author}{\bibfnamefont{B.}~\bibnamefont{Ahmedov}},
  \bibinfo{journal}{Phys. Rev. D} \textbf{\bibinfo{volume}{100}},
  \bibinfo{pages}{024014} (\bibinfo{year}{2019}), \eprint{1904.06207}.

\bibitem[{\citenamefont{Abdujabbarov et~al.}(2016)\citenamefont{Abdujabbarov,
  Juraev, Ahmedov, and Stuchl\'\i{}k}}]{Abdujabbarov:2016efm}
\bibinfo{author}{\bibfnamefont{A.}~\bibnamefont{Abdujabbarov}},
  \bibinfo{author}{\bibfnamefont{B.}~\bibnamefont{Juraev}},
  \bibinfo{author}{\bibfnamefont{B.}~\bibnamefont{Ahmedov}}, \bibnamefont{and}
  \bibinfo{author}{\bibfnamefont{Z.}~\bibnamefont{Stuchl\'\i{}k}},
  \bibinfo{journal}{Astrophys. Space Sci.} \textbf{\bibinfo{volume}{361}},
  \bibinfo{pages}{226} (\bibinfo{year}{2016}).

\bibitem[{\citenamefont{Atamurotov and Ahmedov}(2015)}]{Atamurotov:2015nra}
\bibinfo{author}{\bibfnamefont{F.}~\bibnamefont{Atamurotov}} \bibnamefont{and}
  \bibinfo{author}{\bibfnamefont{B.}~\bibnamefont{Ahmedov}},
  \bibinfo{journal}{Phys. Rev. D} \textbf{\bibinfo{volume}{92}},
  \bibinfo{pages}{084005} (\bibinfo{year}{2015}), \eprint{1507.08131}.

\bibitem[{\citenamefont{Papnoi et~al.}(2014)\citenamefont{Papnoi, Atamurotov,
  Ghosh, and Ahmedov}}]{Papnoi:2014aaa}
\bibinfo{author}{\bibfnamefont{U.}~\bibnamefont{Papnoi}},
  \bibinfo{author}{\bibfnamefont{F.}~\bibnamefont{Atamurotov}},
  \bibinfo{author}{\bibfnamefont{S.~G.} \bibnamefont{Ghosh}}, \bibnamefont{and}
  \bibinfo{author}{\bibfnamefont{B.}~\bibnamefont{Ahmedov}},
  \bibinfo{journal}{Phys. Rev. D} \textbf{\bibinfo{volume}{90}},
  \bibinfo{pages}{024073} (\bibinfo{year}{2014}), \eprint{1407.0834}.

\bibitem[{\citenamefont{Abdujabbarov et~al.}(2013)\citenamefont{Abdujabbarov,
  Atamurotov, Kucukakca, Ahmedov, and Camci}}]{Abdujabbarov:2012bn}
\bibinfo{author}{\bibfnamefont{A.}~\bibnamefont{Abdujabbarov}},
  \bibinfo{author}{\bibfnamefont{F.}~\bibnamefont{Atamurotov}},
  \bibinfo{author}{\bibfnamefont{Y.}~\bibnamefont{Kucukakca}},
  \bibinfo{author}{\bibfnamefont{B.}~\bibnamefont{Ahmedov}}, \bibnamefont{and}
  \bibinfo{author}{\bibfnamefont{U.}~\bibnamefont{Camci}},
  \bibinfo{journal}{Astrophys. Space Sci.} \textbf{\bibinfo{volume}{344}},
  \bibinfo{pages}{429} (\bibinfo{year}{2013}), \eprint{1212.4949}.

\bibitem[{\citenamefont{Atamurotov et~al.}(2013)\citenamefont{Atamurotov,
  Abdujabbarov, and Ahmedov}}]{Atamurotov:2013sca}
\bibinfo{author}{\bibfnamefont{F.}~\bibnamefont{Atamurotov}},
  \bibinfo{author}{\bibfnamefont{A.}~\bibnamefont{Abdujabbarov}},
  \bibnamefont{and} \bibinfo{author}{\bibfnamefont{B.}~\bibnamefont{Ahmedov}},
  \bibinfo{journal}{Phys. Rev. D} \textbf{\bibinfo{volume}{88}},
  \bibinfo{pages}{064004} (\bibinfo{year}{2013}).

\bibitem[{\citenamefont{Cunha and Herdeiro}(2018)}]{Cunha:2018acu}
\bibinfo{author}{\bibfnamefont{P.~V.~P.} \bibnamefont{Cunha}} \bibnamefont{and}
  \bibinfo{author}{\bibfnamefont{C.~A.~R.} \bibnamefont{Herdeiro}},
  \bibinfo{journal}{Gen. Rel. Grav.} \textbf{\bibinfo{volume}{50}},
  \bibinfo{pages}{42} (\bibinfo{year}{2018}), \eprint{1801.00860}.

\bibitem[{\citenamefont{Gralla et~al.}(2019)\citenamefont{Gralla, Holz, and
  Wald}}]{Gralla:2019xty}
\bibinfo{author}{\bibfnamefont{S.~E.} \bibnamefont{Gralla}},
  \bibinfo{author}{\bibfnamefont{D.~E.} \bibnamefont{Holz}}, \bibnamefont{and}
  \bibinfo{author}{\bibfnamefont{R.~M.} \bibnamefont{Wald}},
  \bibinfo{journal}{Phys. Rev. D} \textbf{\bibinfo{volume}{100}},
  \bibinfo{pages}{024018} (\bibinfo{year}{2019}), \eprint{1906.00873}.

\bibitem[{\citenamefont{Perlick et~al.}(2015)\citenamefont{Perlick, Tsupko, and
  Bisnovatyi-Kogan}}]{Perlick:2015vta}
\bibinfo{author}{\bibfnamefont{V.}~\bibnamefont{Perlick}},
  \bibinfo{author}{\bibfnamefont{O.~Y.} \bibnamefont{Tsupko}},
  \bibnamefont{and} \bibinfo{author}{\bibfnamefont{G.~S.}
  \bibnamefont{Bisnovatyi-Kogan}}, \bibinfo{journal}{Phys. Rev. D}
  \textbf{\bibinfo{volume}{92}}, \bibinfo{pages}{104031}
  (\bibinfo{year}{2015}), \eprint{1507.04217}.

\bibitem[{\citenamefont{Nedkova et~al.}(2013)\citenamefont{Nedkova, Tinchev,
  and Yazadjiev}}]{Nedkova:2013msa}
\bibinfo{author}{\bibfnamefont{P.~G.} \bibnamefont{Nedkova}},
  \bibinfo{author}{\bibfnamefont{V.~K.} \bibnamefont{Tinchev}},
  \bibnamefont{and} \bibinfo{author}{\bibfnamefont{S.~S.}
  \bibnamefont{Yazadjiev}}, \bibinfo{journal}{Phys. Rev. D}
  \textbf{\bibinfo{volume}{88}}, \bibinfo{pages}{124019}
  (\bibinfo{year}{2013}), \eprint{1307.7647}.

\bibitem[{\citenamefont{Li and Bambi}(2014)}]{Li:2013jra}
\bibinfo{author}{\bibfnamefont{Z.}~\bibnamefont{Li}} \bibnamefont{and}
  \bibinfo{author}{\bibfnamefont{C.}~\bibnamefont{Bambi}},
  \bibinfo{journal}{JCAP} \textbf{\bibinfo{volume}{01}}, \bibinfo{pages}{041}
  (\bibinfo{year}{2014}), \eprint{1309.1606}.

\bibitem[{\citenamefont{Cunha et~al.}(2017)\citenamefont{Cunha, Herdeiro,
  Kleihaus, Kunz, and Radu}}]{Cunha:2016wzk}
\bibinfo{author}{\bibfnamefont{P.~V.~P.} \bibnamefont{Cunha}},
  \bibinfo{author}{\bibfnamefont{C.~A.~R.} \bibnamefont{Herdeiro}},
  \bibinfo{author}{\bibfnamefont{B.}~\bibnamefont{Kleihaus}},
  \bibinfo{author}{\bibfnamefont{J.}~\bibnamefont{Kunz}}, \bibnamefont{and}
  \bibinfo{author}{\bibfnamefont{E.}~\bibnamefont{Radu}},
  \bibinfo{journal}{Phys. Lett. B} \textbf{\bibinfo{volume}{768}},
  \bibinfo{pages}{373} (\bibinfo{year}{2017}), \eprint{1701.00079}.

\bibitem[{\citenamefont{Johannsen et~al.}(2016)\citenamefont{Johannsen,
  Broderick, Plewa, Chatzopoulos, Doeleman, Eisenhauer, Fish, Genzel, Gerhard,
  and Johnson}}]{Johannsen:2015hib}
\bibinfo{author}{\bibfnamefont{T.}~\bibnamefont{Johannsen}},
  \bibinfo{author}{\bibfnamefont{A.~E.} \bibnamefont{Broderick}},
  \bibinfo{author}{\bibfnamefont{P.~M.} \bibnamefont{Plewa}},
  \bibinfo{author}{\bibfnamefont{S.}~\bibnamefont{Chatzopoulos}},
  \bibinfo{author}{\bibfnamefont{S.~S.} \bibnamefont{Doeleman}},
  \bibinfo{author}{\bibfnamefont{F.}~\bibnamefont{Eisenhauer}},
  \bibinfo{author}{\bibfnamefont{V.~L.} \bibnamefont{Fish}},
  \bibinfo{author}{\bibfnamefont{R.}~\bibnamefont{Genzel}},
  \bibinfo{author}{\bibfnamefont{O.}~\bibnamefont{Gerhard}}, \bibnamefont{and}
  \bibinfo{author}{\bibfnamefont{M.~D.} \bibnamefont{Johnson}},
  \bibinfo{journal}{Phys. Rev. Lett.} \textbf{\bibinfo{volume}{116}},
  \bibinfo{pages}{031101} (\bibinfo{year}{2016}), \eprint{1512.02640}.

\bibitem[{\citenamefont{Johannsen}(2016)}]{Johannsen:2015mdd}
\bibinfo{author}{\bibfnamefont{T.}~\bibnamefont{Johannsen}},
  \bibinfo{journal}{Class. Quant. Grav.} \textbf{\bibinfo{volume}{33}},
  \bibinfo{pages}{113001} (\bibinfo{year}{2016}), \eprint{1512.03818}.

\bibitem[{\citenamefont{Shaikh}(2019)}]{Shaikh:2019fpu}
\bibinfo{author}{\bibfnamefont{R.}~\bibnamefont{Shaikh}},
  \bibinfo{journal}{Phys. Rev. D} \textbf{\bibinfo{volume}{100}},
  \bibinfo{pages}{024028} (\bibinfo{year}{2019}), \eprint{1904.08322}.

\bibitem[{\citenamefont{Allahyari et~al.}(2020)\citenamefont{Allahyari,
  Khodadi, Vagnozzi, and Mota}}]{Allahyari:2019jqz}
\bibinfo{author}{\bibfnamefont{A.}~\bibnamefont{Allahyari}},
  \bibinfo{author}{\bibfnamefont{M.}~\bibnamefont{Khodadi}},
  \bibinfo{author}{\bibfnamefont{S.}~\bibnamefont{Vagnozzi}}, \bibnamefont{and}
  \bibinfo{author}{\bibfnamefont{D.~F.} \bibnamefont{Mota}},
  \bibinfo{journal}{JCAP} \textbf{\bibinfo{volume}{02}}, \bibinfo{pages}{003}
  (\bibinfo{year}{2020}), \eprint{1912.08231}.

\bibitem[{\citenamefont{Yumoto et~al.}(2012)\citenamefont{Yumoto, Nitta, Chiba,
  and Sugiyama}}]{Yumoto:2012kz}
\bibinfo{author}{\bibfnamefont{A.}~\bibnamefont{Yumoto}},
  \bibinfo{author}{\bibfnamefont{D.}~\bibnamefont{Nitta}},
  \bibinfo{author}{\bibfnamefont{T.}~\bibnamefont{Chiba}}, \bibnamefont{and}
  \bibinfo{author}{\bibfnamefont{N.}~\bibnamefont{Sugiyama}},
  \bibinfo{journal}{Phys. Rev. D} \textbf{\bibinfo{volume}{86}},
  \bibinfo{pages}{103001} (\bibinfo{year}{2012}), \eprint{1208.0635}.

\bibitem[{\citenamefont{Cunha et~al.}(2016{\natexlab{a}})\citenamefont{Cunha,
  Herdeiro, Radu, and Runarsson}}]{Cunha:2016bpi}
\bibinfo{author}{\bibfnamefont{P.~V.~P.} \bibnamefont{Cunha}},
  \bibinfo{author}{\bibfnamefont{C.~A.~R.} \bibnamefont{Herdeiro}},
  \bibinfo{author}{\bibfnamefont{E.}~\bibnamefont{Radu}}, \bibnamefont{and}
  \bibinfo{author}{\bibfnamefont{H.~F.} \bibnamefont{Runarsson}},
  \bibinfo{journal}{Int. J. Mod. Phys. D} \textbf{\bibinfo{volume}{25}},
  \bibinfo{pages}{1641021} (\bibinfo{year}{2016}{\natexlab{a}}),
  \eprint{1605.08293}.

\bibitem[{\citenamefont{Moffat}(2015)}]{Moffat:2015kva}
\bibinfo{author}{\bibfnamefont{J.~W.} \bibnamefont{Moffat}},
  \bibinfo{journal}{Eur. Phys. J. C} \textbf{\bibinfo{volume}{75}},
  \bibinfo{pages}{130} (\bibinfo{year}{2015}), \eprint{1502.01677}.

\bibitem[{\citenamefont{Cunha et~al.}(2016{\natexlab{b}})\citenamefont{Cunha,
  Grover, Herdeiro, Radu, Runarsson, and Wittig}}]{Cunha:2016bjh}
\bibinfo{author}{\bibfnamefont{P.~V.~P.} \bibnamefont{Cunha}},
  \bibinfo{author}{\bibfnamefont{J.}~\bibnamefont{Grover}},
  \bibinfo{author}{\bibfnamefont{C.}~\bibnamefont{Herdeiro}},
  \bibinfo{author}{\bibfnamefont{E.}~\bibnamefont{Radu}},
  \bibinfo{author}{\bibfnamefont{H.}~\bibnamefont{Runarsson}},
  \bibnamefont{and} \bibinfo{author}{\bibfnamefont{A.}~\bibnamefont{Wittig}},
  \bibinfo{journal}{Phys. Rev. D} \textbf{\bibinfo{volume}{94}},
  \bibinfo{pages}{104023} (\bibinfo{year}{2016}{\natexlab{b}}),
  \eprint{1609.01340}.

\bibitem[{\citenamefont{Zakharov}(2014)}]{Zakharov:2014lqa}
\bibinfo{author}{\bibfnamefont{A.~F.} \bibnamefont{Zakharov}},
  \bibinfo{journal}{Phys. Rev. D} \textbf{\bibinfo{volume}{90}},
  \bibinfo{pages}{062007} (\bibinfo{year}{2014}), \eprint{1407.7457}.

\bibitem[{\citenamefont{Tsukamoto}(2018)}]{Tsukamoto:2017fxq}
\bibinfo{author}{\bibfnamefont{N.}~\bibnamefont{Tsukamoto}},
  \bibinfo{journal}{Phys. Rev. D} \textbf{\bibinfo{volume}{97}},
  \bibinfo{pages}{064021} (\bibinfo{year}{2018}), \eprint{1708.07427}.

\bibitem[{\citenamefont{Hennigar et~al.}(2018)\citenamefont{Hennigar, Poshteh,
  and Mann}}]{Hennigar:2018hza}
\bibinfo{author}{\bibfnamefont{R.~A.} \bibnamefont{Hennigar}},
  \bibinfo{author}{\bibfnamefont{M.~B.~J.} \bibnamefont{Poshteh}},
  \bibnamefont{and} \bibinfo{author}{\bibfnamefont{R.~B.} \bibnamefont{Mann}},
  \bibinfo{journal}{Phys. Rev. D} \textbf{\bibinfo{volume}{97}},
  \bibinfo{pages}{064041} (\bibinfo{year}{2018}), \eprint{1801.03223}.

\bibitem[{\citenamefont{Chakhchi et~al.}(2022)\citenamefont{Chakhchi,
  El~Moumni, and Masmar}}]{Chakhchi:2022fls}
\bibinfo{author}{\bibfnamefont{L.}~\bibnamefont{Chakhchi}},
  \bibinfo{author}{\bibfnamefont{H.}~\bibnamefont{El~Moumni}},
  \bibnamefont{and} \bibinfo{author}{\bibfnamefont{K.}~\bibnamefont{Masmar}},
  \bibinfo{journal}{Phys. Rev. D} \textbf{\bibinfo{volume}{105}},
  \bibinfo{pages}{064031} (\bibinfo{year}{2022}).

\bibitem[{\citenamefont{Pantig and
  \"Ovg\"un}(2022{\natexlab{c}})}]{pantig2022testing}
\bibinfo{author}{\bibfnamefont{R.~C.} \bibnamefont{Pantig}} \bibnamefont{and}
  \bibinfo{author}{\bibfnamefont{A.}~\bibnamefont{\"Ovg\"un}}
  (\bibinfo{year}{2022}{\natexlab{c}}), \eprint{2206.02161},
  \urlprefix\url{https://arxiv.org/abs/2206.02161}.

\bibitem[{\citenamefont{Pantig and Rodulfo}(2020{\natexlab{b}})}]{Pantig2020b}
\bibinfo{author}{\bibfnamefont{R.~C.} \bibnamefont{Pantig}} \bibnamefont{and}
  \bibinfo{author}{\bibfnamefont{E.~T.} \bibnamefont{Rodulfo}},
  \bibinfo{journal}{Chinese J. Phys.} \textbf{\bibinfo{volume}{68}},
  \bibinfo{pages}{236} (\bibinfo{year}{2020}{\natexlab{b}}).

\bibitem[{\citenamefont{{Ingram} and {Done}}(2010)}]{Ingram2010MNRAS}
\bibinfo{author}{\bibfnamefont{A.}~\bibnamefont{{Ingram}}} \bibnamefont{and}
  \bibinfo{author}{\bibfnamefont{C.}~\bibnamefont{{Done}}},
  \bibinfo{journal}{Mon.Not.R.Astron.Soc} \textbf{\bibinfo{volume}{405}},
  \bibinfo{pages}{2447} (\bibinfo{year}{2010}), \eprint{0907.5485}.

\bibitem[{\citenamefont{{Schaab} and {Weigel}}(1999)}]{Schaab1999MNRAS}
\bibinfo{author}{\bibfnamefont{C.}~\bibnamefont{{Schaab}}} \bibnamefont{and}
  \bibinfo{author}{\bibfnamefont{M.~K.} \bibnamefont{{Weigel}}},
  \bibinfo{journal}{Mon. Not. R. Astron. Soc.} \textbf{\bibinfo{volume}{308}},
  \bibinfo{pages}{718} (\bibinfo{year}{1999}), \eprint{astro-ph/9904211}.

\bibitem[{\citenamefont{{T{\"o}r{\"o}k} and
  {Stuchl{\'\i}k}}(2005)}]{Torok2005AA}
\bibinfo{author}{\bibfnamefont{G.}~\bibnamefont{{T{\"o}r{\"o}k}}}
  \bibnamefont{and}
  \bibinfo{author}{\bibfnamefont{Z.}~\bibnamefont{{Stuchl{\'\i}k}}},
  \bibinfo{journal}{Astron. Astrophys} \textbf{\bibinfo{volume}{437}},
  \bibinfo{pages}{775} (\bibinfo{year}{2005}), \eprint{astro-ph/0502127}.

\bibitem[{\citenamefont{{Ingram} et~al.}(2016)\citenamefont{{Ingram}, {van der
  Klis}, {Middleton}, {Done}, {Altamirano}, {Heil}, {Uttley}, and
  {Axelsson}}}]{Ingram2016MNRAS}
\bibinfo{author}{\bibfnamefont{A.}~\bibnamefont{{Ingram}}},
  \bibinfo{author}{\bibfnamefont{M.}~\bibnamefont{{van der Klis}}},
  \bibinfo{author}{\bibfnamefont{M.}~\bibnamefont{{Middleton}}},
  \bibinfo{author}{\bibfnamefont{C.}~\bibnamefont{{Done}}},
  \bibinfo{author}{\bibfnamefont{D.}~\bibnamefont{{Altamirano}}},
  \bibinfo{author}{\bibfnamefont{L.}~\bibnamefont{{Heil}}},
  \bibinfo{author}{\bibfnamefont{P.}~\bibnamefont{{Uttley}}}, \bibnamefont{and}
  \bibinfo{author}{\bibfnamefont{M.}~\bibnamefont{{Axelsson}}},
  \bibinfo{journal}{Mon. Not. R. Astron. Soc.} \textbf{\bibinfo{volume}{461}},
  \bibinfo{pages}{1967} (\bibinfo{year}{2016}), \eprint{1607.02866}.

\bibitem[{\citenamefont{{Stuchl{\'\i}k}
  et~al.}(2013)\citenamefont{{Stuchl{\'\i}k}, {Kotrlov{\'a}}, and
  {T{\"o}r{\"o}k}}}]{Stuchlik2013AA}
\bibinfo{author}{\bibfnamefont{Z.}~\bibnamefont{{Stuchl{\'\i}k}}},
  \bibinfo{author}{\bibfnamefont{A.}~\bibnamefont{{Kotrlov{\'a}}}},
  \bibnamefont{and}
  \bibinfo{author}{\bibfnamefont{G.}~\bibnamefont{{T{\"o}r{\"o}k}}},
  \bibinfo{journal}{Astron. Astrophys} \textbf{\bibinfo{volume}{552}},
  \bibinfo{eid}{A10} (\bibinfo{year}{2013}), \eprint{1305.3552}.

\bibitem[{\citenamefont{{Stella} and {Vietri}}(1998)}]{Stella1998ApJL}
\bibinfo{author}{\bibfnamefont{L.}~\bibnamefont{{Stella}}} \bibnamefont{and}
  \bibinfo{author}{\bibfnamefont{M.}~\bibnamefont{{Vietri}}},
  \bibinfo{journal}{The Astrophys. Jour. Lett.} \textbf{\bibinfo{volume}{492}},
  \bibinfo{pages}{L59} (\bibinfo{year}{1998}), \eprint{astro-ph/9709085}.

\bibitem[{\citenamefont{{Rezzolla}
  et~al.}(2003{\natexlab{a}})\citenamefont{{Rezzolla}, {Yoshida}, {Maccarone},
  and {Zanotti}}}]{Rezzolla_qpo_03a}
\bibinfo{author}{\bibfnamefont{L.}~\bibnamefont{{Rezzolla}}},
  \bibinfo{author}{\bibfnamefont{S.}~\bibnamefont{{Yoshida}}},
  \bibinfo{author}{\bibfnamefont{T.~J.} \bibnamefont{{Maccarone}}},
  \bibnamefont{and}
  \bibinfo{author}{\bibfnamefont{O.}~\bibnamefont{{Zanotti}}},
  \bibinfo{journal}{Mon. Not. R. Astron. Soc.} \textbf{\bibinfo{volume}{344}},
  \bibinfo{pages}{L37} (\bibinfo{year}{2003}{\natexlab{a}}),
  \eprint{arXiv:astro-ph/0307487}.

\bibitem[{\citenamefont{{Rezzolla}
  et~al.}(2003{\natexlab{b}})\citenamefont{{Rezzolla}, {Yoshida}, and
  {Zanotti}}}]{Rezzolla_qpo_03b}
\bibinfo{author}{\bibfnamefont{L.}~\bibnamefont{{Rezzolla}}},
  \bibinfo{author}{\bibfnamefont{S.}~\bibnamefont{{Yoshida}}},
  \bibnamefont{and}
  \bibinfo{author}{\bibfnamefont{O.}~\bibnamefont{{Zanotti}}},
  \bibinfo{journal}{Mon. Not. R. Astron. Soc.} \textbf{\bibinfo{volume}{344}},
  \bibinfo{pages}{978} (\bibinfo{year}{2003}{\natexlab{b}}),
  \eprint{arXiv:astro-ph/0307488}.

\bibitem[{\citenamefont{{German{\`a}}}(2017)}]{Germana2017PhRvD}
\bibinfo{author}{\bibfnamefont{C.}~\bibnamefont{{German{\`a}}}},
  \bibinfo{journal}{Phys.Rev.D} \textbf{\bibinfo{volume}{96}},
  \bibinfo{eid}{103015} (\bibinfo{year}{2017}), \eprint{1711.01626}.

\bibitem[{\citenamefont{{T{\"o}r{\"o}k}
  et~al.}(2019)\citenamefont{{T{\"o}r{\"o}k}, {Goluchov{\'a}},
  {{\v{S}}r{\'a}mkov{\'a}}, {Urbanec}, and {Straub}}}]{Torok2019MNRAS}
\bibinfo{author}{\bibfnamefont{G.}~\bibnamefont{{T{\"o}r{\"o}k}}},
  \bibinfo{author}{\bibfnamefont{K.}~\bibnamefont{{Goluchov{\'a}}}},
  \bibinfo{author}{\bibfnamefont{E.}~\bibnamefont{{{\v{S}}r{\'a}mkov{\'a}}}},
  \bibinfo{author}{\bibfnamefont{M.}~\bibnamefont{{Urbanec}}},
  \bibnamefont{and} \bibinfo{author}{\bibfnamefont{O.}~\bibnamefont{{Straub}}},
  \bibinfo{journal}{Mon. Not. R. Astron. Soc.} \textbf{\bibinfo{volume}{488}},
  \bibinfo{pages}{3896} (\bibinfo{year}{2019}), \eprint{1907.05174}.

\bibitem[{\citenamefont{{Zdunik} et~al.}(2000)\citenamefont{{Zdunik},
  {Haensel}, {Gondek-Rosi{\'n}ska}, and {Gourgoulhon}}}]{Zdunik2000AA}
\bibinfo{author}{\bibfnamefont{J.~L.} \bibnamefont{{Zdunik}}},
  \bibinfo{author}{\bibfnamefont{P.}~\bibnamefont{{Haensel}}},
  \bibinfo{author}{\bibfnamefont{D.}~\bibnamefont{{Gondek-Rosi{\'n}ska}}},
  \bibnamefont{and}
  \bibinfo{author}{\bibfnamefont{E.}~\bibnamefont{{Gourgoulhon}}},
  \bibinfo{journal}{Astron.Astrophys.} \textbf{\bibinfo{volume}{356}},
  \bibinfo{pages}{612} (\bibinfo{year}{2000}), \eprint{astro-ph/0002394}.

\bibitem[{\citenamefont{{van der Klis}}(2000)}]{Klis2000ARAA}
\bibinfo{author}{\bibfnamefont{M.}~\bibnamefont{{van der Klis}}},
  \bibinfo{journal}{Annu. Rev. Astron. Astrophys}
  \textbf{\bibinfo{volume}{38}}, \bibinfo{pages}{717} (\bibinfo{year}{2000}),
  \eprint{astro-ph/0001167}.

\bibitem[{\citenamefont{Bambi}(2017)}]{Bambi17e}
\bibinfo{author}{\bibfnamefont{C.}~\bibnamefont{Bambi}},
  \emph{\bibinfo{title}{Black Holes: A Laboratory for Testing Strong Gravity}}
  (\bibinfo{publisher}{Springer, Singapore}, \bibinfo{year}{2017}).

\bibitem[{\citenamefont{{Stuchl{\'\i}k} and
  {Kolo{\v{s}}}}(2015)}]{Stuchlik2015MNRAS}
\bibinfo{author}{\bibfnamefont{Z.}~\bibnamefont{{Stuchl{\'\i}k}}}
  \bibnamefont{and}
  \bibinfo{author}{\bibfnamefont{M.}~\bibnamefont{{Kolo{\v{s}}}}},
  \bibinfo{journal}{Mon. Not. R. Astron. Soc.} \textbf{\bibinfo{volume}{451}},
  \bibinfo{pages}{2575} (\bibinfo{year}{2015}), \eprint{1603.07339}.

\bibitem[{\citenamefont{{Silbergleit} et~al.}(2001)\citenamefont{{Silbergleit},
  {Wagoner}, and {Ortega-Rodr{\'\i}guez}}}]{Silbergleit2001ApJ}
\bibinfo{author}{\bibfnamefont{A.~S.} \bibnamefont{{Silbergleit}}},
  \bibinfo{author}{\bibfnamefont{R.~V.} \bibnamefont{{Wagoner}}},
  \bibnamefont{and}
  \bibinfo{author}{\bibfnamefont{M.}~\bibnamefont{{Ortega-Rodr{\'\i}guez}}},
  \bibinfo{journal}{Astrophys. J} \textbf{\bibinfo{volume}{548}},
  \bibinfo{pages}{335} (\bibinfo{year}{2001}), \eprint{astro-ph/0004114}.

\bibitem[{\citenamefont{{Wagoner} et~al.}(2001)\citenamefont{{Wagoner},
  {Silbergleit}, and {Ortega-Rodr{\'\i}guez}}}]{Wagoner2001ApJL}
\bibinfo{author}{\bibfnamefont{R.~V.} \bibnamefont{{Wagoner}}},
  \bibinfo{author}{\bibfnamefont{A.~S.} \bibnamefont{{Silbergleit}}},
  \bibnamefont{and}
  \bibinfo{author}{\bibfnamefont{M.}~\bibnamefont{{Ortega-Rodr{\'\i}guez}}},
  \bibinfo{journal}{Astrophys.J.Lett} \textbf{\bibinfo{volume}{559}},
  \bibinfo{pages}{L25} (\bibinfo{year}{2001}), \eprint{astro-ph/0107168}.

\bibitem[{\citenamefont{{Rayimbaev} et~al.}(2022)\citenamefont{{Rayimbaev},
  {Majeed}, {Jamil}, {Jusufi}, and {Wang}}}]{Rayimbaev2022PDU}
\bibinfo{author}{\bibfnamefont{J.}~\bibnamefont{{Rayimbaev}}},
  \bibinfo{author}{\bibfnamefont{B.}~\bibnamefont{{Majeed}}},
  \bibinfo{author}{\bibfnamefont{M.}~\bibnamefont{{Jamil}}},
  \bibinfo{author}{\bibfnamefont{K.}~\bibnamefont{{Jusufi}}}, \bibnamefont{and}
  \bibinfo{author}{\bibfnamefont{A.}~\bibnamefont{{Wang}}},
  \bibinfo{journal}{Physics of the Dark Universe}
  \textbf{\bibinfo{volume}{35}}, \bibinfo{eid}{100930} (\bibinfo{year}{2022}),
  \eprint{2202.11509}.

\bibitem[{\citenamefont{{Stuchl{\'\i}k} and
  {Vrba}}(2021{\natexlab{a}})}]{Vrba2021Univ}
\bibinfo{author}{\bibfnamefont{Z.}~\bibnamefont{{Stuchl{\'\i}k}}}
  \bibnamefont{and} \bibinfo{author}{\bibfnamefont{J.}~\bibnamefont{{Vrba}}},
  \bibinfo{journal}{Universe} \textbf{\bibinfo{volume}{7}},
  \bibinfo{pages}{279} (\bibinfo{year}{2021}{\natexlab{a}}),
  \eprint{2108.09562}.

\bibitem[{\citenamefont{{Stuchl{\'\i}k} and
  {Vrba}}(2021{\natexlab{b}})}]{Vrba2021EPJP}
\bibinfo{author}{\bibfnamefont{Z.}~\bibnamefont{{Stuchl{\'\i}k}}}
  \bibnamefont{and} \bibinfo{author}{\bibfnamefont{J.}~\bibnamefont{{Vrba}}},
  \bibinfo{journal}{European Physical Journal Plus}
  \textbf{\bibinfo{volume}{136}}, \bibinfo{eid}{1127}
  (\bibinfo{year}{2021}{\natexlab{b}}), \eprint{2110.10569}.

\bibitem[{\citenamefont{{Stuchl{\'\i}k} and
  {Vrba}}(2021{\natexlab{c}})}]{Vrba2021JCAP}
\bibinfo{author}{\bibfnamefont{Z.}~\bibnamefont{{Stuchl{\'\i}k}}}
  \bibnamefont{and} \bibinfo{author}{\bibfnamefont{J.}~\bibnamefont{{Vrba}}},
  \bibinfo{journal}{J. Cosmol. Astropart. Phys}
  \textbf{\bibinfo{volume}{2021}}, \bibinfo{eid}{059}
  (\bibinfo{year}{2021}{\natexlab{c}}), \eprint{2110.07411}.

\bibitem[{\citenamefont{{Rayimbaev}
  et~al.}(2021{\natexlab{a}})\citenamefont{{Rayimbaev}, {Shaymatov}, and
  {Jamil}}}]{Rayimbaev2021EPJCQPO}
\bibinfo{author}{\bibfnamefont{J.}~\bibnamefont{{Rayimbaev}}},
  \bibinfo{author}{\bibfnamefont{S.}~\bibnamefont{{Shaymatov}}},
  \bibnamefont{and} \bibinfo{author}{\bibfnamefont{M.}~\bibnamefont{{Jamil}}},
  \bibinfo{journal}{European Physical Journal C} \textbf{\bibinfo{volume}{81}},
  \bibinfo{eid}{699} (\bibinfo{year}{2021}{\natexlab{a}}), \eprint{2107.13436}.

\bibitem[{\citenamefont{{Rayimbaev}
  et~al.}(2021{\natexlab{b}})\citenamefont{{Rayimbaev}, {Tadjimuratov},
  {Abdujabbarov}, {Ahmedov}, and {Khudoyberdieva}}}]{Rayimbaev2021GalaxQPO}
\bibinfo{author}{\bibfnamefont{J.}~\bibnamefont{{Rayimbaev}}},
  \bibinfo{author}{\bibfnamefont{P.}~\bibnamefont{{Tadjimuratov}}},
  \bibinfo{author}{\bibfnamefont{A.}~\bibnamefont{{Abdujabbarov}}},
  \bibinfo{author}{\bibfnamefont{B.}~\bibnamefont{{Ahmedov}}},
  \bibnamefont{and}
  \bibinfo{author}{\bibfnamefont{M.}~\bibnamefont{{Khudoyberdieva}}},
  \bibinfo{journal}{Galaxies} \textbf{\bibinfo{volume}{9}}, \bibinfo{pages}{75}
  (\bibinfo{year}{2021}{\natexlab{b}}), \eprint{2010.12863}.

\bibitem[{\citenamefont{{Rayimbaev}
  et~al.}(2021{\natexlab{c}})\citenamefont{{Rayimbaev}, {Abdujabbarov}, and
  {Wen-Biao}}}]{Rayimbaev2021PhRvDQPO}
\bibinfo{author}{\bibfnamefont{J.}~\bibnamefont{{Rayimbaev}}},
  \bibinfo{author}{\bibfnamefont{A.}~\bibnamefont{{Abdujabbarov}}},
  \bibnamefont{and}
  \bibinfo{author}{\bibfnamefont{H.}~\bibnamefont{{Wen-Biao}}},
  \bibinfo{journal}{Phys.Rev.D} \textbf{\bibinfo{volume}{103}},
  \bibinfo{eid}{104070} (\bibinfo{year}{2021}{\natexlab{c}}).

\bibitem[{\citenamefont{{Stuchl{\'\i}k}
  et~al.}(2011)\citenamefont{{Stuchl{\'\i}k}, {Kotrlov{\'a}}, and
  {T{\"o}r{\"o}k}}}]{Stuchlik2011AA}
\bibinfo{author}{\bibfnamefont{Z.}~\bibnamefont{{Stuchl{\'\i}k}}},
  \bibinfo{author}{\bibfnamefont{A.}~\bibnamefont{{Kotrlov{\'a}}}},
  \bibnamefont{and}
  \bibinfo{author}{\bibfnamefont{G.}~\bibnamefont{{T{\"o}r{\"o}k}}},
  \bibinfo{journal}{Astron.Astrophys.} \textbf{\bibinfo{volume}{525}},
  \bibinfo{eid}{A82} (\bibinfo{year}{2011}), \eprint{1010.1951}.

\bibitem[{\citenamefont{{T\"or\"ok, G.} et~al.}(2011)\citenamefont{{T\"or\"ok,
  G.}, {Kotrlov\'a, A.}, {Sr\'amkov\'a, E.}, and {Stuchl\'{\i}k,
  Z.}}}]{Torok2011AA}
\bibinfo{author}{\bibnamefont{{T\"or\"ok, G.}}},
  \bibinfo{author}{\bibnamefont{{Kotrlov\'a, A.}}},
  \bibinfo{author}{\bibnamefont{{Sr\'amkov\'a, E.}}}, \bibnamefont{and}
  \bibinfo{author}{\bibnamefont{{Stuchl\'{\i}k, Z.}}},
  \bibinfo{journal}{Astron.Astrophys.} \textbf{\bibinfo{volume}{531}},
  \bibinfo{pages}{A59} (\bibinfo{year}{2011}),
  \urlprefix\url{https://doi.org/10.1051/0004-6361/201015549}.

\bibitem[{\citenamefont{{Rayimbaev}
  et~al.}(2021{\natexlab{d}})\citenamefont{{Rayimbaev}, {Tadjimuratov},
  {Abdujabbarov}, {Ahmedov}, and {Khudoyberdieva}}}]{Rayimbaev2021Galax}
\bibinfo{author}{\bibfnamefont{J.}~\bibnamefont{{Rayimbaev}}},
  \bibinfo{author}{\bibfnamefont{P.}~\bibnamefont{{Tadjimuratov}}},
  \bibinfo{author}{\bibfnamefont{A.}~\bibnamefont{{Abdujabbarov}}},
  \bibinfo{author}{\bibfnamefont{B.}~\bibnamefont{{Ahmedov}}},
  \bibnamefont{and}
  \bibinfo{author}{\bibfnamefont{M.}~\bibnamefont{{Khudoyberdieva}}},
  \bibinfo{journal}{Galaxies} \textbf{\bibinfo{volume}{9}}, \bibinfo{pages}{75}
  (\bibinfo{year}{2021}{\natexlab{d}}), \eprint{2010.12863}.

\bibitem[{\citenamefont{Rayimbaev et~al.}(2020)\citenamefont{Rayimbaev,
  Abdujabbarov, Jamil, Ahmedov, and Han}}]{Rayimbaev2020PhysRevDRGI}
\bibinfo{author}{\bibfnamefont{J.}~\bibnamefont{Rayimbaev}},
  \bibinfo{author}{\bibfnamefont{A.}~\bibnamefont{Abdujabbarov}},
  \bibinfo{author}{\bibfnamefont{M.}~\bibnamefont{Jamil}},
  \bibinfo{author}{\bibfnamefont{B.}~\bibnamefont{Ahmedov}}, \bibnamefont{and}
  \bibinfo{author}{\bibfnamefont{W.-B.} \bibnamefont{Han}},
  \bibinfo{journal}{Phys. Rev. D} \textbf{\bibinfo{volume}{102}},
  \bibinfo{pages}{084016} (\bibinfo{year}{2020}),
  \urlprefix\url{https://link.aps.org/doi/10.1103/PhysRevD.102.084016}.

\bibitem[{\citenamefont{Bokhari et~al.}(2020)\citenamefont{Bokhari, Rayimbaev,
  and Ahmedov}}]{BokhariPhysRevD2020}
\bibinfo{author}{\bibfnamefont{A.~H.} \bibnamefont{Bokhari}},
  \bibinfo{author}{\bibfnamefont{J.}~\bibnamefont{Rayimbaev}},
  \bibnamefont{and} \bibinfo{author}{\bibfnamefont{B.}~\bibnamefont{Ahmedov}},
  \bibinfo{journal}{Phys. Rev. D} \textbf{\bibinfo{volume}{102}},
  \bibinfo{pages}{124078} (\bibinfo{year}{2020}),
  \urlprefix\url{https://link.aps.org/doi/10.1103/PhysRevD.102.124078}.

\bibitem[{\citenamefont{Juraeva et~al.}(2021)\citenamefont{Juraeva, Rayimbaev,
  Abdujabbarov, Ahmedov, and Palvanov}}]{JuraevaEPJC2021}
\bibinfo{author}{\bibfnamefont{N.}~\bibnamefont{Juraeva}},
  \bibinfo{author}{\bibfnamefont{J.}~\bibnamefont{Rayimbaev}},
  \bibinfo{author}{\bibfnamefont{A.}~\bibnamefont{Abdujabbarov}},
  \bibinfo{author}{\bibfnamefont{B.}~\bibnamefont{Ahmedov}}, \bibnamefont{and}
  \bibinfo{author}{\bibfnamefont{S.}~\bibnamefont{Palvanov}},
  \bibinfo{journal}{The European Physical Journal C}
  \textbf{\bibinfo{volume}{81}}, \bibinfo{pages}{124078}
  (\bibinfo{year}{2021}),
  \urlprefix\url{https://doi.org/10.1140/epjc/s10052-021-08876-5}.

\bibitem[{\citenamefont{Bardeen}(1968)}]{Bardeen68}
\bibinfo{author}{\bibfnamefont{J.}~\bibnamefont{Bardeen}}, in
  \emph{\bibinfo{booktitle}{Proceedings of GR5}}, edited by
  \bibinfo{editor}{\bibfnamefont{C.}~\bibnamefont{DeWitt}} \bibnamefont{and}
  \bibinfo{editor}{\bibfnamefont{B.}~\bibnamefont{DeWitt}},
  \bibinfo{organization}{Tbilisi, USSR} (\bibinfo{publisher}{Gordon and
  Breach}, \bibinfo{year}{1968}), p. \bibinfo{pages}{174}.

\bibitem[{\citenamefont{Stella}(2001)}]{Stella2001AIPC}
\bibinfo{author}{\bibfnamefont{L.}~\bibnamefont{Stella}}, \bibinfo{journal}{AIP
  Conf. Proc.} \textbf{\bibinfo{volume}{599}}, \bibinfo{pages}{365}
  (\bibinfo{year}{2001}), \eprint{astro-ph/0011395}.

\bibitem[{\citenamefont{Ingram and Motta}(2014)}]{Ingram2014MNRAS}
\bibinfo{author}{\bibfnamefont{A.}~\bibnamefont{Ingram}} \bibnamefont{and}
  \bibinfo{author}{\bibfnamefont{S.}~\bibnamefont{Motta}},
  \bibinfo{journal}{Mon. Not. Roy. Astron. Soc.}
  \textbf{\bibinfo{volume}{444}}, \bibinfo{pages}{2065} (\bibinfo{year}{2014}),
  \eprint{1408.0884}.

\bibitem[{\citenamefont{{Abramowicz} and
  {Klu{\'z}niak}}(2001)}]{Abramowicz2001AA}
\bibinfo{author}{\bibfnamefont{M.~A.} \bibnamefont{{Abramowicz}}}
  \bibnamefont{and}
  \bibinfo{author}{\bibfnamefont{W.}~\bibnamefont{{Klu{\'z}niak}}},
  \bibinfo{journal}{Astronomy and Astrophysics} \textbf{\bibinfo{volume}{374}},
  \bibinfo{pages}{L19} (\bibinfo{year}{2001}), \eprint{astro-ph/0105077}.

\bibitem[{\citenamefont{{Kato}}(2004)}]{Kato2004PASJ}
\bibinfo{author}{\bibfnamefont{S.}~\bibnamefont{{Kato}}},
  \bibinfo{journal}{Publications of the Astronomical Society of Japan}
  \textbf{\bibinfo{volume}{56}}, \bibinfo{pages}{905} (\bibinfo{year}{2004}),
  \eprint{astro-ph/0409051}.

\bibitem[{\citenamefont{{Kato}}(2008)}]{Kato2008PASJ}
\bibinfo{author}{\bibfnamefont{S.}~\bibnamefont{{Kato}}},
  \bibinfo{journal}{Publications of the Astronomical Society of Japan}
  \textbf{\bibinfo{volume}{60}}, \bibinfo{pages}{889} (\bibinfo{year}{2008}),
  \eprint{0803.2384}.

\bibitem[{\citenamefont{{Reid} et~al.}(2014)\citenamefont{{Reid}, {McClintock},
  {Steiner}, {Steeghs}, {Remillard}, {Dhawan}, and {Narayan}}}]{Reid2014ApJ}
\bibinfo{author}{\bibfnamefont{M.~J.} \bibnamefont{{Reid}}},
  \bibinfo{author}{\bibfnamefont{J.~E.} \bibnamefont{{McClintock}}},
  \bibinfo{author}{\bibfnamefont{J.~F.} \bibnamefont{{Steiner}}},
  \bibinfo{author}{\bibfnamefont{D.}~\bibnamefont{{Steeghs}}},
  \bibinfo{author}{\bibfnamefont{R.~A.} \bibnamefont{{Remillard}}},
  \bibinfo{author}{\bibfnamefont{V.}~\bibnamefont{{Dhawan}}}, \bibnamefont{and}
  \bibinfo{author}{\bibfnamefont{R.}~\bibnamefont{{Narayan}}},
  \bibinfo{journal}{Astrophys. J.} \textbf{\bibinfo{volume}{796}},
  \bibinfo{eid}{2} (\bibinfo{year}{2014}), \eprint{1409.2453}.

\bibitem[{\citenamefont{{Stuchl{\'\i}k} and
  {Kolo{\v{s}}}}(2016)}]{Stuchlik2016AA}
\bibinfo{author}{\bibfnamefont{Z.}~\bibnamefont{{Stuchl{\'\i}k}}}
  \bibnamefont{and}
  \bibinfo{author}{\bibfnamefont{M.}~\bibnamefont{{Kolo{\v{s}}}}},
  \bibinfo{journal}{Astronomy \& Astrophysics} \textbf{\bibinfo{volume}{586}},
  \bibinfo{eid}{A130} (\bibinfo{year}{2016}), \eprint{1603.07366}.

\bibitem[{\citenamefont{{Miller} et~al.}(2001)\citenamefont{{Miller},
  {Wijnands}, {Homan}, {Belloni}, {Pooley}, {Corbel}, {Kouveliotou}, {van der
  Klis}, and {Lewin}}}]{Miller2001ApJ}
\bibinfo{author}{\bibfnamefont{J.~M.} \bibnamefont{{Miller}}},
  \bibinfo{author}{\bibfnamefont{R.}~\bibnamefont{{Wijnands}}},
  \bibinfo{author}{\bibfnamefont{J.}~\bibnamefont{{Homan}}},
  \bibinfo{author}{\bibfnamefont{T.}~\bibnamefont{{Belloni}}},
  \bibinfo{author}{\bibfnamefont{D.}~\bibnamefont{{Pooley}}},
  \bibinfo{author}{\bibfnamefont{S.}~\bibnamefont{{Corbel}}},
  \bibinfo{author}{\bibfnamefont{C.}~\bibnamefont{{Kouveliotou}}},
  \bibinfo{author}{\bibfnamefont{M.}~\bibnamefont{{van der Klis}}},
  \bibnamefont{and} \bibinfo{author}{\bibfnamefont{W.~H.~G.}
  \bibnamefont{{Lewin}}}, \bibinfo{journal}{Astrophys. J.}
  \textbf{\bibinfo{volume}{563}}, \bibinfo{pages}{928} (\bibinfo{year}{2001}),
  \eprint{astro-ph/0105371}.

\bibitem[{\citenamefont{{Remillard} et~al.}(2006)\citenamefont{{Remillard},
  {McClintock}, {Orosz}, and {Levine}}}]{Remillard2006ApJ}
\bibinfo{author}{\bibfnamefont{R.~A.} \bibnamefont{{Remillard}}},
  \bibinfo{author}{\bibfnamefont{J.~E.} \bibnamefont{{McClintock}}},
  \bibinfo{author}{\bibfnamefont{J.~A.} \bibnamefont{{Orosz}}},
  \bibnamefont{and} \bibinfo{author}{\bibfnamefont{A.~M.}
  \bibnamefont{{Levine}}}, \bibinfo{journal}{Astrophys. J.}
  \textbf{\bibinfo{volume}{637}}, \bibinfo{pages}{1002} (\bibinfo{year}{2006}),
  \eprint{astro-ph/0407025}.

\bibitem[{\citenamefont{{Abramowicz} et~al.}(2004)\citenamefont{{Abramowicz},
  {Kluzniak}, {Stuchl{\'\i}k}, and {T{\"o}r{\"o}k}}}]{Abramowicz2004ragt}
\bibinfo{author}{\bibfnamefont{M.~A.} \bibnamefont{{Abramowicz}}},
  \bibinfo{author}{\bibfnamefont{W.}~\bibnamefont{{Kluzniak}}},
  \bibinfo{author}{\bibfnamefont{Z.}~\bibnamefont{{Stuchl{\'\i}k}}},
  \bibnamefont{and}
  \bibinfo{author}{\bibfnamefont{G.}~\bibnamefont{{T{\"o}r{\"o}k}}}, in
  \emph{\bibinfo{booktitle}{RAGtime 4/5: Workshops on black holes and neutron
  stars}} (\bibinfo{year}{2004}), pp. \bibinfo{pages}{1--23}.

\bibitem[{\citenamefont{Takizawa et~al.}(2020)\citenamefont{Takizawa, Ono, and
  Asada}}]{Takizawa:2020egm}
\bibinfo{author}{\bibfnamefont{K.}~\bibnamefont{Takizawa}},
  \bibinfo{author}{\bibfnamefont{T.}~\bibnamefont{Ono}}, \bibnamefont{and}
  \bibinfo{author}{\bibfnamefont{H.}~\bibnamefont{Asada}},
  \bibinfo{journal}{Phys. Rev. D} \textbf{\bibinfo{volume}{101}},
  \bibinfo{pages}{104032} (\bibinfo{year}{2020}), \eprint{2001.03290}.

\bibitem[{\citenamefont{Ono and Asada}(2019)}]{Ono:2019hkw}
\bibinfo{author}{\bibfnamefont{T.}~\bibnamefont{Ono}} \bibnamefont{and}
  \bibinfo{author}{\bibfnamefont{H.}~\bibnamefont{Asada}},
  \bibinfo{journal}{Universe} \textbf{\bibinfo{volume}{5}},
  \bibinfo{pages}{218} (\bibinfo{year}{2019}), \eprint{1906.02414}.

\bibitem[{\citenamefont{Nakashi et~al.}(2019)\citenamefont{Nakashi, Kobayashi,
  Ueda, and Saida}}]{Nakashi2019}
\bibinfo{author}{\bibfnamefont{K.}~\bibnamefont{Nakashi}},
  \bibinfo{author}{\bibfnamefont{S.}~\bibnamefont{Kobayashi}},
  \bibinfo{author}{\bibfnamefont{S.}~\bibnamefont{Ueda}}, \bibnamefont{and}
  \bibinfo{author}{\bibfnamefont{H.}~\bibnamefont{Saida}},
  \bibinfo{journal}{Prog. Theor. Exp. Phys. 2019}
  \textbf{\bibinfo{volume}{2019}}, \bibinfo{pages}{073E02}
  (\bibinfo{year}{2019}).

\bibitem[{\citenamefont{Arakida}(2013)}]{Arakida:2012ya}
\bibinfo{author}{\bibfnamefont{H.}~\bibnamefont{Arakida}},
  \bibinfo{journal}{Int. J. Theor. Phys.} \textbf{\bibinfo{volume}{52}},
  \bibinfo{pages}{1408} (\bibinfo{year}{2013}), \eprint{1212.6289}.

\bibitem[{\citenamefont{Islam}(1983)}]{Islam:1983rxp}
\bibinfo{author}{\bibfnamefont{J.~N.} \bibnamefont{Islam}},
  \bibinfo{journal}{Phys. Lett. A} \textbf{\bibinfo{volume}{97}},
  \bibinfo{pages}{239} (\bibinfo{year}{1983}).

\bibitem[{\citenamefont{Freire et~al.}(2001)\citenamefont{Freire, Bezerra, and
  Lima}}]{Freire:2001hmq}
\bibinfo{author}{\bibfnamefont{W.~H.~C.} \bibnamefont{Freire}},
  \bibinfo{author}{\bibfnamefont{V.~B.} \bibnamefont{Bezerra}},
  \bibnamefont{and} \bibinfo{author}{\bibfnamefont{J.~A.~S.}
  \bibnamefont{Lima}}, \bibinfo{journal}{Gen. Rel. Grav.}
  \textbf{\bibinfo{volume}{33}}, \bibinfo{pages}{1407} (\bibinfo{year}{2001}),
  \eprint{gr-qc/0201036}.

\bibitem[{\citenamefont{Will}(2018)}]{will}
\bibinfo{author}{\bibfnamefont{C.~M.} \bibnamefont{Will}},
  \bibinfo{journal}{Cambridge University Press}  (\bibinfo{year}{2018}),
  \eprint{10.1017/9781316338612}.

\bibitem[{\citenamefont{Perlick et~al.}(2018)\citenamefont{Perlick, Tsupko, and
  Bisnovatyi-Kogan}}]{Perlick:2018}
\bibinfo{author}{\bibfnamefont{V.}~\bibnamefont{Perlick}},
  \bibinfo{author}{\bibfnamefont{O.~Y.} \bibnamefont{Tsupko}},
  \bibnamefont{and} \bibinfo{author}{\bibfnamefont{G.~S.}
  \bibnamefont{Bisnovatyi-Kogan}}, \bibinfo{journal}{Phys. Rev. D}
  \textbf{\bibinfo{volume}{97}}, \bibinfo{pages}{104062}
  (\bibinfo{year}{2018}).

\bibitem[{\citenamefont{Aghanim et~al.}(2020)}]{Planck:2018vyg}
\bibinfo{author}{\bibfnamefont{N.}~\bibnamefont{Aghanim}} \bibnamefont{et~al.}
  (\bibinfo{collaboration}{Planck}), \bibinfo{journal}{Astron. Astrophys.}
  \textbf{\bibinfo{volume}{641}}, \bibinfo{pages}{A6} (\bibinfo{year}{2020}),
  \bibinfo{note}{[Erratum: Astron.Astrophys. 652, C4 (2021)]},
  \eprint{1807.06209}.

\end{thebibliography}

\end{document}